\input epsf
\documentstyle[12pt,preprint]{aastex}

\def\bib{\par\noindent\hangindent=3mm\hangafter=1}
\def\mv{M_V}

\def\mk{M_K}

\def\lsol{\mbox{L}_\odot}

\def\msol{\mbox{M}_\odot}

\def\pc{\mbox{pc}}
\def\pc3{\mbox{pc}^{-3}}

\def\mvol{\mbox{M}_\odot\, \mbox{pc}^{-3}}
\def\mlvol{(\log\mbox{M}_\odot)^{-1}\, \mbox{pc}^{-3}}
\def\mden{\mbox{M}_\odot^{-1}\, \mbox{pc}^{-3}}

\def\kms{\,\mbox{km.s}^{-1}}

\def\teff{T_{\rm eff}}
\def\mh{\rm [M/H]}
\def\nw{\mbox{nW\,m}^{-2}\mbox{sr}^{-1}}
\def\simgr{\,\hbox{\hbox{$ > $}\kern -0.8em \lower 1.0ex\hbox{$\sim$}}\,}
\def\simle{\,\hbox{\hbox{$ < $}\kern -0.8em \lower 1.0ex\hbox{$\sim$}}\,}

\begin{document}

\title{GALACTIC STELLAR AND SUBSTELLAR INITIAL MASS FUNCTION}

\author{ Gilles Chabrier}
\affil{Ecole Normale Sup\'erieure de Lyon,\\
Centre de Recherche Astrophysique de Lyon (UMR CNRS 5574),
 69364 Lyon Cedex 07, France\\
(chabrier@ens-lyon.fr)}

\vskip 10.cm

\indent{\bf Short title:} THE GALACTIC MASS FUNCTION
\vskip 1.cm

\newpage

\centerline{ABSTRACT}

We review recent determinations of the present day mass function (PDMF) and initial mass functions (IMF) in various
components of the Galaxy, disk, spheroid, young and globular clusters and in conditions characteristic of early star formation.
As a general feature, the IMF is found to depend weakly
on the environment and to be well described by a power-law form for $m\ga 1\,\msol$
and a lognormal form below, except possibly for early star formation conditions.
The disk IMF for single objects has a characteristic mass around
$m_c\sim 0.08\,\msol$ and a variance in logarithmic mass $\sigma \sim 0.7$, whereas the
IMF for multiple systems has $m_c\sim 0.2\,\msol$ and $\sigma \sim 0.6$.
The extension of the single MF into the brown dwarf regime is in good agreement with present estimates of
L- and T-dwarf densities and
yields a disk brown dwarf number density comparable to the stellar one $n_{BD}\sim n_\star \sim 0.1\,\pc3$.
The IMF of young clusters is found to be consistent with the disk field IMF, providing the
same correction for unresolved binaries, confirming the fact that young
star clusters and disk field stars represent the same stellar population.
Dynamical effects, yielding depletion of the lowest-mass objects, are found to become
consequential for ages $\ga$ 130 Myr.
The spheroid IMF relies on much less robust grounds. The large metallicity spread in
the local subdwarf photometric sample, in particular, remains puzzling. Recent observations
suggest that there is a continuous kinematic shear between the thick-disk population,
present in local samples, and the genuine spheroid one. This enables
us to derive only an upper limit for the spheroid mass density and IMF. Within
all the uncertainties, this latter is found to be similar to the one derived for
globular clusters, and is well represented also by a lognormal form
with a characteristic mass slightly larger than for the disk, $m_c\sim 0.2$-0.3 $\msol$,
excluding a significant population of brown dwarfs in globular clusters and in the spheroid. The IMF characteristic of early star formation at large redshift remains undetermined, but different observational constraints
suggest that it does not extend below $\sim 1\,\msol$. These results suggest a characteristic
mass for star formation which decreases with time, from conditions prevailing at large redshift
to conditions characteristic of the spheroid (or thick-disk), to present-day conditions. These conclusions,
however, remain speculative, given the large uncertainties in  the spheroid and early star IMF
determinations.

These IMFs allow a reasonably robust determination of the
Galactic present-day and initial stellar and brown dwarf contents. They also have important galactic implications beyond
the Milky Way in yielding more accurate mass-to-light ratio determinations.
The  mass-to-light ratios obtained with the disk and the spheroid IMF
yield values 1.8 to 1.4 smaller than for a Salpeter IMF, respectively, in agreement
with various  recent dynamical determinations.
This general IMF determination is examined in the context of star formation theory. None of the theories
based on a Jeans-type mechanism, where fragmentation is due only to gravity, can fulfill all the
observational constraints on star formation and predict
a large number of substellar objects.
On the other hand, recent numerical simulations of compressible turbulence, in particular in
super-Alfv\'enic conditions, seem to reproduce both qualitatively and quantitatively the
stellar and substellar IMF, and thus provide an appealing theoretical foundation.
In this picture, star formation is induced by the dissipation of large scale
turbulence to smaller scales, through radiative MHD shocks, producing filamentary
structures. These shocks produce local, non-equilibrium structures with large density contrasts,
which collapse eventually in gravitationally bound objects under the combined influence of turbulence
and gravity. The concept of a
single Jeans mass is replaced by a distribution of local
Jeans masses, representative of the lognormal probability density function
of the turbulent gas. Objects below the mean thermal Jeans mass still have
a possibility to collapse, although with a decreasing probability.


\newpage


{\bf 1 Introduction}
\vskip 2pt
\indent Historical Perspective
\dotfill\quad\ 5
\vskip 2pt
\indent Definitions
\dotfill\quad\ 6
\vskip 2pt
\indent Mass-magnitude relationships
\dotfill\quad\ 11
\vskip 5pt

{\bf 2 The Galactic disk mass function}
\vskip 2pt
\indent{2.1 The field mass function}
\dotfill\quad\ 13

\indent \indent {2.1.1 The disk stellar luminosity function}
\dotfill\quad\ 13
\vskip 2pt

\indent \indent {2.1.2 The disk stellar mass function}
\dotfill\quad\ 15
\vskip 2pt

\indent \indent {2.1.3 Correction for binaries. The disk system mass function}
\dotfill\quad\ 17
\vskip 2pt

\indent \indent {2.1.4 The disk brown dwarf mass function}
\dotfill\quad\ 19
\vskip 2pt
\vskip 2pt

\indent {2.2 The young cluster mass function}
\dotfill\quad\ 21
\vskip 2pt

\indent {2.3 The planetary mass function}
\dotfill\quad\ 26
\vskip 2pt
\vskip 5pt

{\bf 3 The Galactic spheroid mass function}
\dotfill\quad\ 27
\vskip 5pt

{\bf 4 The globular cluster mass function}
\dotfill\quad\ 33
\vskip 5pt

{\bf 5 The dark halo and early star mass function}
\dotfill\quad\ 36
\vskip 5pt

{\bf 6 Galactic mass budget. Mass-to-light ratios}
\dotfill\quad\ 41
\vskip 5pt

{\bf 7 The initial mass function theory}
\dotfill\quad\ 42
\vskip 5pt

{\bf 8 Summary and conclusion}
\dotfill\quad\ 53
\vskip 5pt

\newpage

\section{INTRODUCTION}

\subsection{Historical Perspective}

Since the pioneering paper of Salpeter (1955),
several fundamental reviews on the Galactic stellar mass function (MF) have been written
by, in particular, Schmidt (1959), Miller \& Scalo (1979, hereafter MS79), Scalo (1986). A shorter, more
recent discussion is given by Kroupa (2002). The determination of the stellar MF is
a cornerstone in astrophysics, for the stellar mass distribution determines the
evolution, surface brightness, chemical enrichment, and baryonic content of galaxies. 
Determinating whether this MF has been constant along the evolution of the universe, or varies with redshift, bears crucial consequences on the so-called cosmic star formation, i.e. on the universe light and matter evolution.
Furthermore, the knowledge of the MF  in our Galaxy yields the complete census of its
stellar and substellar population, and provides an essential diagnostic to understand the formation of star-like objects.
As emphasized by Scalo (1986), the stellar and substellar mass distribution is the
link between stellar and galactic evolution.

As too rarely stressed, there is no direct observational determination of the MF.
What is observed is the individual or integrated light of objects, i.e. the
luminosity function (LF) or the surface brightness. Transformation of this
observable quantity into the MF thus relies on theories of stellar evolution,
and more precisely, on the relationship between mass, age and light, i.e. mass-age-luminosity relations.

Until recently, only the LFs of giants and sun-like stars, i.e. objects
with mass $m\ga 1\,\msol$, were observed with enough precision to derive
stellar MFs. These latter were presented as power law approximations, $dN/dm \propto m^{-\alpha}$, as initially
suggested by Salpeter (1955), with an exponent close to the Salpeter value $\alpha=2.35$.
A departure from this monotonic behaviour, with a flattening of the MF below
$\sim 1\,\msol$, was first proposed by MS79, suggesting a
lognormal form.
The tremendous progress realized within the past few years from the observational side, from both ground-based and space-based surveys, now
probes the M-dwarf stellar distribution
down to the bottom of the main sequence (MS). Moreover, over a hundred brown dwarfs have now been discovered, both in the Galactic field and in young clusters,
down to a few Jupiter masses,
providing important constraints on the census of substellar objects in the Galaxy. Not mentioning the ongoing detection of planets orbiting stars outside
our solar system. All these recent discoveries show unambiguously that the stellar
MF extends well below the hydrogen- and probably the deuterium-burning limit, and urge a revised determination of the stellar and substellar census in the Galaxy, and thus of its MF.
In the meantime, the general theory of low-mass star and brown dwarf evolution has now
reached a mature state, allowing a reasonably robust description of the mechanical and
thermal properties of these complex objects, and of their observational signatures.

It is the purpose of this review to summarize these recent discoveries, to examine
which lessons from the Milky Way can be applied to a more general galactic and cosmological context, and to
determine our present knowledge of the Galactic MF and our present understanding of star formation.
Detailed discussions on the MF of massive stars ($m\ga 1\,\msol$) have been developed in the
remarkable reviews of MS79 and Scalo (1986) and we orient the reader to these
papers for these objects. The present review will focus on the low-mass part of
the MF in various regions of the Galaxy, and its extension into the substellar regime.
Low-mass stars (hereafter stars with mass $m< 1\,\msol$) have effective temperatures
$\teff \la 6000$ K, which implies eventually formation of molecules in their atmosphere. Below
$\teff\sim 4000$ K, their
spectral energy distribution strongly departs from a blackbody distribution, and peaks generally in the visible or near-infrared,
with yellow to red colors (see Chabrier \& Baraffe 2000 for a recent review).
These objects live for a Hubble time, or longer, and provide the overwhelming majority of the galactic stellar contents.

The organization of the paper is as follows: in \S1, we summarize the various definitions used in the
present review and we briefly summarize our present knowledge, and the remaining uncertainties, of 
the mass-magnitude relationship. Sections 2-5 are devoted respectively to the determination of the
Galactic field and young clusters, Galactic spheroid, globular cluster and dark halo and early star MFs. The 
stellar and substellar Galactic mass budget and the cosmological implications are presented in \S6.
Examination of our present understanding
of star formation is discussed in \S7, while \S8 is devoted to the conclusion.

\subsection{Definitions}

\subsubsection{Mass function}

The MF was originally defined by Salpeter (1995) as the number of stars $N$
 in a volume of space $V$
observed at a time $t$ 
per logarithmic mass interval
$d\log m$:

\begin{eqnarray}
\xi(\log m)={d(N/V)\over d\log m}={dn\over d\log m}
\label{imflog}
\end{eqnarray}

\noindent where $n=N/V$ is the stellar number-density, in pc$^{-3}$ in the following.

This definition was used also by MS79\footnote{Note that Miller \& Scalo use the stellar {\it surface}-density, in pc$^{-2}$, in their definition of the MF, to be divided by the respective
Galactic scale heights of the various stellar populations to get the volume density in the solar neighborhood.} and Scalo (1986). Since the formation
of star-like objects is now observed to take place over five orders of magnitude
in mass, from about 100 to 10$^{-3}\,\msol$, such a logarithmic definition of
the MF seems to be the most satisfactory representation of the mass distribution
in the Galaxy.
Conversely, Scalo (1986) defines the {\it mass spectrum} as the number density distribution
per mass interval $dn/dm$ with the obvious relation:

\begin{eqnarray}
\xi(m)={dn\over dm}={1\over m\,({\rm Ln} \,10)}\xi(\log m)
\label{imflin}
\end{eqnarray}

With these definitions, if the MF is approximated as a power law, the exponents
are usually denoted respectively $x$ and $\alpha$, with $\xi(\log m)\propto m^{-x}$ and $\xi(m)\propto m^{-\alpha}$, $x=\alpha-1$. The original
Salpeter value is $x=1.35$, $\alpha=2.35$.

Stars eventually evolve off the main sequence (MS) after a certain age, so that the
present-day MF (PDMF) of MS stars, which can be determined from the observed
present-day LF, differs from the so-called {\it initial}
mass function (IMF), i.e. the number of stars which were originally created per mass-interval in the Galaxy. Indeed, stars with masses above the minimum so-called
"turn-off" mass will have evolved as red giants and white dwarfs, or neutron stars or black holes as the end product of type-II supernovae explosion, depending on the initial MS mass. Here the minimum turn-off mass is defined as
the mass for which the age at which the star starts evolving off the MS on
the giant branch equals the age of the Galaxy (or the age of a given cluster). For an age $\tau_D\approx
10$ Gyr, about the age of the Galactic disk, this corresponds to a mass
$m_{TO}\approx 0.9\,\msol$ for solar metallicity. The determination of the PDMF
thus involves the stellar formation rate (SFR) $b(t)$, i.e. the number of stars
(more generically star-like objects) formed per time interval along galactic evolution. For this reason, the quantity to be considered to link the PDMF and IMF is the so-called {\it stellar creation function},
as introduced by MS79.

\subsubsection{Creation function}

The creation function $C(\log\,m,t)$ is defined as the number of stars per unit
volume formed in the mass range $[\log m,\log m+d\log m]$ during the time
interval $[t,t+dt]$. Given this definition, the total number of star-like objects per unit volume {\it ever formed} in the Galaxy reads:

\begin{eqnarray}
n_{tot}=\int_{\log (m_{inf})}^{\log (m_{sup})}\int_0^{\tau_G} C(\log m,t)\,d\log m\,dt
\label{ntot}
\end{eqnarray}

\noindent where $m_{inf}$ and $m_{sup}$ denote respectively the minimum and maximum mass for the formation of star-like objects, and $\tau_G$ denotes the age of the
Galaxy\footnote{Or the age of a given cluster if one wants to determine a cluster stellar content.}.

The creation function is related to the total birthrate $B(t)$, i.e. the total number-density of star-like objects ever formed per {\it unit time}, as:

\begin{eqnarray}
B(t)=\int_{\log (m_{inf})}^{\log (m_{sup})} C(\log m,t)\,d\log m
\label{rate}
\end{eqnarray}

Following MS79, we refer to the star-like formation rate (SFR), as the ratio of the
absolute birthrate at time $t$ over the average birthrate:

\begin{eqnarray}
b(t)={B(t)\over {1\over \tau_G}\int_0^{\tau_G} B(t)\,dt}
\label{sfr}
\end{eqnarray}

\noindent so that $\int_0^{\tau_G} b(t)\,dt=\tau_G$.

It is generally admitted that the creation function can be separated into the
product of a function of mass - the mass function - and a function of time -
the formation rate. The underlying physical hypothesis is that the MF, the issue of
the physical process which drives star formation per mass interval, does not depend on time. In fact, as will be illustrated later on, time may play a role
in this mechanism in determining some characteristic mass, but
without affecting the generic form of
$\xi(m)$. Under such a condition of separability of mass and time, the creation
function $C(m,t)$ can be rewritten:

\begin{eqnarray}
C(\log m,t)=\xi (\log m)\times{B(t)\over \int_0^{\tau_G} B(t)\,dt}
=\xi (\log m)\times { b(t) \over \tau_G}
\label{creation}
\end{eqnarray}

The IMF, i.e. the total number-density of star-like objects {\it ever formed} per unit log mass, thus reads:

\begin{eqnarray}
\xi (\log m)=\int_0^{\tau_G} C(\log m,t)\,dt
\label{xi}
\end{eqnarray}

From the definitions (\ref{ntot}) and (\ref{sfr}), the IMF and the SFR are related to
the total number-density of star-like objects ever formed in the Galaxy by:

\begin{eqnarray}
n_{tot}(t=\tau_G)={1\over \tau_G}\int_0^{\tau_G}b(t)dt \times \int_{\log (m_{inf})}^{\log (m_{sup})}  \xi(\log m)d\log m
=\int_{\log (m_{inf})}^{\log (m_{sup})}  \xi(\log m)d\log m
\label{ntot2}
\end{eqnarray}

As noted by MS79, all stars with MS lifetimes greater than the age of the
Galaxy are still on the MS. In that case the PDMF and the IMF are equivalent.
This holds for brown dwarfs (BD) too. Brown dwarfs have
unlimited lifetimes so that all BDs ever formed in the Galaxy still exist today,
regardless on when they were formed, and the BD PDMF
is the BD IMF.
For stars with MS lifetimes $\tau_{MS}$ less than the age of the Galaxy, only
those within the last $\tau_{MS}$ are observed today as MS stars. In that case
the PDMF $\phi_{MS}(\log \,m)$ and the IMF $\xi (\log m)$ are different, and - using the separability
condition for the creation function - obey the condition (MS79):

\begin{eqnarray}
\phi_{MS}(\log \,m)={\xi (\log m)\over \tau_G} \times \int_{\tau_G-\tau_{MS}}^{\tau_G} b(t)\,dt\,\,\,\,\,\,\,\,\,\, \tau_{MS}<\tau_G
\label{phi}
\end{eqnarray}

\subsubsection{Functional forms}

The most widely used functional form for the MF is the power law, as suggested
originally by Salpeter (1955):

\begin{eqnarray}
\xi(\log\,m)=A\,m^{-x}
\label{IMF1}
\end{eqnarray}

This form is believed to adequately describe the IMF of massive stars in our Galaxy,  $m\ga 1\,\msol$, with an
exponent $x\simeq 1.7$ (Scalo 1986, Table VII), for a standard fraction of observationally unresolved binaries (Kroupa 2001).
Uncertainties remain, however, in the exact value of the exponent,
and a Salpeter exponent, $x=1.3$, seems to be more consistent with the measured light
from high-z galaxies. This issue
will be discussed in \S6.
 The PDMF of massive stars has been calculated
by Scalo (1986, Table IV), and the corresponding volume-density distribution is adequately fitted by the following 3-segment power law:

\begin{eqnarray}
x=4.37,\,\,\,\,\,0\,\,\le \log\,m\le 0.54 \nonumber\\
\,x=3.53,\,\,\,\,\,0.54\le \log\,m\le 1.26\\
\,x=2.11,\,\,\,\,\,1.26\le \log\,m\le 1.80 \nonumber
\label{PDMF1}
\end{eqnarray}

\noindent with the respective normalization constants $A=0.044$ $\mlvol$, $A=0.015$ $\mlvol$ and $A=2.5\times 10^{-4}$ $\mlvol$. Note that this MF denotes the {\it volume}-density of objects $\pc3$
per interval of $\log\,m$, where the surface density has been transformed in
volume density using the Scalo (1986) mass-dependent scale heights.
The distinctive property of a power law MF is that it has no preferred mass
scale, as will be discussed in \S7.

The second widely used form is the normal, or Gaussian distribution, as suggested by MS79:

\begin{eqnarray}
\xi(\log \,m)={A\over \sqrt{2\pi} \sigma} \,\exp\{-{(\log \, m\,\,-\,\,\log \, m_c)^2\over 2\,\sigma^2}\}
\label{IMF2}
\end{eqnarray}

\noindent where $\log \,m_c$ and $\sigma^2=\langle (\log\, m-\langle \log \, m\rangle)^2\rangle$ denote respectively the mean mass and the
variance in $\log \,m$.

A third, more general form, is the so-called "generalized Rosin-Rammler" function:

\begin{eqnarray}
\xi(\log \,m)=A\,m^{-x}\, \exp\{-({ {B \over m}})^{\beta}\},\,\,\,\,\,\,\,\,\,\beta > 0
\label{IMF3}
\end{eqnarray}

This form recovers asymptotically
a power-law at large $m$, $\xi(\log \,m)_{m\rightarrow \infty}\rightarrow m^{-x}$ and resembles a log-normal form in the other limit, with
a peak value at $m_p=B({\beta \over x})^{(1/\beta)}$. The case $\beta =0$
corresponds to the power law.

In terms of statistical physics, the MF can be interpreted as a probability density function $p(m)=\xi(m)/n_{tot}$ and thus a probability density:

\begin{eqnarray}
\int_{m_{inf}}^{m_{sup}} p(m)\,dm=1
\label{proba}
\end{eqnarray}

\noindent with the probability for a star to have a mass $\in [m_{inf},m]$:

\begin{eqnarray}
P(m)=\int_{m_{inf}}^{m} p(x)\,dx={1\over n_{tot}}\int_{m_{inf}}^{m} \xi(x)\,dx
\label{proba2}
\end{eqnarray}

\subsection{Mass-magnitude relationships}

The only possible direct determination of a stellar mass is by use of Kepler's third law in a binary system, providing a long enough time basis to get the appropriate dynamical information. As shown below, the statistics of such a sample is largely insufficient to allow a reasonable estimate of the MF, but it certainly provides stringent
constraints for the models.
Therefore, as mentioned earlier, the only possible way
to determine a PDMF is by transformation of an observed LF $\Phi=dN/dM$, i.e. the number of stars $N$ per absolute magnitude interval $dM$, into a
MF. This involves the derivative of a mass-luminosity relationship, for a given age $\tau$, or
preferentially of a mass-magnitude relationship (MMR) which applies directly in the
observed magnitude and avoids the use of often ill-determined bolometric
corrections:

\begin{eqnarray}
 {dn\over dm}(m)_\tau=({dn\over dM_\lambda(m)})\times ({dm\over dM_\lambda(m)})^{-1}_\tau
\label{lfmf}
\end{eqnarray}

\noindent where $M_\lambda$ denotes the absolute magnitude in a given bandpass.
Another way to proceed is to attribute a mass to each star of the sample, which
avoids involving explicitly the derivative of the MMR. In practice,
both methods should yield similar results.

A first compilation of mass-luminosity data in the M-dwarf domain was published by Popper (1980), and was subsequently
extended by Henry \& McCarthy (1993), who
used speckle interferometry to
obtain MMRs in the V, J, H and K-bands. The determination of the V-magnitude was improved subsequently
with the HST (Henry et al. 1999), reducing appreciably the uncertainty in the $m$-$\mv$ relation.
This sample has been improved significantly recently by Delfosse et al. (2000) and S\'egransan et al. (2003b). Combining adaptative optic images and accurate radial velocities, these authors
determined the MMR of about 20 objects between
$\sim 0.6$ and $\sim 0.09$ $\msol$ in the aforementioned bands with mass accuracies of 0.2 to 5\%.

The MMRs derived from the Baraffe et al. (1998, hereafter BCAH98) models reproduce the Delfosse et
al. (2000) and S\'egransan et al. (2003b) data over the entire aforementioned mass range in the J, H and K-bands within less than 1-$\sigma$ (see Figure 3 of Delfosse et al. 2000).
The agreement is not as good in the V-band, with a systematic
offset of a few tenths of a magnitude between theory and observation below $\sim 0.3\,\msol$, $\mv \ga 12$. In term of mass determination for a given $\mv$, however, the effect remains modest, with a maximum 15\% error on the mass determined with the
theoretical MMR around magnitude $\mv\sim 12$-13 (Chabrier 2001, Fig. 1 and 2). Using the VLTI, S\'egransan et al. (2003a) obtained accurate radius measurements for some of the aforementioned
very-low-mass objects. The theoretical calculations (BACH98) agree within 1\% or less for $m\le 0.5\,\msol$.
No BD eclipsing binary has been detected yet so that theoretical masses cannot
be confronted directly with observation for BDs (assuming the age of the system is
well determined, a mandatory condition for BDs). However, the observation
of multiple systems, believed to be coeval, with dynamically determined total
mass, and with components extending well into the BD domain, provides stringent
constraints on the theory (White et al. 1999). The recent observations of different color-magnitude
diagrams of young clusters which extend down to Jupiter-like objects (Zapatero Osorio et al. 2000, B\'ejar et al. 2001, Lucas et al. 2001, Mart\'\i n et al. 2001) provide
an other precious constraint. The BCAH98 models successfully reproduce all these observations along one isochrone (see references above). One must remain cautious, however,
about the exact accuracy of the models in the substellar domain, given the lack of precise constraints.
A recent analysis by Dobbie et al. (2002a) seems to suggest an uncertainty of about $10^{-2}\,\msol=10\,M_{Jup}$ ($M_{Jup}\equiv $ Jupiter mass $\approx 10^{-3}\,\msol$) in this domain.
Although not drastic for the present MF determinations, this uncertainty illustrates the level of accuracy
to be reached eventually in the description of the physical properties of substellar objects, i.e. about a Jupiter mass,
one thousandth of a solar mass !

The lack of M-dwarf binary detection for the globular cluster or Galactic spheroid population prevents to test the MMR for metal-depleted stellar
abundances. A stringent observational constraint, however, stems for the observation of several cluster sequences down to the bottom of the MS with the
HST cameras, both in the optical and the infrared domains. Globular clusters provide a precious
test to confront the theory with observation since the metallicity, the distance
and the extinction are determined relatively accurately from the brightest stars,
which leaves no free parameter to adjust the theory to observation. The
Baraffe et al. (1997, hereafter BCAH97) models reproduce with excellent accuracy the
observed sequences, in both optical and infrared colors, of clusters with metallicity ranging from
$\mh=-2$ to $\mh=-1.0$ (Pulone et al. 1998, King et al. 1998, DeMarchi et al. 2000, BCAH97), whereas the agreement near the bottom of the
MS starts deteriorating in the optical for the more metal-rich clusters ($\mh>-1.0$)
(BCAH98, Bedin et al. 2001), as mentioned above for solar metallicity.
As discussed at length in BCAH98, this shortcoming in the theory stems very likely from a
 missing opacity source in the optical,
due to still uncomplete treatment of metal/molecular line absorption.
This shortcoming translates into theoretical sequences in optical colors which
lie about 0.5 mag blueward of the observed sequences in $\mv$ vs (V-I) for metallicities $\mh \ga -1.0$ (BCAH98, Figure 1 of Chabrier
et al. 2000).
This would affect appreciably the determination of the
absolute magnitude of an object from its observed color, as done for example in the determination of the LF from
photometric surveys, but, as mentioned earlier, it affects only marginally the determination of a mass from an observed $\mv$ magnitude. This uncertainty has
been quantified in
Chabrier (2001),
who shows that the mass inferred from such a theoretical $m$-$\mv$ relation is about 15 \%
smaller than the one determined observationally by Delfosse et al. (2000).
In terms of MF determination, this uncertainty remains within the observational
Poisson error bars.

All these successful confrontations of theory with observation for low-mass
stars and BDs, based on consistent evolutionary calculations between the emergent spectrum and the atmospheric
and interior thermal profiles, i.e. consistent magnitude-color-mass-age relations, give us reasonable confidence in the MMR derived from these models and
thus on the inferred MFs from observed LFs.

\section{ The Galactic disk mass function}

\subsection{ The field mass function}

\subsubsection{The disk stellar luminosity function}

The determination of the low-mass star LF is a difficult task. First, the observed sample may
be altered by various spurious effects: in a magnitude limited sample, the
so-called Malmquist bias leads to an overestimate of the local stellar space
density; for surveys that reach large distances, corrections due to the Galactic
structure should be taken into account; the exact completeness of the sample may
be harsh to determine with precision, in particular when estimating the BD
space density. Last but not least, all surveys have limited angular resolution
so that a certain fraction of the systems are unresolved.
An extensive discussion of these various biases has been given
in the papers by Kroupa, Tout \& Gilmore
(1991, 1993) and Kroupa (1995).

The LF requires the determination of the distance of the objects.
The easiest way to determine the distance is by knowing the trigonometric parallax, which implies a search within near distances from the Sun, typically $d\le 20$ pc for the bright part of the LF ($\mv<9.5$), which defines the
{\it Gliese Catalogue of Nearby Stars}, a few parsecs for the faint end. For the faintest M-dwarfs, the estimated completeness distance is $r_{compl}\approx 5$ pc (Henry et al. 1997). This
yields the so-called {\it nearby} LF $\Phi_{near}$. The main caveat of the nearby LF is that, given the
limited distance, it covers only a limited volume and thus a limited sample of objects. This yields
important statistical undeterminations at faint magnitudes ($M_V\ga 12$).
 On the other hand, a fundamental advantage of the nearby LF, besides
the reduced error on the distance, and thus on the magnitude, is the accurate identification of binary systems.
A V-band nearby LF can be derived by combining Hipparcos parallax data (ESA 1997), which
is essentially complete for $\mv < 12$ at r=10 pc, and the sample of nearby stars 
with ground-based parallaxes for $\mv>12$ to a completeness distance r=5.2 pc (Dahn et al. 1986).
Henry \& MCarthy (1990) used speckle interferometry to resolve companions of every known M-dwarf within 5 pc
and obtained the complete M-dwarf LF $\Phi_{near}$ in the H and K bands. Their sample recovers the Dahn et al. (1986)
one, plus one previously unresolved companion (GL 866B).
Reid and Gizis (1997) and more recently Reid, Gizis and Hawley (2002) extended this determination to a larger volume and determined a nearby LF based on a volume sample within about 8 pc.
It turns out that, down to the limit of completeness
claimed for this sample, $\mv \sim 14$, the two LFs agree reasonably
well.

Other determinations of the disk LF are based on photographic surveys, which extend to
$d\approx 100$-$200$ pc from the Sun, and thus encompass a significantly larger amount of stars.
However, photometric LFs $\Phi_{phot}$ suffer in general from significant Malmquist bias and, as mentioned above,
the low spatial resolution of photographic surveys does not allow the resolution of binaries at faint magnitudes.
An extensive analysis of the different nearby and photometric LFs has been conducted by Kroupa
(1995).
As shown by this author, most of the discrepancy between photometric and nearby LFs
for $M_V>12$ results from Malmquist bias and unresolved binary systems in the low-spatial resolution photographic surveys (see also Reid \& Gizis (1997)
for an alternative point of view). 
A recent determination of $\Phi_{phot}$ has been obtained with the HST
(Gould, Bahcall \& Flynn 1997, hereafter GBF97), which extends to an apparent magnitude $I\la 24$. The Malmquist bias is negligible because all stars down to $\sim 0.1\,\msol$ are seen through to the edge
of the thick disk. A major caveat of any photometric LF, however, is that the determination of the distance relies on a photometric determination from a color-magnitude diagram. The former analysis of the HST data (GBF97)
used for the entire sample a color-magnitude transformation characteristic of
stars with solar-abundances. A significant fraction of the sample probed by the
HST, however, lies at galactic scale height $\| z\| \ga 1$ kpc (Zheng et al. 2001, Fig. 2) and thus belongs to the thick-disk population, and is likely to have metal-depleted abundances
$-0.5 \la \mh \la 0$. Assuming solar metallicity for the entire sample results in an underestimate of the absolute
magnitude for a given color (the lower the metallicity, the fainter the
absolute magnitude for a given color), and thus an overestimate of the distance and an underestimate of the number density,
in particular near the faint end of the LF. An extension and a reanalysis of the HST sample, taking
into account a statistically weighted metallicity gradient along the Galactic scale height, and a related color-magnitude-metallicity
relationship, yields a revised
$\Phi_{HST}$ (Zheng et al. 2001), with indeed a larger number of M-dwarfs at faint absolute magnitude. However,
because of its limited angular resolution
($< 0.1^{\prime \prime}$),
the HST misses all the binaries, and the LF must be corrected from this
caveat to yield a single LF (\S2.1.3 below).

In practice, the determination of the MF from the LF implies the knowledge of
each star chemical composition, since the colors and the magnitude
depend on the metallicity. This metallicity spread translates into a spread in
the LF and in the MF. Analysis of the Hipparcos color-magnitude diagram, however,
indicates that $\sim 90\%$ of the thin-disk stars have abundances $\mh=0 \pm 0.2$ (Reid 1999), so the spread of metallicity in the solar neighborhood should not affect
significantly the derivation of the MF through the MMR.

The magnitude of an object, however, varies with age, so the determination
of its mass through a theoretical MMR necessitates the knowledge of
its age. The luminosity of MS stars above $m\sim 0.7\,\msol$ starts increasing substantially after
$\sim 10$ Gyr, about the age of the Galactic disk. On the other hand, objects below $m\sim 0.13\,\msol$ (for a solar
composition) take more than 5$\times 10^8$ years to reach the main sequence (see Table 1 of Chabrier \& Baraffe
2000). Therefore, for a constant SFR and a age of the disk $\tau_D=10$ Gyr, at most $\sim$ 5\% of the nearby stars
in the mass range $0.13\le m/\msol \le 0.7$ might still be contracting on the pre-main sequence,
an uncertainty well within the statistical observational ones. Within this mass-range,
the position of the star is fixed in the mass-luminosity diagram. Below 0.1 $\msol$,
and in particular in the BD domain, age variations must be taken into account,
within a given SFR, for a proper determination of the MF from the observed star or BD
counts (Chabrier 2002).

\subsubsection{The disk stellar mass function}

Recently, Chabrier (2001) has determined the Galactic disk M-dwarf MF from
the 5-pc and 8-pc $\Phi_{near}$. He has shown that, although still rising down to
the H-burning limit, the  IMF $\xi(\log \,m)$ starts shallowing with respect to the Scalo or Salpeter value below $\sim$1 $\msol$ and flattens off below $\sim$0.3 $\msol$,
as noted previously by MS79 and Kroupa et al. (1993). Combining the M-dwarf MF
with the Scalo (1986) power law for masses above 1 $\msol$, and
fulfilling the so-called continuity condition for stars with $\tau_{TO}\approx
\tau_G$ (MS79), i.e. $m\simeq 0.9\,\msol$, Chabrier (2001) showed that the
MF is well described over the entire stellar mass range, from about 100 $\msol$
to 0.1 $\msol$, by any of the functional forms mentioned in \S 1.2.3, i.e. a two-segment power law, a lognormal form or an exponential (Rosin-Rammler) form.
This analysis has been completed by Chabrier (2003), who has calculated the MF from the nearby LF $\Phi_{near}$
obtained both in the V-band (Dahn et al. 1986) and in the K-band (Henry \& McCarthy 1990).
Figure \ref{pasp_MF1_col.ps} displays
such a comparison. The conversion of the V-band LF into an MF was done using the Delfosse et al. (2000)
$m$-$\mv$ relation, which fits the observed data, whereas the BCAH98 $m$-$\mk$ relation was used to convert the K-band LF. We note the
very good agreement between the two determinations, which establishes the consistency of the two
observed samples and the validity of the mass-magnitude relationships.
The $\sim$1.5-$\sigma$ difference in the mass range
$\log m \sim$ -0.5 to -0.6, i.e. $\mv\sim 12$-13, reflects the remaining uncertainties either in the MMR or in the LF $\Phi_{near}$. The MF
derived from the new V-band LF or Reid et al. (2002), not displayed
in the figure, closely resembles the one derived from the Henry \& McCarthy (1990) K-band LF. 
The solid line displays an analytic form which gives a fairly good representation of the results.
The uncertainties in the MF are illustrated by the surrounding dashed lines. This analytic form for the
disk MF for single objects below 1 $\msol$, within these
uncertainties, is given by the following lognormal form (Chabrier 2003):

\begin{eqnarray}
\xi(\log \,m)_{m\le 1}=0.158^{+0.051}_{-0.046}\times \exp\Bigl\{-{(\log \, m\,\,-\,\,\log \, 0.079^{-0.016}_{+0.021})^2\over 2\times (0.69_{+0.05}^{-0.01})^2}\Bigr\}\,\,\,\mlvol
\label{IMFdisk}
\end{eqnarray}

The derivation of this MF from the Hipparcos and local sample provides the
normalization at 0.7 $\msol$: $({dn\over dm})_{0.7}=
3.8\,\times 10^{-2}\,\mden$, with at most a 5\% uncertainty.
Age effects above 0.7 $\msol$ are illustrated in the figure by the empty circles and empty squares which
display the MF obtained with the MMR for $t=10$ Gyr and 1 Gyr, respectively, whereas the
solid circles correspond to $t=5$ Gyr, the average age for the Galactic thin-disk\footnote{The vast majority
of stars in the Galactic mid-plane belong to the old-disk ($h\sim 300$ pc, $\tau \sim$ 5 Gyr), and about
20\% to the young disk ($h\sim 100$ pc, $\tau \sim$ 1 Gyr) (see e.g. Gilmore \& Reid 1983).}. As mentioned previously,
age effects become negligible below $m=0.7$ $\msol$.
The dotted line displays part of the 4-segment power-law MF derived by Kroupa (2002). This MF slightly
overestimates the M-dwarf density.

Note that eq.[\ref{IMFdisk}] yields the Scalo (1986) normalization for 5 Gyr at 1 $\msol$, $({dn\over dm})_{1.0}=
1.9\,\times 10^{-2}\,\mden$, which corresponds to a condition $\mv=4.72$ for $m=1.0\,\msol$ at 5 Gyr.
As shown by Scalo (1986) and illustrated in Figure \ref{pasp_MF1_col.ps}, 1 $\msol$ is about the limit for which
 the disk PDMF and IMF start to differ appreciably, so that only the $m>1\,\msol$
 power-law part of the MF will differ, depending whether the IMF ($x=1.3\pm 0.3$) or the PDMF
 ($x$ given by eq.[11]) is considered.
This yields
the global disk PDMF and IMF, as summarized in Table 1.
As mentioned earlier, substantial uncertainty remains in the value of $x$ at large masses for the IMF. As will be discussed
in \S7, observations of high-z galaxies, which constraint the fraction
of very massive stars to solar-type stars, seem to favor a Salpeter slope at large
masses and to exclude a Scalo slope.
For these reasons, we elected to take a Salpeter exponent for the IMF above 1 $\msol$, with the
aforementioned $\sim 0.3$ uncertainty.

As shown by Chabrier (2001) and Kroupa (2001), the low-mass part of the MF can be described by
a one- or two-segment power law. Extension of these segments into the BD domain, however,
severely overestimates the number of BD detections (Chabrier 2002) thus requiring an
other, different power-law segment in this regime. Such many-segment power law functions,
implying as many characteristic masses, seem difficult to reconcile with reasonable scenarios
of star formation. A power-law at large masses and a lognormal form in the low-mass range,
implying one single characteristic mass, on the other hand, seems to be supported by physically-motivated
scenarios, as will be discussed in \S7. For this reason, the PDMF and IMF displayed in Table 1
seem to be favored over other functional forms.

Figure \ref{pasp_MF1_col.ps} also displays the IMF derived from the observed J-band LF of the
Galactic bulge (Zoccali et al. 2000) down to its completeness limit ($J\sim 24$, $M_J\sim 9$,
$m\sim 0.15\,\msol$) with the BCAH98 MMR, for an age 10 Gyr and a solar
metallicity (empty triangles). The bulge IMF is normalized to the disk value at 0.7 $\msol$.
We note the remarkable agreement between the bulge and the disk IMF.

\subsubsection{Correction for binaries. The disk system mass function.}

As demonstrated by the detailed study of Kroupa, Tout \& Gilmore (1991, 1993),
correction for unresolved binaries can lead to a major revision of the IMF below
about 1 $\msol$. This study demonstrated that a large part of the disagreement between photometric and parallax surveys stemmed from unresolved binaries and Malmquist bias.
The disagreement between the MF inferred from the aforementioned nearby LF and from
the  photometric {\it HST} LF (GBF97), however, had remained a controversial,
unsettled issue until recently
(see e.g. Figure 1 of M\'era, Chabrier \& Schaeffer 1998).
As mentioned in \S2.1.1, an important source of the disagreement was the assumption by GBF97
of a solar composition for the stars in the HST sample for the photometric determination of the distance.
This yields
an underestimate of the faint part of the LF, as demonstrated in Figure 4 of
Zheng et al. (2001). The other source of discrepancy was the contribution from
unresolved binaries in the HST LF. Indeed,
as mentioned earlier, because of its angular resolution, the HST resolves only $\sim 1\%$ of the
multiple systems (see GBF97). Since about half of
the stars are known to be in multiple systems, the HST LF misses essentially
all companions. Recently, Chabrier (2003) 
has reconducted a detailed analysis of the bias due to unresolved binaries with the new (Zheng et al. 2001)
HST LF. This author has shown that the MF derived from the revised HST LF (i) is very similar to the
local so-called {\it system} MF, i.e. the MF derived from the local LF once the companions
of all identified multiple systems have been merged into unresolved systems, (ii) is consistent with the single
MF (eq.[\ref{IMFdisk}])  providing a binary fraction $X\approx 50\%$ among M-dwarfs, with the mass of both
the single objects and the companions originating from the same single MF (eq.[\ref{IMFdisk}]).
This multiplicity rate implies that about $\sim 30\%$ of M-dwarfs have a {\it stellar} (M-dwarf) companion, whereas
about $\sim 20\%$ have a {\it substellar} (BD) companion (Chabrier 2003), a result in agreement with present-day determinations
of the M-dwarf binary fraction in the solar neighborhood (Marchal et al. 2003) and of BD companions of M-dwarfs (Gizis et al. 2001, Close et al. 2003).
This system MF can be parametrized by the same type of lognormal form as the single MF (eq.[\ref{IMFdisk}]),
with the same normalization at 1 $\msol$, with the coefficients (Chabrier 2003):

\begin{eqnarray}
\xi(\log \,m)_{m\le 1}=0.086\times \exp\Bigl\{-{(\log \, m\,\,-\,\,\log \, 0.22)^2\over 2\times 0.57^2}\Bigr\}\,\,\,\mlvol
\label{IMFsys}
\end{eqnarray}

\noindent and is displayed by the long-dash line in Figure (\ref{pasp_MFbin_col.ps}).

These calculations show that the disk stellar IMF determined from
either the nearby geometric (parallax) LF or the HST photometric LF are consistent, and that the previous
source of disagreement was due to {\it two effects}, namely (i) incorrect color-magnitude determined parallaxes, due to the
fact that a substantial fraction of the HST M-dwarf sample belongs to the metal-depleted, thick-disk
population
and (ii) unresolved binaries. As discussed in Chabrier (2003),
these results yield a reinterpretation of the so-called "brown dwarf desert". This latter expresses the lack of BD companions
to solar-type stars (G-K), as compared with stellar or planetary companions, at separation of less than 5 AU (Marcy, Cochran \& Mayor 2000).
Indeed, proper motion data from Hipparcos have revealed that a significant fraction of low-mass companions in the
substellar regime have low-inclination and thus larger, possibly stellar masses (Marcy et al. 2000,
Halbwachs et al. 2000). Correction for this inclination yields a deficit of small-separation BD companions,
the so-called BD desert, suggesting that the MF of substellar companions to solar-type stars, at least at separations less than 5 AU, 
differs significantly from the one determined for the field.
The present calculations, however, show that this "desert" should be reinterpreted as a lack of high-mass ratio ($q=m_2/m_1 \la 0.1$)
systems, and does not preclude a substantial fraction of BDs as companions
of M-dwarfs or other BDs, as suggested by recent analysis (Marchal et al. 2003, Burgasser et al. 2003, Close et al. 2003).

\subsubsection{Disk brown dwarf mass function}

As shown in \S2.1.2, the IMF (eq.[\ref{IMFdisk}]) gives a good representation of the
stellar regime in the disk down to $\log \, m\sim -0.9$ ($m\sim 0.12\,\msol$),
where all objects have reached the MS. This IMF, which closely resembles the IMF2 derived in Chabrier (2001),
gives also a good description of the star counts in the deep field of
the ESO Imaging Survey (EIS) (Groenewegen et al. 2002), better than the power-law
forms of Kroupa (2001) or IMF1 of Chabrier (2001). It has been shown also
to agree fairly well with the L-dwarf and T-dwarf BD detections of various
field surveys (Chabrier 2002). Figure \ref{pasp_LFcounts.ps} displays the predicted BD
luminosity functions (BDLF) in the K-magnitude and in terms of fundamental
parameters ($T_{eff},\, L/L_\odot$) from the bottom of the MS
over the entire BD domain. 
These BDLFs were obtained from Monte Carlo simulations, with mass, age and
distance probability distributions as described in Chabrier (2002).
Only the case of a constant SFR has been considered presently.
The various dashed and dotted lines display the relative contributions of
BDs ($m\le 0.072\,\msol$ for solar metallicity, Chabrier \& Baraffe 1997), T-dwarfs, identified as faint objects with
(J-H)$<0.5$, (H-K)$<0.5$ (Kirkpatrick et al. 2000) and objects below the deuterium-burning limit ($m\le 0.012 \,\msol$).
The predictions are compared with various available data, namely the
nearby K-band LF (Henry \& McCarthy 1990), converted into a bolometric
LF on the bottom panel with the M-dwarf bolometric corrections of Tinney et al. (1993) and Leggett
et al. (1996), the L-dwarf density estimate of Gizis et al. (2000), and
the L-dwarf and T-dwarf estimated densities of Kirkpatrick et al. (1999, 2000)
and Burgasser (2001). It is important to mention that the $V_{max}$
and thus the explored volume and density determinations
for BD surveys are a very delicate task, affected by numerous uncertainties 
(see Burgasser 2001). A $\sim$ 1 mag uncertainty in the maximum limit
of detection translates into a factor of $\sim 4$ in the $V_{max}$ and thus in
the estimated density $\Phi=\Sigma V_{max}^{-1}$. Not mentioning
difficult completeness corrections for such surveys. Furthermore, the observational $T_{eff}$ and bolometric
correction determinations remain 
presently ill-determined for BDs.
On the other hand, theoretical models of BD cooling, although now
in a mature state, are still far from including all complex processes such
as dust sedimentation, cloud diffusion, or non-equilibrium chemistry.
For all these reasons, the present results should be considered with
caution. The BDLFs calculated with the IMF (eq.[\ref{IMFdisk}]) yield a
very good agreement with the determinations of Kirkpatrick et al. (1999, 2000)
but seem to overestimate by a factor of about 3 the density of L-dwarfs
obtained by Gizis et al. (2000) and the number of bright T-dwarfs observed
by Burgasser (2001).
The decreasing number of L-dwarfs in the Kirkpatrick et al. (1999) survey at bright magnitudes
is due to their color selection ($J$-$K$ $>$ 1.3).
Given all the aforementioned uncertainties, the comparison
between the observed and predicted LFs can be considered as satisfactory.
This assesses the validity of the present disk IMF determination in the BD regime.

The factor $\sim 3$ overestimate of the predicted LF, if confirmed, might stem from various
plausible explanations. First, this might indicate too high
a normalization of the IMF near the bottom of the MS, due to the presence of
hot BDs, misidentified as MS very-low-mass stars, in
the faintest bins of the nearby LF. However, as seen from the top panel
of Figure \ref{pasp_LFcounts.ps}, the contribution of young BDs to
the local LF is zero for $\mk \le 9$, which corresponds to a $\sim$0.12 $\msol$ MS M dwarf, and thus does not
affect the MF normalization at this mass.
An alternative, similar explanation would be the presence of a statistically significant number of very-low-mass stars younger
than 10$^8$ yr, still contracting on the PMS, in the local sample. This implies a small scale height for these objects.
Indeed, for a constant SFR and a young-disk age $\tau\simeq$ 1 Gyr, the probability to find an object with $t<10^8$ yr
is $\sim$10\%, for a homogeneous sample.
Only redoing the same observations in a few hundred million years could help resolving this issue, a rather challenging task !
A second possible explanation could be unresolved BD binaries.
The dotted line in the bottom panel of Figure \ref{pasp_LFcounts.ps} 
displays the LF obtained with the IMF (eq. [\ref{IMFsys}]), illustrating the effect due to
BD unresolved systems.
The effect of unresolved binaries on the BDLF is much more dramatic than
on the stellar LF. This stems from the much steeper
mass-magnitude relationship at a given age in the BD regime.
At 1 Gyr, a factor of 2 in mass corresponds 
to about $\sim$ 2 mag difference in the stellar regime,
against $\sim$ 4 mag or more in the BD regime.
On the other hand, remember that the difference between the single (eq.[\ref{IMFdisk}]) and
system (eq.[\ref{IMFsys}]) MFs assumes a binary correction $X\approx 50\%$ among {\it stellar} objects.
Extrapolating these corrections into the BD domain assumes that the binary rate in star formation does not
depend on the mass of the primary. If the present discrepancy between theory and observation is confirmed, it might indicate
that this frequency is significantly smaller in the BD regime (with $X\la 20\%$) due for example to the fact
that very-low-mass systems cannot form with large mass ratio ($q=m_2/m_1<< 1$) or with large orbital separations (see e.g. Burgasser et al. 2003\footnote{Note that the BD binaries observed by Burgasser et al. (2003) have an orbital separation $>1$ A.U., and do not include BD systems with smaller separation such as PPL15}). Again, long time basis observations are mandatory
to nail down this issue.
A third, appealing explanation for the factor 3 discrepancy might be substantial incompleteness of the present BD surveys
resulting from selection effects.
Salim et al. (2003) estimate that some 40\% of bright L-dwarfs are missed
because they lie close to the Galactic plane, a region avoided by most
searches. This correction factor would bring the present theoretical predictions
in perfect agreement with the observational BD census.
Finally, the present factor $\sim 3$ disagreement between the predicted
and observationally-derived counts might just reflect the remaining imperfections in
BD cooling models.

It is interesting to note the bimodal form of the stellar+BD LFs.
The stellar LF peaks near the bottom of the MF, because of the rising IMF. The LF then decreases severely, because of the
steepness of the MMR below $\sim 0.2\,\msol$, where a very small mass
range translates into a large magnitude interval, a consequence of
the decreasing nuclear support to halt contraction (Chabrier \& Baraffe
1997, 2000). The brightest, i.e. youngest and most massive BDs, start
contributing near $\log (L/\lsol) \simeq -3$ but the bulk of the BD population
dominates the LF only $\sim$1.5 mag fainter. The BDLF thus rises again,
a direct consequence of the cumulative effects of the increasing number of BDs and of BD cooling.
The decreasing IMF eventually yields decreasing BD densities for
$\log (L/\lsol) \la -6$, $\teff \la 400$ K.
The observational confirmation of the dip in the BDLF near $\log (L/\lsol) \simeq -4$, $\mk \simeq 13$ would be an interesting confirmation of the present general theory (IMF and BD cooling).
The dotted line in the middle panel of Figure (\ref{pasp_LFcounts.ps}) displays the results obtained with a power-law
MF $\xi(\log m)=m^x$ with $x=0$ ($\alpha =1$) for the same normalization as IMF (eq.[\ref{IMFdisk}]) at 0.1 $\msol$.
As already noted by Chabrier (2002), such an MF yields very similar results in the L-dwarf and hot T-dwarf range,
but predicts more cool T-dwarfs. Note that the same power-law MF ($x=0$), but with a normalization $\xi(\log m=-1)=0.08\,\mlvol$,
as in Reid et al. (1999), instead of 0.156 $\mlvol$ (eq.[\ref{IMFdisk}]), would bring predicted and observationally-derived LFs in very good agreement. Such
a low normalization near the bottom of the MS, however, is likely to be due to incompleteness of the 8-pc sample at faint magnitudes, as can be seen easily
from a comparison of the Reid \& Gizis (1997) revised K-band LF and the Henry \& McCarthy (1990) one
(displayed at the top of Figure \ref{pasp_LFcounts.ps}) (see also Chabrier 2001, Fig. 2).

\subsection{ The young cluster mass function}

In principle, star forming regions (age $\la 1 Myr$, e.g. $\rho$-Oph, IC348, Trapezium,
Taurus-Auriga, Chameleon I, Serpens) or young clusters (age $\sim$ 10-200 Myr, e.g. Pleiades) are particularly favorable targets to determine the very initial stellar
mass function. Indeed, 
(i) all objects in the cluster are likely to be coeval within a limited
range, except possibly in star forming regions, where the spread in ages for
cluster members can be comparable with the age of the cluster, (ii) young objects are brighter for a given mass
which makes more easy the detection of very-low-mass objects, (iii) young clusters are less dynamically evolved
than older open clusters and thus subtend a more compact region of the sky, yielding smaller foreground and
background contamination.
In practice, however, the determination of the IMF is hampered by several difficulties,
both from the observational and theoretical sides. First of all, membership of the objects to the
cluster must be assessed unambiguously, which implies either accurate
spectroscopic observations (lithium absorption, H$_\alpha$ emission, mm-excess,...)
for the regions of star formation or accurate proper motion measurements for the young clusters.
Second of all, extinction and differential reddening caused by the surrounding dust
 in star forming regions
modifies both the intrinsic magnitude and the colors of each individual object, preventing  direct mass-magnitude-color determinations
and making photometric determinations very uncertain, not mentioning
the amount of accretion which varies significantly from an object to an other (see e.g. Comer\'on et al. 2003). Moreover, the near-IR excess of embedded young clusters associated with hot circumstellar dusty disk complicates significantly the
interpretation of near-IR luminosity into {\it stellar} luminosity functions.
Third of all,  {\it some} dynamical evaporation may have taken place, rejecting low-mass objects to the
periphery of the cluster, where contamination from field stars become important. This
holds even for very young clusters ($< 1$ Myr) which contain O stars, like e.g. the Orion Nebula cluster (Kroupa, Aarseth \& Hurley 2001).
Finally, there is presently no appropriate effective temperature calibration  for gravities characteristic of PMS M-dwarfs,
yielding people to rely on empirical $\teff$-spectral type ($Sp$) determinations, as discussed below.

From the theoretical point of view, accurate models must include gravity effects, which, for young objects,
affect both the spectrum and the evolution (Baraffe et al. 2002). As shown by these authors, {\it no} theoretical model
is presently reliable for ages younger than $\sim 10^6$ yr. At such young ages, the evolution is severely
affected by several uncertainties, like e.g. the unknown convection efficiency (and thus mixing length parameter), the accretion rate, the
deuterium abundance, not mentioning the fact that at these ages the models are affected by the (arbitrary) initial conditions. As shown by Baraffe et al. (2002), the
evolution along the contracting PMS phase for $t \la 10^6$ yr depends not only on
the (unknown) efficiency of convection but also, for the coolest objects, on the
formation of molecular hydrogen H$_2$ in the atmosphere. Both effects affect
significantly the evolution. Therefore, assuming a constant $\teff$ evolution for a given mass in a
HR diagram for young, very-low-mass objects, as done sometimes in the literature,
may lead to inaccurate mass determinations and the inferred IMFs
must be considered with great caution.
In fact, 3D calculations
are necessary to determine accurately the entropy profile of objects in the initial accreting, gravitational contracting phase,
for 1D collapse calculations yield erroneous results (Hartmann et al. 1997, Hartmann 2003, Baraffe et al. 2003). No such consistent calculation and thus no reliable
temperature and mass calibration exists today for low-mass PMS stars.
Only for ages $t\ga 10^6$ yr, do these uncertainties disappear, or at least become less important, and can
reasonably reliable PMS models be calculated (Baraffe et al. 2002).
Finally, as pointed out by Luhman et al. (2000), what
is really observed in star forming clusters is not the IMF but the creation function (see \S1.2.2). The same underlying IMF convolved with different age
distributions will yield different LFs, a result which can be misinterpreted as originating from different IMFs.
Conversely, assuming a single, median age for objects in star-forming regions,
where the typical age spread can reach a few Myr, yields an IMF
of limited validity, given the
strong age-dependence of the mass-luminosity relationship for PMS and young stars.
Without an independent estimate of the age distribution of the cluster members, the creation function, and thus the IMF can not really
be determined. Not mentioning the fact that no one knows whether the star forming process in the star forming regions or very young clusters is finished or is still going on, and thus whether
the mass function is really the {\it initial} mass function.
For all these theoretical and observational
reasons, the exact determination of the IMF of star forming regions or very young ($\la 10^6$ yr) clusters remains presently speculative
 and IMF determinations claimed so far in the literature are of limited significance. Only general features,
such as ratios of substellar over stellar objects, can be considered as reasonably reliable indicators.
On the other hand, star forming regions are certainly very useful
testbeds to study the various processes of star formation (accretion, multiplicity,  collisions, rotation,...).

As mentioned above, no accurate $\teff$ calibration exists today for PMS low-mass stars.
An interesting, although empirical method, however,  has
been suggested by Luhman (1999), based on the analysis of the GG-tau system by White et al. (1999), to calibrate the effective temperature from the spectral type of 
young  low-mass stars. Since such objects have gravities between
M-giants and M-dwarfs, Luhman (1999) derived a $\teff$-Sp relation intermediate between those of giants and dwarfs,
based on the 6500 to 9000 $\AA$ observed spectra.
When using this relationship, the four components of the quadruple system
GG-tau, which extend from 1.2 $\msol$ to about $0.03\,\msol$, lie on a common isochrone of the BCAH98 models, for the correct, dynamically determined mass of the system.
Luhman (1999) further showed that the spectra of PMS stars in IC 348 are better fitted by an average of dwarf and giant spectra
of the same spectral type, and the combination of BCAH98 isochrones and Luhman's (1999) $\teff$-scale
provides the best fit to the IC 348 cluster locus (Luhman 1999, Najita et al. 2000). However, one
must remains cautious with such an agreement, which could
happen to be coincidental, and with the aforementioned empirical $\teff$-$Sp$ relation. As noted by Najita et al. (2000),
this latter does not apply at cool temperatures
in spectral regions shaped by water band absorption in the infrared, because of the dissociation of water due to backwarming effect, which yields eventually
a {\it cooler} temperature scale for M-dwarf PMS gravities below 3000 K.
Moreover, this intermediate $\teff$-$Sp$ relationship is approximated by a simple linear fit. Such a linear dependence
between spectral type and effective temperature is unlikely to be adequate over a large range of spectral types.

All these limitations being kept in mind, the results of Luhman et al. (2000)
point to an interesting suggestion.
Using the aforementioned $\teff$-$Sp$ relationship, i.e. a consistent methodology for the analysis of various observations,
these authors derived what is supposed to be
the IMF of several star forming clusters, namely $IC348$, $\rho$-Oph or the Trapezium,
and showed that these IMFs are very similar, except for the Taurus star forming region which exhibits a significant deficit of BDs, as confirmed recently by the larger survey of Brice\~no et al. (2002).
Although, as mentioned earlier, these MF determinations, in spite of the effort of these authors, are of limited reliability,
 given the very young age of these clusters ($<$ 1 Myr),
they seem to indicate a (moderately) rising MF in the substellar regime down to about
the deuterium burning limit. This is confirmed by Najita et al. (2000), who conducted a very careful study of IC348 in the infrared with the HST, extending 4 mag below the previous K-band
study of Luhman (1999). Using BCAH98 models and Luhman (1999) temperature scale, these authors obtain a
MF $\xi(log m) \propto m^{0.5}$ in the mass range $0.015  \la m/\msol \la 0.7$, based on the four lowest mass bins of the sample.

More robust determinations can be derived from the observations of older,
so-called young open clusters like the Pleiades ($\tau \approx 100$ Myr).
Figure \ref{pasp_MFPleiades_col.ps} displays the MF obtained for the Pleiades, from various observed LFs covering a significant fraction of the cluster area
(Hambly et al. 1999, Dobbie et al. 2002b, Moraux et al. 2003),
using the BCAH98 and Chabrier et al. (2000) models, which accurately reproduce the observed magnitude-color diagrams.
Membership to the cluster has been assessed by proper motion measurements and follow-up observations in the near-IR. The field single object IMF (eq.[\ref{IMFdisk}]) and system IMF (eq.[\ref{IMFsys}]) derived in \S2.1 are displayed for comparison (dashed lines). As seen in the figure, the MF derived from the observed LF
is adequately reproduced by the disk {\it system} MF (eq.[\ref{IMFsys}]), suggesting that the difference between the
Pleiades MF and the field single MF stems primarily from unresolved companions, assuming the same kind of correction for binaries as for the field, and possibly from a moderate dynamical evaporation of very-low-mass objects
(see also Moraux et al. 2003)\footnote{The difference between the observed and theoretical MF
at large masses stems from incompleteness in the Hambly et al. (1999) survey, completed recently by Adams et al. (2001).}. 

Figure \ref{pasp_MFclusters.ps} displays the MFs obtained for various similar clusters, with ages ranging from $\sim$ 5 Myr to $\sim 170$ Myr.
Again, membership of the very low mass objects to the clusters has been assessed by accurate proper motion measurements (better than 10 mas/yr) or by follow-up spectroscopy (see references in the figure caption).
Sigma-Orionis cluster, in spite of its relatively young age, exhibits negligible extinction (Zapatero Osorio et al. 2000), and masses can be inferred
from MMRs at the proper age. 
It is important to note that these observations do not consider the effect of binarity, so that the derived MFs reflect the system MFs. The field system MF (eq.[\ref{IMFsys}]) is superposed for each cluster, for comparison (long-dash line).
We verified that this system MF reproduces also very well the MF derived for
the young cluster IC348 (Luhman et al. 2003) down to $\sim 0.01\,M_\odot$.

Figure \ref{pasp_MFclusters.ps} clearly points to a similar underlying IMF between open clusters and the Galactic field,
extending below the H-burning limit. Differences are likely to arise from unresolved binaries and from uncertainties in the mass determination of the
lowest mass objects, due to uncertainties in the theoretical models at
young ages (see Baraffe et al. 2002) and in the treatment of dust
formation (Chabrier et al. 2000 and discussion in \S1.3).
The same figure illustrates also the effect of dynamical evolution,
 which affects dominantly the lowest-mass objects.
As suggested by the present analysis,
dynamical evolution starts to affect significantly the IMF of the clusters
somewhere between the ages of the Pleiades ($\sim 120$ Myr) and the age of
M35 ($\sim 175$) Myr.
This estimate is in agreement with recent N-body 
star cluster simulations, which predict very limited evaporation ($\la 10\%$)
within 100 Myr (de la Fuente Marcos \& de la Fuente Marcos 2000). It is thus not surprising that deep
surveys of the Hyades ($\sim 800$ Myr) have failed to discover any BD (Gizis et al. 1999, Dobbie et al. 2002c).
These results corroborate the traditional view that the Galactic field has
been populated from the evaporation of young, dense ($n\ga 10^3$ pc$^{-3}$) clusters as the one considered presently, with the same
underlying IMF.

The amount of evaporation of a cluster, and thus the departure from the IMF,
can be determined by the approximate number of objects and total mass in a mass range domain relative to a
well determined value, which provides the normalization, i.e.:

\begin{eqnarray}
\Delta \, N(< m_{norm})={ \Bigl( \int_{m_{min}}^{m_{norm}} \xi(\log m) d\log m \Bigr)_{IMF}- \Bigl( \int_{m_{min}}^{m_{norm}} \xi(\log m) d\log m \Bigr)_{PDMF} \over \Bigl( \int_{m_{min}}^{m_{norm}}  \xi(\log m) d\log m\Bigr)_{IMF}}
\label{evol}
\end{eqnarray}

\noindent for the number density and equivalently for the mass density. From Figure \ref{pasp_MFclusters.ps}, we get $\Delta \, N_{sys}(\la 0.4\,\msol) \approx 60\%$, $ \Delta \, M_{sys}(\la 0.4\,\msol) \approx 35\%$ for M35 at $\sim 175$ Myr.

On the other hand, as will be discussed in \S7, small variations between the low-mass parts of the IMFs in various clusters
might stem from various levels of turbulence related to the cluster mean density (see e.g. Myers 1998).
If confirmed, the difference in BD detection for instance between isolated regions like Taurus ($n\sim $ 1-10 pc$^{-3}$) and high-density
 star-forming regions like e.g. Orion, Ophiucus or Trapezium ($n\sim 10^3$-10$^4$ pc$^{-3}$) might thus reflect the importance of the level of turbulence in the cloud on the low-mass end of the IMF.
As shown in Figure \ref{pasp_MFclusters.ps}, however, for a mean density above $\ga 10^3$
pc$^{-3}$, these effects do not seem to yield drastically different IMFs, suggesting a universal, dominant process in star formation both in young clusters and in the Galactic field.

\subsection{The planetary mass function}

Over a hundred planets orbiting stars outside the solar system have now been discovered, with periods $P\la 1500$ days. A statistical analysis of their
mass distribution, corrected for the uncertainty due to the inclination $sin \, i$ of the orbital plane on the sky, has been
established by different authors (Zucker \& Mazeh 2001, Jorissen, Mayor \& Udry 2001). The resulting planetary mass 
distribution  $dN/dm_p$
peaks around $\sim$ 1-2 $M_{Jup}$, due to present detection limits of radial velocity surveys to detect smaller objects,
and decreases rapidly to reach essentially 0 at $\sim 10\,M_{Jup}$, with only 4
systems extending up to 16 $M_{Jup}$. This distribution corresponds to
a relatively flat MF $\xi(\log m)\approx$constant below $\sim 10\,M_{Jup}$
(Zucker \& Mazeh 2001).
This mass distribution differs completely from the stellar+BD mass function derived in \S2.1.4,
and clearly points to a different population of substellar objects, namely {\it planetary} companions of stars, which formed from
a different mechanism than the one yielding the IMF (eq.[\ref{IMFdisk}]). 
It is interesting to note that these planets are now found with a large distribution of eccentricities,
from near-zero to large eccentricity, suggesting complex mechanisms of dynamical interactions (Udry et al. 2003).
As mentioned above, the overwhelming
majority of these planetary companions have masses near $\sim 1\, M_{Jup}$, significantly below the deuterium-burning
minimum mass $\simeq 12\,M_{Jup}$ (Saumon et al. 1996,
Chabrier et al. 2000b), confirming the fact that this mechanism is unlikely to play any specific role in either stellar or
planet formation. Notice that this cannot be due to an observational bias since the majority of these surveys are also intended
to detect BD companions of solar-type stars and could easily detect companions up to $\sim 100 \,M_{Jup}$.

Zucker \& Mazeh (2001) compared the distribution of {\it stellar companions} of G and K stars with the aforementioned distribution of exoplanets to these solar-type stars. This comparison highlights the lack of objects in the mass range $\sim 10$-100 $M_{Jup}$,
compared with substantial fraction of companions on both sides of this mass range. This defines the so-called "brown dwarf"
desert, which prompted Zucker \& Mazeh to identify two distinct populations of objects, originating from two distinct process of
formation, namely the stellar companions and the planetary companions. It would be interesting to extend this argument to other
stellar populations, in particular to M dwarfs. Indeed, as mentioned in \S2.1.3, the "brown dwarf desert" illustrates the lack of companions of solar-type stars in the BD regime, i.e. with large mass ratio $q=m_2/m_1 < 0.1$, compared with {\it planetary} companions,
 and does not preclude a substantial fraction of
BD companions of smaller mass stars, i.e. of M dwarfs, as suggested by recent determinations (Gizis et al. 2001, Close et al. 2003).
Therefore, there should not be any BD desert in the M-dwarf domain.
However, if the fraction of planets appears to be the same around M-dwarfs as around solar-type stars, their mass distribution should be
quantitatively and qualitatively different from the IMF (eq.[\ref{IMFdisk}]). While this latter decreases with decreasing mass below
about the hydrogen burning limit (in logarithmic scale) (Figure \ref{pasp_MF1_col.ps}), the former one should be rising, or flat, below
$\sim 10\,M_{Jup}$. It is thus interesting to search for substellar companions around M-dwarfs to find out whether the
fraction of companions rises or decreases below a certain mass. Since, given our present ignorance of the exact formation history of stars, BDs or planets, it
is impossible to distinguish BDs from planets\footnote{The only difference between a giant planet like Jupiter or Saturn and a BD
is the presence of a rock+ice core, of several Earth masses, at the very center of these planets, reminiscent of the protoplanetary disk from which they were
formed. But the only {\it indirect} clue about the presence of this core stems from the very accurate determinations of the
gravitational moments of the planet, from Voyager, Pionneer and Galileo. Such data are obviously unavailable for the exoplanets.},
a rising distribution of low-mass companions around M-dwarfs would be the signature of a population of planets around these stars.

A comparable argument to identify two very distinct populations could stem eventually from the evaluation of their space densities,
by comparing the space density of exoplanets with the density of BDs with masses
$m<10\,M_{Jup}$. From IMF (eq.[\ref{IMFdisk}]), this latter is $\sim 0.025$ pc$^{-3}$. Most of the planets discovered today orbit solar-type stars. About 7\% of these stars have a planetary companion, of mass $m_p\ga 0.5\,M_{Jup}$ and orbital
period $P\la 4$ yr. This yields a density of planets around solar-type stars $n_P\approx 0.07\times 1.0\;10^{-2}\approx 10^{-3}\,$ pc$^{-3}$. This is obviously a lower limit since (i) only solar-type stars have been surveyed
with enough accuracy (ii) only giant planets are accessible to present detections and (iii) only planets with periods shorter than
$\sim 1500$ days have been detected\footnote{As
pointed out by M. Mayor, in the aforementioned mass-period range, our solar system has no giant planet !}.
If the same fraction of planetary companions applies to M dwarf, for example, $n_P$ rises by about a factor of 10
and the density of planetary companions becomes comparable with the density of low-mass BDs. Present detections,
however, do not allow robust conclusions to distinguish BDs from planets from their estimated space densities.

\section{ The spheroid mass function}

In this paper, we define the spheroid as the Galactic component described by a De Vaucouleurs $\rho(r) \propto e^{-r^{1/4}}$ or a Hubble $\rho(r) \propto 1/r^3$ density
profile. It is often called also the stellar halo, in opposition to the isothermal dark halo. 

The direct determination of the spheroid LF is a very difficult task, since the population of the thick-disk, with a scale height $\sim$1-1.5 kpc contributes appreciably to star counts up to at least $V\sim 20$, $I\sim 19$.
Furthermore, a major difficulty of photometric surveys
at large magnitude is to distinguish stars from galaxies.
These difficulties are in principle circumvented with the HST, which can distinguish stars from galaxies to a limit magnitude $I\simeq 23$, avoiding serious contamination from disk stars  (Gould, Flynn \& Bahcall 1998, hereafter GFB98), but the small field of the HST
yields too small statistics to derive a robust LF. One thus relies on ground-based observations, where the spheroid population in the solar neighborhood is identified from its kinematic properties. The most recent determination of the subdwarf sample of Luyten's LHS catalogue has been obtained by Dahn et al. (1995), updated recently (Dahn \& Harris 2002, private communication). The 298 stars 
in the sample
have a tangential velocity with respect to the local standard of rest 
$v_{\rm T}> 220$ km\, s$^{-1}$, a strong indication of a stellar halo population,  and most of them have a determined parallax. The volume-density is determined
by the usual $1/V_{max}$ method, where the volume limit is set by both
apparent magnitude and proper motion observational constraints. This yields
a $1/\xi = 2.35$ correction factor to account for stars excluded from the
sample by selection criteria (Dahn, private communication). The faint end of this LF has been confirmed recently by Gizis \& Reid (1999) and  by the photometric and kinematic
identification of halo stars in the fourth Catalogue of Nearby Stars (Fuchs \& Jahreiss 1998).
An other recent, photometric, determination is based on a reduced proper motion analysis of the Revised NLTT Catalog, which contains about 5000 halo stars to a completeness limit $V\sim 18$ (Gould 2003).

These kinematically determined samples must be corrected for incompleteness.
As pointed out by Bahcall and Casertano (1986, BC86), the completeness correction factor depends on the assumed spheroid
kinematics. Following GFB98, we adopt the 3-component Galactic model of Casertano, Ratnatunga and Bahcall (1990), which includes a thin disk,
a thick disk and a spheroid component. This model
yields an excellent agreement with the spheroid RR Lyrae population (GFB98, Gould \& Popowski 1998). With this spheroid kinematic model, GFB98 estimate the completeness factor for spheroid subdwarfs with $v_{\rm T}> 220$ km\,s$^{-1}$ to be
$\xi =0.54$. Following the same procedure as these authors, we thus multiply the Dahn et al. (1995 + 2002 private communication)
data by a factor $1/(2.35\times 0.54)=0.79$.
Figure \ref{pasp_LFsph.ps} displays the four aforementioned subdwarf LFs, all corrected for incompleteness with factors
consistent with the aforementioned Casertano et al. (1990) kinematic model.
Significant differences exist between these various determinations.
The most recent analysis (Dahn et al. 1995, 2002, Gould 2003)
predict a significantly larger number of halo subdwarfs, for $\mv \ga 8$, than the previous
BC86 determination once the
$1/\xi = 3.0$ BC86 correction factor has been applied\footnote{Note that the
BC86 spheroid LF displayed in Figures 6 and 7 of Dahn et al. (1995) was
{\it not} corrected by this factor (Dahn, private communication).}. 
This points to a larger incompleteness factor than admitted in the BC86 analysis.
As noted by Gizis and Reid (1999), the BC86 LF, based on Eggen's (1979, 1980) survey of
southern ($\delta < 30$) stars with $\mu > 0.7^{''}$ yr$^{-1}$, might be
underestimated by $\sim 30\%$, due to incompleteness of Eggen's sample in the Galactic Plane. 

Differences between the HST and nearby LFs might be due to the HST small field of view.
On the other hand, it is generally admitted that the spheroid
is substantially flattened, with $q\sim 0.7$, so that most of the local subdwarfs
would reside close to the disk, and this population would not be included in
the HST sample (see e.g. Digby et al. 2003).
Sommer-Larsen \& Zhen (1990) estimate this subdwarf fraction
to be about 40\%. For this reason the local normalization of the spheroid
subdwarf density from the number-density observed at large distances from
the plane, as done with the HST (GFB98), is a very uncertain task.
We note also some difference, at the $\sim 2\sigma$ level, between the Dahn et al. (1995, 2002) LF and the NLTT one (Gould 2003), this latter rising more steeply and peaking at a $\sim 1$ mag brighter magnitude.
The reason for such a difference is unclear. It might stem from the
limited statistics in the Dahn et al. survey ($\sim$ 10 to 30 stars per bin in the
$\mv=$9-12 range) or from the simple color-magnitude relations adopted by Gould (2003).
On the other hand,
Dahn et al. used a purely kinematic criterion to select halo objects in their sample.
As acknowledged by these authors themselves, this undoubtedly rejects bona fide spheroid subdwarfs due to their
directional locations in the sky. Such an even small correction might be consequential in the last bins. Uncompleteness of the LHS Catalogue at faint magnitude would also affect the faint part of the LF.
All these uncertainties must be kept in mind when considering the present results.

The spheroid population can also be identified photometrically, which strongly correlates with metallicity.
Figure \ref{pasp_CMDsph.ps}
displays the 114 spheroid stars identified in the Dahn et al. (1995) survey in a $\mv$-($V$-$I$) color-magnitude diagram as well as the observed thin-disk M-dwarf sequence (Monet et al. 1992, small dots) and
superimposed to the observations five 10 Gyr isochrones with metallicities $\mh=$-2.0, -1.5, -1.3, -1.0 and -0.5, respectively.
Recall that these isochrones reproduce accurately the observed sequences of various globular clusters of comparable
metallicity, except for the more metal-rich ones ($\mh \ga -1.0$) (see BCAH97 and \S1.3). This figure clearly shows that the
kinematically-identified spheroid subdwarf population covers a wide range of metallicities, from $\sim$1\% solar to near-solar, with an
average value $\langle \mh \rangle \simeq$ -1.0 to -1.3, i.e. $[Fe/H]\simeq$ -1.7 to -1.4\footnote{For metal depleted objects, a metallicity $\mh$ corresponds to an iron to hydrogen
abundance $[Fe/H]\simeq \mh-0.35$, due to the $\alpha$-element enrichment (see BCAH97).} (see also Fuchs, Jahreiss \& Wielen 1999).
Such a large dispersion remains unexplained and is at odd with a burst of star formation
in the spheroid $\sim$ 10 to 12 Gyr ago.
Accretion during the star orbital motion across the disk is unlikely. A Bondi-Hoyle
accretion rate, most likely an upper limit except possibly during the early stages
of evolution, yields $\dot m_{acc}\approx 2\pi(Gm)^2nm_H/v^3\approx 2.6\times 10^{-19}
(m/\msol)^2\times (n/1\,{\rm cm}^{-3})(v/220\,\kms)^{-3}\,\msol$yr$^{-1}$, i.e. $m_{acc}\la 10^{-9}\,\msol$, for subdwarf masses, in 10 Gyr (here $n$ is the density of the ISM and $v$ the
velocity of the star).
An alternative possibility is a metallicity and velocity gradient along the spheroid vertical structure above the disk.
In that case, the subdwarfs discovered with the HST should be more metal-depleted than the one in the local sample. Recent observations (Gilmore et al. 2002) have detected
a substantial population of stars a few kpc above the Galactic disk with kinematic
properties (rotational velocity and velocity dispersion) intermediate between the canonical thick disk and the spheroid. These authors interpret this "vertical shear" as an
extension of the thick disk, caused by the ancient merging of a nearby galaxy. This interpretation confirms the previous
analysis of Fuchs et al. (1999) and is supported by the recent
analysis of Fuhrmann (2002) who finds that the majority of subdwarfs within 25 pc from the Sun with large space velocities
($(U^2+V^2+W^2)^{1/2} \ga 100\,\kms$) have a chemical composition characteristic of the thick disk ($[Fe/H]\la -0.5, [Fe/Mg]\approx -0.5$).
If this interpretation is confirmed, this implies a substantial revision of the thick-disk and spheroid models. In that case, the local subdwarf sample and the one observed with the
HST probe two different stellar populations.
In particular, as discussed by BC86, 
the inclusion of a few stars with high-velocity
belonging to this extended thick-disk population in the local genuine spheroid subdwarf sample will yield a severe overestimate of the
supposed spheroid density.
Until this issue is solved, we will assume that the NLTT (Gould 2003) or LHS (Dahn et al. 1995, 2002) samples
are representative of the spheroid one\footnote{Note that the NLTT LF of Gould (2003)
is in agreement with the one derived recently from a detailed reduced proper motion analysis of the Sloan and SuperCosmos
surveys (Digby et al. 2003), presumably probing the genuine spheroid subdwarf population.}, keeping in mind that these samples may include a fraction of thick-disk stars with high dispersion velocities.
Such an assumption
yields the {\it maximum} mass contribution and local normalization of the Galactic spheroidal component.

A correct analysis of the subdwarf metallicity would require a statistical approach but the metallicity probability distribution for these stars is presently unknown, and the derivation of such a
distribution from a two-color criterion only is of weak significance. For this reason, we have converted the observed LFs of Figure \ref{pasp_LFsph.ps} into MFs, based on the BCAH97
mass-$\mv$ relationships, assuming that all stars have a given metallicity.
In order to estimate the uncertainty on the MF due to possible metallicity variations, we have used
$m$-$\mv$ relationships for $\mh=$ -1.5, -1.0 and -0.5, respectively.
Figure \ref{pasp_MFsph_col.ps} displays the MF derived from the NLTT LF for these
three metallicities. To illustrate the uncertainty due to the different LFs, the MF derived from
the Dahn et al. (1995, 2002) LF is also shown, for $\mh=$-1.0 (dot-line).
Interestingly enough,
the differences between the MFs derived for the three different metallicities remain modest, a consequence of the limited effect of metallicity, in the aforementioned range,
on the slope $d\mv/dm$ of the MMR (BCAH97).
The effect is dominant at the low-mass end of the MF: the lower the metallicity the steeper the MF.
The wiggly behaviour of the MF derived from the Dahn et al. LF, with a peak around
$\log m=-0.2$ followed by a dip, stems from the flattening behaviour of both their LF and the m-$\mv$ relation in the
$\mv \approx 8$-10 range.

The MF is reasonably well described by the following lognormal form below 0.7 $\msol$, illustrated by the solid line in Figure (\ref{pasp_MFsph_col.ps}):

\begin{eqnarray}
\xi(\log \,m)= 3.6\times 10^{-4}\,\,\exp\bigl\{-{[\log \, m\,\,-\,\,\log (0.22\pm 0.05)]^2\over 2\times 0.33^2}\bigr\}, \, \, \, \, \,\,\,m\le 0.7\,\msol
\label{IMFsph}
\end{eqnarray}

Although an IMF similar to the disk one (eq.[\ref{IMFdisk}] ) can not be totally excluded, in particular
if the Dahn et al. (1995, 2002) LF happens to be more correct than the Gould (2003) one, equation [\ref{IMFsph}] gives a
better representation of the data.
For comparison, the IMF derived from the HST LF decreases as a straight line $\xi(\log \, m) \propto m^{0.25}$ below 0.7 $\msol$ (GFB98),
a consequence of the much smaller number of faint subdwarfs detected by the HST, as noted previously
(Figure {\ref{pasp_LFsph.ps}).

For an age $t>10$ Gyr, a lower limit for the spheroid, stellar evolution either on or off the MS affects objects
with mass $m\ga 0.7\, \msol$, i.e. $\mv \la 6$. Objects brighter than this magnitude must then
be ignored for the IMF determination and normalization. The shape of the IMF above the
turn-off mass ($\la 0.9\, \msol$ for $t>10$ Gyr), is undetermined. 
Various analysis
of the high-mass part of the IMF in the LMC, SMC and in various spheroidal galaxies (Massey 1998 and references therein) seem to be
consistent with a Salpeter slope, for all these metal-depleted environments. We thus elected to prolongate the IMF (eq.[\ref{IMFsph}]) by such a power-law,
with a common normalization at 0.7 $\msol$, yielding the global spheroid IMF given in Table 2.

Equation [\ref{IMFsph}] yields a normalization at 0.70 $\msol$ : $\xi(\log m)_{0.7}=(1.13\pm 0.5)\times 10^{-4}\,\mlvol$.
This yields a spheroid main-sequence star number-density $n_{MS}\simeq (2.4\pm 0.1)\times 10^{-4}\,\pc3$ 
and mass density $\rho_{MS}\simeq (6.6\pm 0.7)\times 10^{-5}\;\mvol$.
Note that this value is more than twice larger than the determination of GFB98, due to the different IMFs, as mentioned
above. Note that we assumed that the power-law form extends to 0.7 $\msol$. Given
the unknown slope of the IMF for spheroid stars in this mass range, we could as well have chosen
0.9 $\msol$ for the limit of the power-law part of the IMF and have extended the lognormal form to this limit. The difference in the
derived densities, however, is small and largely within the present uncertainties of the IMF.
Integration of this IMF in the substellar regime
yields a negligible BD number-density $n_{BD}\la 3\times 10^{-5}\,\pc3$ 
and mass-density contribution $\rho_{BD_{sph}}\la 2\times 10^{-6}\,\mvol$.

The mass density of the spheroid must include upper main sequence and evolved stars ($0.7\la m/\msol \la 0.9$) and remnants, with progenitor masses
above 0.9 $\msol$. Integration of the presently derived IMF yields for these contributions,
respectively, $\rho_{ev}\approx 0.8\times 10^{-5}\,\mvol$ and $n_{rem}\approx (2.7\pm 1.2)\times 10^{-5}\,\pc3$ (assuming a Scalo coefficient, $x=1.7$ for $m>0.7\,\msol$, as in
GFB98, yields $n_{rem}\approx 1.9\times 10^{-5}\,\pc3$),
i.e., for an average WD mass $\langle m_{WD}\rangle=0.65\,\msol$, a remnant WD mass-density $\rho_{WD}\approx
(1.8\pm 0.8)\times 10^{-5}\,\mvol$, similar to previous determinations (GFB98).
This yields the spheroid total stellar mass density $\rho_{sph}\approx (9.4\pm 1.0)\times 10^{-5} \;\mvol$, about $1\%$ of the local dark matter density, and a
local stellar+BD normalization $\rho_{sph}/\rho_{disk}\approx 1/600$ (see Table 3),
in agreement with estimates based on the fourth Catalogue of Nearby Stars (Fuchs \& Jahreiss 1998).
The corresponding microlensing optical depth towards the LMC is $\tau_{sph}\simeq 10^{-9}$.
As discussed earlier, this represents an {\it upper limit} for the true spheroid mass density, since
a fraction of the local subdwarf population identified in Figure {\ref{pasp_LFsph.ps} might belong to
the high-velocity tail
of the extended thick-disk population, of which local normalization is about two orders of magnitude larger.

The present determinations yield, for an average WD mass upper limit 1.4 $\msol$
 a {\it maximum}
spheroid WD mass-density $\rho_{WD}\la (3.8\pm 1.7)\times 10^{-5}\,\mvol$,
i.e. less than 0.7\% of the dark
matter local density.
As mentioned above, however, the normalization of the spheroid MF is not as straightforward as one would wish since
all the BC86, NLTT, HST and parallax
surveys are affected by completeness correction factors, of which determination depends on the assumed
Galactic model and corresponding asymmetric
drift and velocity dispersion for the population identified as the spheroid one.
Since the afore-derived normalization at 0.7 $\msol$ is directly proportional to the
$1/\xi$ correction factor in the LF,
the detection of a genuine spheroid WD population
exceeding 1\% of the dark matter density would imply a correction factor larger by a factor $\ga$2 near
$\mv \sim 6$. A more plausible explanation, as discussed above, is that the thick-disk population
extends well above the $\sim 1$ kpc scale height and includes a partially pressure-supported population. The identified high-kinematic WD population would thus
be the remnant of the high-mass tail  of this relic population dating from the early epoch of the disk formation. An alternative explanation, finally, would be a radically different IMF for $m>0.9\,\msol$ in the halo, an issue addressed in the next section.

Note that the IMF (eq.[\ref{IMFsph}]) is not corrected for binaries. The fraction of subdwarfs in binary systems is presently unknown but it is
probably smaller than the one in the disk (Gizis \& Reid 2000).
Since, as examined in the next section, the spheroid and globular cluster populations seem to originate from a similar IMF, we
assume that the binary fraction is similar, i.e. $\sim$15-20\% (Albrow et al. 2001), and thus small enough not to affect
significantly the present determination.
Note that even if the binary fraction is similar to the disk one, this should not
affect the present normalization at 0.7 $\msol$ by more than $\sim 10\%$.
However, as illustrated in \S2 for the disk and for the young cluster populations, a binary fraction in globular clusters and in the spheroid comparable to the disk one ($\sim 50\%$)
would bring the spheroid IMF in reasonable agreement with the
disk one. An issue of prime importance for assessing the dependence of the IMF upon metallicity.

Although, as mentioned above, the determination of the spheroid IMF and density, and the very identification of the spheroid
population itself rely on much weaker grounds than for the disk,
the following conclusions seem to be reasonably robust : (i) the dynamical contribution of the spheroid to the Galactic mass budget is negligible, (ii) the IMF is very likely
lognormal below $\sim$ 0.9 $\msol$, with a characteristic mass near $\sim 0.2$-0.3 $\msol$, above the one inferred for the disk IMF, {\it if} indeed the binary fraction
in the spheroid is significantly smaller than the one in the disk,
(iii) the main contribution to the spheroid mass budget comes largely ($\sim 75\%$) from main sequence
stars (see Table 3) and the spheroid BD population is negligible,
(iv) an identified WD population with halo-like kinematic properties exceeding 1\% of the dark matter
density would imply either a completely different spheroid kinematic model, suggesting
that present surveys are severely incomplete in the identification of the spheroid
population, or a different original population, implying a thick-disk population with low angular momentum support extending well above the plane,
or an IMF peaked in the WD-progenitor mass range.

\section{ The Globular Cluster mass function}

Globular clusters provide a particularly interesting testbed to investigate the stellar MF. They provide a homogeneous
sample of MS stars with the same age, chemical composition and reddening, their distance is relatively
well determined, allowing straightforward determinations of the stellar LF, and the binary fraction is negligible
($\sim 10$-20$\%$) (Albrow et al. 2001), so that the correction due to unresolved binaries on the LF is unsignificant. From the theoretical point of view, as mentioned in \S1.3, accurate evolutionary models exist which reproduce the observed
color-magnitude diagrams of various clusters with metallicity $\mh \le -1.0$ both in optical and infrared colors, down
to the bottom of the main sequence (BCAH97, Pulone et al. 1998, King et al. 1998, DeMarchi et al. 2000), with the limitations in the optical mentioned in \S1.3
for more metal-rich clusters ($\mh \ga -1.0$). As discussed in \S1.3, however, the consequences of
this shortcoming on the determination of the MF remain modest.

The major problem to determine the IMF of globular clusters is the inclusion of its dynamical history, from present-day
observations.
Dynamical evolution arises from the fact that
(i) N-body systems evolve towards  energy equipartition and gravitational equilibrium by expelling less massive objects to the cluster periphery, while the most
massive ones accumulate towards the center, (ii) interactions with the Galactic potential, interstellar clouds or other clusters
along the orbit leads to evaporation of the cluster with time. Both effects lead to a mass segregation of stars with time and space.
The characteristic timescale for mass segregation for a cluster of total mass $M_{tot}$ is about the cluster mean dynamical relaxation time, i.e. its relaxation time near the half-mass radius $R_h$ (in pc) (Meylan 1987) : 

\begin{eqnarray}
t_{relax} \simeq 9\times 10^5 \, {M_{tot}^{1/2} \over \langle m \rangle } \, {R_h^{3/2} \over \log (0.4\, {M_{tot}\over \langle m \rangle}) } \,\,{\rm yr}
\end{eqnarray}

This relaxation time is only an approximate dynamical time, since the relaxation timescale strongly varies with mass
(as obvious from the dependence on $\langle m \rangle$) and distance from the core, but it gives an estimate for the dynamical
timescale of the cluster. It is clear, in particular, that the relaxation time is shortest near the center, where the most massive stars (and thus larger 
$\langle m \rangle$) accumulate. The relaxation time near the core can be written (Meylan 1987)  :

\begin{eqnarray}
t_{core} \simeq 1.5\times 10^7 {1 \over \langle m_0 \rangle } {v_c R_{c}^{2} \over \log (0.5\, {M_{tot}\over \langle m \rangle}) }\, \,{\rm yr}
\end{eqnarray}

\noindent where $R_c$ is the core radius (in pc), $v_c$ the velocity scale (in km s$^{-1}$), and $\langle m_0 \rangle$ the mean mass
of stars in thermal equilibrium in the central parts.

This dynamical issue has been addressed in particular by Paresce and De Marchi (2000).
These authors used standard multimass Michie-King models to quantify this effect on the presently observed LF
as a function of radial position from the center.
They found that mass segregation can affect significantly the regions either inner or beyond the half-light radius $r_h$, but that
near $r_h$, the deviations from the cluster global MF are unsignificant (Figure 4 of Paresce \& DeMarchi 2000). Therefore, for
clusters whose LF has been measured at significant distance from $r_h$, mass segregation must be accounted for
to determine the global MF from the local one. Otherwise the global MF will appear steeper than it really is.
To estimate the effect of tidal disruption, Paresce and DeMarchi (2000) examined the evolution of the ratio $\Delta \log N$ of
lower to higher mass stars in the observed cluster sequences, similar to eq.[\ref{evol}]. Indeed, this parameter is likely to be the most relevant one
to describe the region affected by external and internal dynamics. The value of $\Delta \log N$ for twelve clusters was found to
exhibit no specific trend in particular with the cluster disruption time, an indication of tidal disruption effects. In other words Paresce and De Marchi found no obvious dependence of the twelve deep LFs on the cluster dynamical history, in spite of the very different cluster
conditions. With the noticeable exception of NGC6712 (De Marchi et al. 1999) which is probably close to complete disruption. These results are consistent with the fact that
the 12 clusters examined by Paresce and De Marchi (2000) are located well inside the survival boundaries of the vital diagrams
obtained from numerical simulations (Gnedin \& Ostriker 1997). This suggests that the clusters probably remained undisturbed
in their internal parts. Therefore the MF measured near the half-light radius for these clusters should resemble very closely
the IMF.

The most striking conclusion of the study of Paresce and De Marchi (2000) is that (i) a single power-law MF can not reproduce both
the bright part and the faint part of the
observed LFs, (ii) the PDMFs derived for all the clusters are consistent with the same lognormal form peaked at $m_c= 0.33\pm 0.03\,\msol$, with a standard deviation $\sigma =0.34 \pm 0.04$, the error bars illustrating the variations between all clusters: 

\begin{eqnarray}
\xi(\log \,m)\propto  \,\exp\{-{(\log \, m\,\,-\,\,\log \,( 0.33\pm 0.3))^2\over 2\times (0.34 \pm 0.04)^2}\}, \, \, \, \, \,\,\,\,m\le 0.9\,\msol
\label{IMFGC}
\end{eqnarray}

\noindent The limit $\sim 0.8$-0.9 $\msol$ corresponds to the turn-off mass for an age $t\approx 10$ Gyr for metal-depleted environments.

This MF is displayed by the dash-line in Figure \ref{pasp_MFGC_col.ps}, superimposed to the MFs derived in the present paper with the BCAH97 MMRs of appropriate metallicity
from the 12 cluster LFs
observed both in optical (WFPC) and infrared (NICMOS) colors with the
HST by Paresce and De Marchi (2000). These observations are in excellent agreement with observations of the same clusters
by other groups. As cautiously stressed by Paresce and De Marchi (2000), however, only 4 cluster sequences extend significantly beyond
the peak of the LF, which corresponds to a mass $m\approx 0.3\,\msol$. Future large, deep field surveys, for example with the HST
Advanced Camera (ACS), are necessary to make sure that all the cluster IMFs are adequately reproduced by the aforementioned
IMF. Since this latter, however, adequately reproduces the bright part of the LF, i.e. the upper part of the IMF, there seems to be
no reason why significant departures would occur at the lower part. Except if - contrarily to what seems to have been established
from either Michie-King
models or Fokker-Planck calculations -
mass segregation or tidal shocks affect significantly the shape of the IMF even near the half-light radius, yielding a deficiency of low-mass stars (see e.g. Baumgardt \& Makino 2003).
Or, similarly, if the half-{\it mass} radius does not correspond to the observed half-{\it light} radius.
The dotted lines on Figure \ref{pasp_MFGC_col.ps} illustrate the IMF derived in the
previous section for the
spheroid population (eq.[\ref{IMFsph}]), with the characteristic mass shifted by 1 to 2 $\sigma$'s (i.e. $m_c=0.22$ to $0.32$ $\msol$).
The agreement between the globular cluster
and the spheroid IMF is striking, exhibiting
similar standard deviations $\sigma$ and characteristic masses, within 2-$\sigma$. The slightly larger characteristic mass for globular cluster
stems most likely from some dynamical evolution, yielding some evaporation of the objects
with mass $\la 0.3\,\msol$, even near the half-mass radius.
This similarity between the globular cluster and the spheroid MF
corroborates the traditional view that globular star clusters and spheroid stars originate from
the same stellar population (Fall \& Rees 1985).
The low mass-to-light ratio of globular clusters
compared to comparatively old systems like elliptical galaxies or bulges of spiral galaxies
stems from the dynamical evolution of the clusters, depriving them from their low-mass stellar content.
As mentioned earlier, the observed binary fraction in globular clusters is too small ($<20\%)$ (Albrow et al. 2001) to affect significantly the IMF. Note that this fraction has been
determined in the core, and thus should represent an {\it upper} limit for the cluster
initial binary fraction.
Therefore, the globular cluster IMF (eq.[\ref{IMFGC}]) and spheroid IMF (eq.[\ref{IMFsph}])
are genuinely different from the disk field IMF (eq.[\ref{IMFdisk}]), illustrated by the long-dash line at the top of
Figure \ref{pasp_MFGC_col.ps}, providing, as mentioned above, (i) that the cluster MF near the half-mass radius has not been
affected significantly by dynamical evolution and (ii) that the binary fraction is small. If true, this difference suggests that the IMF characteristic mass
is larger for metal-depleted environment, an issue examined in the next section.

\section{ The dark halo and early star mass function}

\subsection{The dark halo mass function}

Various constraints exist on the IMF of the Galactic isothermal dark halo ($\rho(r) \propto 1/r^2$), and
thus on its baryonic mass content.

\indent (i) Star-count observations of the HDF exclude the presence of
a significant dark halo main sequence stellar population (Bahcall et al. 1994,
M\'era, Chabrier \& Schaeffer 1996, Elson, Santiago \& Gilmore 1996, Graff \& Freese 1996, Chabrier \& M\'era 1997). This implies that the IMF can not extend below $m\simeq 0.8\,\msol$, for a halo age $\tau_H\sim 13$ Gyr.

\indent (ii) {\it One} red giant, HE 0107-5140, has been detected recently with the Hamburg/ESO survey (HES), with
an iron abundance $[Fe/H]=-5.3\pm 0.2$ (Christlieb et al. 2002). Note that $[Fe/H]$ is a very good tracer of
the composition/metallicity of the surrounding environment, contrarily to C, N or O which can be self-processed
by the star through the CNO-cycle. The inferred
mass and effective temperature are $m\approx 0.8\,\msol$, $\teff=5100\pm 150$ K, respectively. Given the magnitude limit of the survey ($B_{lim}\sim 17.5$),
its detection could be possible at a distance of about 11 kpc, near the edge of
the spheroid. Previous surveys were limited to brighter
magnitudes, within the inner part of the Galactic halo, so that the lack of detection of very metal-depleted ($\mh \ll -4$)
stars in the halo population might be an artefact due to too faint detection limits.
The faintest giants in the HES survey extend up to 20 kpc or more. Further
analysis of the survey should tell us whether or not it reveals the tip of the red giant branch of
a Pop III stellar population.

\indent (iii) The existence of a significant remnant population in the dark halo is not a completely settled issue yet.
As mentioned in the previous section, the maximum contribution from spheroid and/or dark halo WDs
predicted by the IMF (eq.[\ref{IMFsph}])
represents at most $\sim 0.5\%$ of the dark matter density, i.e. $\rho_{WD}\approx 4\times 10^{-5}\,\mvol$, so that the
unambiguous detection of a genuine halo WD population with a significantly larger density
would imply that the halo IMF differs significantly from this form and peaks in the
1-10 $\msol$ mass range.
Microlensing observations of dark matter baryonic candidates in the halo, however, remain controversial.
The MACHO observations (Alcock et al. 2000) yield a microlensing optical depth, based
on 13 to 17 events, $\tau=1.2^{+0.4}_{-0.3}\times 10^{-7}$, with a total mass in the
objects within 50 kpc $M_{50}=9^{+4}_{-3}\times 10^{10}\,\msol$, i.e. $\la 20\%$ of
the dynamical mass. For a standard isothermal halo model with a velocity dispersion $v_\perp = 220\,\kms$,
the event time distribution corresponds to
a peak in the range $\sim 0.5\pm 0.4\,\msol$. Since M-dwarf stars are excluded as a significant dark
 halo population (point (i) above), this implies halo WDs. 
The EROS project, exploring a larger field around
the disks of the LMC and SMC, derived an upper limit contribution to the dark matter of 25\%
for objects in the mass range $2\times 10^{-7}\le m/\msol \le 1$ at the 95\% confidence level (Afonso et al. 2003).
Interestingly enough, the only events detected today towards the SMC have been shown to
belong to the SMC population. One thus can not exclude that events detected towards the LMC are mainly due to self-lensing events, as pointed out originally by Sahu (1994), or that some events such as supernovae or very long-period variables have been misidentified
as microlensing events.

Another important constraint on the dark halo population stems from the abundances of helium and heavy elements, which point
to a primordial WD mass fraction in the halo
$\rho_{WD}\la 0.1\times \rho_{dyn}\la 10^{-4}\,\mvol$ (Gibson \& Mould 1997,
Fields et al. 2000). This is confirmed by recent nucleosynthesis calculations of zero
or near-zero metallicity low-mass and intermediate-mass stars which show that helium burning
and CNO-cycle processes material (in particular C and O) to the surface (Fujimoto et al. 2000, Siess et al. 2002).

Several detections of faint, cool, high-velocity WDs in the solar neighborhood, based on either spectroscopic, kinematic or
photometic identifications, have been claimed recently (Oppenheimer et al. 2001). These detections, however, remain controversial, and are based on a
limited number of objects. Indeed, if the extended thick-disk suggested by Gilmore et al. (2002) and
Fuhrmann (2002) is confirmed, with kinematic properties intermediate between the standard thick-disk
and the spheroid ones, a substantial fraction of the high-proper motion
WDs discovered by Oppenheimer et al. (2001) might indeed belong to this population.
 As shown by Chabrier (1999), one needs large field ($> 1$ sq.deg.) surveys at faint magnitude ($V,R,I > 20$) to really nail down this issue and
derive a reasonably robust estimate of the halo WD density.

It seems thus clear that the present dark halo contains only a negligible fraction of the Galactic baryonic mass.

\subsection{The early star mass function}

This brings us to the hypothetical determination of the IMF of primordial stars, formed at large redshift.
Only indirect information on such early star formation processes can be inferred from various observational constraints
and from galactic evolution (see e.g. Larson 1998).

\indent (i) The modest increase of metallicity along Galactic
history, from $\mh\approx -2$, characteristic of the spheroid, to $\mh=0$, but the 
scarcity of very-metal-depleted $\mh <<-4.0$ stars in the Milky Way as well as in other galaxies, the so-called
G-dwarf problem, or conversely
the similarity of the massive (oxygen producing) stars over low-mass star ratio
 between spheroid and disk,
imply
that relatively few LMS were formed when the metallicity was very
low at early times.

\indent (ii) 
Observations of young galaxies at $z>1$ in the submillimetre and far-IR domains rule out a Salpeter IMF extending down to the H-burning minimum mass and
suggest a top-heavy IMF, with a cut-off near $\sim 0.7\,\msol$,
to produce massive stars without producing low-mass stars of which light would
remain visible to the present time (Dwek et al. 1998, Blain et al. 1999).

\indent (iii) 
The observed abundances of heavy elements in clusters of
galaxies require an increase by a factor of $\sim 3$ in the total mass of heavy elements
than predicted by
a Salpeter IMF. Thus a comparable increase in the ratio
of heavy elements per solar-mass produced by high-mass stars relative to the number of low-mass stars formed.

\indent (iv)
A top-heavy IMF at early times of galactic evolution
increases the number of SNII per visible stars, providing more
excess thermal energy  and thus a larger amount of hot
gas and heavy elements ejected in the IGM from
the bound clusters of galaxies. This is consistent with the fact that
most of the heavy elements in clusters are in the IGM rather than in
galaxies.

\indent (v)
Increasing M/L ratio and Mg/H and Mg/Fe abundances with mass
are observed in early-type galaxies (Worthey, Faber \& Gonz\'alez 1992).
This points
to a larger relative contribution from massive stars, i.e. a dominant
high-mass mode formation, and more mass locked in remnants.

\indent (vi)
Recent observations of the large scale polarization of the cosmic microwave
background measured by the WMAP satellite require a mean optical depth to Thomson
scattering $\tau_e \sim 0.17$, suggesting that reionization of the universe must have
begun at large redshift ($z\ga 10$). A possible (but not unique) solution is a top heavy
IMF for primordial, nearly metal-free stars (Ciardi, Ferrara \& White 2003, Cen 2003).

Although certainly not conclusive, all these independent constraints (to be considered with caution) point to an early-type IMF with a minimum low-mass cut-off
$\ga 1\,\msol$.
On the other hand, [$\alpha$-element/Fe] ratios measured in the intergalactic hot gas
seem to be only slightly oversolar, which implies a significant contribution from type Ia SN,
suggesting a constant Salpeter-like slope of the high-mass tail ($m\ga 1\,\msol$) of the IMF.
Indeed, an IMF with a Scalo slope ($\xi(\log m)\propto m^{-1.7}$) seems to underestimate the fraction of very massive stars to solar-type stars in
high-z field galaxies,
producing too much long-wavelength light by the present epoch
(Lilly et al. 1996, Madau, Pozzetti \& Dickinson 1998)\footnote{Note, however, that these results depend on the correction due to dust extinction and should be considered
with due caution.}. 

Indeed, the thermal Jeans mass strongly depends
on the temperature ($\propto T^{3/2}$) and more weakly on the pressure ($\propto P^{-1/2})$.
Although there is no reason for this latter to have changed significantly during the
universe evolution, the temperature did evolve significantly. As pointed out by Larson (1998), the very minimum ambient temperature of the medium is given by
the cosmic background radiation $2.73(1+z)$ K so that the thermal Jeans mass, i.e. the
minimum mass for gravitationally-bound objects, increases with redshift. Whether this
mass is the very characteristic mass in star formation, or whether a {\it distribution} of Jeans masses is more relevant, will be examined in \S 7.
It is also important to note that, in the absence of
a significant fraction of metals, the cooling, and thus fragmentation,
of the cloud proceeds via collisional excitation and radiative deexcitation of H$_2$, which
can not cool below 85 K (first rotational level of H$_2$) (see e.g. Abel et al. 2000, Nakamura \& Umemura 2002, Bromm et al. 2002).

Given all this general context, it is interesting to examine the signature of a primordial
IMF biased towards large masses ($>1\,\msol$), and more specifically towards WD progenitors,
i.e. with a characteristic mass in the AGB-mass range, like the following form :

\begin{eqnarray}
\xi(\log \, m)=A\,m^{-1.9} \,\exp\{-({3.2\,\msol  \over m})^{1.6}\}
\label{IMFhalo}
\end{eqnarray}

\noindent which is adequately represented also by a form similar to the one
used previously for the disk and the spheroid, namely:

\begin{eqnarray}
\xi(\log \,m)\propto \exp\bigl\{-{[\log \, m\,\,-\,\,\log (3.5)]^2\over 2\times 0.2^2}\bigr\}, \, \, \, \, \,\,\,m\le 4.0\,\msol\\
\xi(\log \,m)\propto m^{-1.7},\,\,\,\,\,\,\,\,\,\,\,\,\,\,\,\,\,\,\,\,\,\,\,\,\,\,\,\,\,\,\,\,\,\,\,\,\,\,\,\,\,\,\,\,\,\,\,\,\,\,\,\,\,\,\,\,\,\,\,\,\, \, \, \, \,m\ge 4.0\,\msol
\label{IMFhalo2}
\end{eqnarray}

This IMF is similar to the one suggested by Chabrier, Segretain \& M\'era (1996) and Adams \& Laughlin (1996), with a cut-off below $\sim 1\,\msol$, but it extends now with a power-law tail $\xi(\log \, m)_{m>> 1} \propto \,\sim m^{-1.7}$ to produce a larger number of SNII progenitors, as discussed above. 
As shown in the next section, such a remnant-dominated IMF increases the mass-to-light ratio and the relative contribution of WD progenitors to the total mass.
Madau \& Pozzetti (2000) have examined the constraint on the extragalactic background light $I_{EBL}$ received today on Earth, produced by a burst at time $t_F$ of primordial stars formed with such an IMF.
They found out that, for a cosmological
model $(h;\Omega_m;\Omega_\Lambda)=(0.65;0.30;0.70)$, 
a mass fraction of primordial stars produced by the IMF of Chabrier et al. (1996) as high as
$\Omega_\star\,h^2=\simeq 0.30 \times (\Omega_B\,h^2)$, where $\Omega_B\,h^2=0.0193$ is the BBN baryon density,
$ \Omega_\star={\rho_\star\over \rho_c}$
and $\rho_C=3H_0^2/8\pi G$ is the universe critical density, is compatible with
the observed upper limit for diffuse background
light today $I_{EBL_{obs}}\simeq 100\,\nw$ (Hauser \& Dwek 2001), providing these stars
formed at a redshift $z\ga 5$.
This is obviously an upper limit since the contributions from subsequent star formation episodes must be
added, and these calculations must be considered as purely indicative. Whether the
inferred chemical enrichment and light production at high redshift is compatible
with observations remains to be determined accurately. But
they illustrate the fact that a substantial fraction of baryons could be trapped in a primordial generation
of intermediate-mass stars whose remnants would be present today in galactic halos or in the intergalactic medium,
providing they formed at high enough redshift.
As noted by Madau \& Pozzetti (1999), the returned fraction of gas in that case is about 80\% so that the corresponding
WD mass fraction today would be $< 10\%$, consistent with the values discussed in \S5.1. 

\section{Galactic mass budget. Mass-to-light ratios}

The IMF has different implications in the general process of galaxy formation
and evolution, depending on the considered mass range.
The chemical enrichment of the galaxies and of the intergalactic
medium (IGM), i.e. their heavy elements content and the energy feedback
produced by SNII, depend primarily on stars with $m\ga 10\,\msol$,
whereas their luminosity results mostly from the stars from about 1
to a few $\msol$, and most of the mass is contained in objects with
$m\le 1\,\msol$. The relative mass fractions of these different
quantities thus bear important consequences for the evolution of
the galaxies and their observational signatures (colors, magnitudes).
Galactic evolution models generally assume that the IMF is universal
and thus does not evolve with time. Given the arguments presented in
the previous sections, however, the low-mass cut-off of the IMF
may have evolved from the conditions of
early star formation, at high redshift, to the ones prevailing in today spiral
galaxy disks, affecting the evolution of mass-to-light ratios (M/L)
from early-type galaxies to present-day disk galaxies.
Figure \ref{pasp_ML_col.ps} displays the evolution of the M/L ratio in optical and infrared bandpasses (from Bruzual \& Charlot 2003)\footnote{These values
correspond to a birthrate parameter $b=0$, where $b={\rm SFR}/\langle {\rm SFR} \rangle$ is defined as the ratio between the present and past average star formation rate. This corresponds to a burst of star formation at a time t=0 and is
thus appropriate for elliptical galaxies. Spiral galaxies are characterized by
values of $b\ne 0$ ($b\ga 0.8$ for late spirals) (Kennicutt et al. 1994), which corresponds to an exponentially
decreasing SFR $\propto e^{-t/\tau}$, yielding M/L ratios decreasing with increasing
$b$ from the
$b=0$ value.}
obtained with the disk,
spheroid and early star top heavy IMFs derived in the previous sections, from 1 to 13 Gyr (i.e. redshift $z\sim 6$ to $z=0$).
The results obtained with a Salpeter IMF  over the entire mass-range 0.1-100 $\msol$ are shown
for comparison (dotted line).
Fot $t\ge 10$ Gyr, the disk (eq.[\ref{IMFdisk}]) and spheroid (eq.[\ref{IMFsph}]) IMFs yield M/L ratios a factor
1.8 and 1.4 smaller, respectively, than the ones obtained with a Salpeter IMF. This result is
in excellent agreement with the values determined
in disk galaxies and required for these latter to reproduce the observed Tully-Fischer
relation (Sommer-Larsen \& Dolgov 2001, Portinari et al. 2003).
As noted by Portinari et al. (2003), a Salpeter slope for the high-mass tail of
the IMF predicts too high gas-to-luminosity fraction and
metal yields in spiral discs. Observations tend to favor a high-mass slope
somewhere between the Salpeter ($x=1.35$) and the Scalo ($x=1.7$) value. This is
within the previously mentioned uncertainty of the high-mass part of the IMF (see Table 1), not to mention remaining uncertainties
in stellar yield determinations.
 
Tables 3 and 4 display
the relative number- and mass-fractions over the entire BD+stellar regime obtained with the IMFs derived in the present review,
for various mass ranges representing BDs ($m<0.072\,\msol$), low-mass
stars (LMS) ($0.072<m/\msol \le 1$), intermediate-mass stars (IMS) ($1<m/\msol \le 9$) and
high-mass stars (HMS) ($9<m/\msol$), respectively.
These fractions are defined as :

\begin{eqnarray}
{\mathcal N}={\int_{\log (m_{min})}^{\log (m_{max})} \xi(\log\, m)\,d\log m \over \int_{0.001}^{100} \xi(\log \,m)\,d\log m}\\\nonumber \\
{\mathcal M}={\int_{\log (m_{min})}^{\log (m_{max})} m \,\xi(\log \,m)\,d\log m \over \int_{0.001}^{100} m \,\xi(\log \,m)\,d\log m}
\end{eqnarray}

Tables 3 and 4 give also the inferred present and initial Galactic number and mass-densities.
As in Chabrier (2001), we take for the disk
a white dwarf density $n_{WD}\simeq 5.5\pm 0.8 \,\times 10^{-3}$ pc$^{-3}$ (Holberg, Oswalt \& Sion 2002)
with an average mass $\langle m_{WD}\rangle =0.65\,\msol$, i.e.
a white dwarf mass density $\rho_{WD}\simeq 3.7\pm 0.5 \,\times 10^{-3}\,\mvol$,
a neutron star density $n_{NS}\simeq 10^{-3}$ pc$^{-3}$ (Popov et al. 2000) with a mass $\langle m_{NS}\rangle=1.4\,\msol$ and a red giant contribution $n_{RG}\simeq 0.3\times 10^{-3}$ pc$^{-3}$, $\rho_{RG}\simeq 0.6\times 10^{-3}\,\mvol$ (Haywood et al. 1997).
Recent determinations suggest a {\it thick-disk} local normalization
of $\sim 15$-20\%, significantly larger than previous determinations (Soubiran et al. 2003,
Fuhrmann 2002).

\section{ The initial mass function theory}

Several clues to understand star formation can be deduced from
the IMFs determined in the previous sections and from observations of
star forming regions:

\indent (i)
Star formation in the Galactic disk and in young clusters extend well below the
hydrogen-burning limit ($=0.072\,\msol$)
and very likely below the deuterium-burning limit ($\simeq 0.012\,\msol$), with a number-density
of brown dwarfs comparable to the stellar one.

\indent (ii) 
The shape of the IMF seems to be similar in very diverse environments, pointing
 to a power-law form for
large masses and a lognormal distribution at low-masses, below $\sim 1\,\msol$.
Within the present (admitly large) uncertainties concerning the spheroid
and primordial star IMF determinations, there is a hint
for a characteristic mass decreasing with time, from a few $\msol$ or more for early star formation
conditions at large redshift to $\sim$0.2-0.3$\,\msol$ for the spheroid and metal-depleted globular clusters and $\sim$0.1 $\msol$ for the disk field and
young clusters. This assumes a small ($\la 20\%$) fraction of binaries for the spheroid and
the globular clusters. If not, the disk IMF is very likely representative of the spheroid
conditions as well. If real, this trend might reflect the effect of the increasing ambient temperature, $T_{min}=2.73(1+z)$ K, with increasing
redshift formation, or simply the decreasing ability of the cloud to cool and fragment
to smaller scales with decreasing metal abundances.
The high mass parts of these IMFs, however, seem to be very similar, consistent with a Salpeter power law, within a $\sim \pm 0.3$
remaining uncertainty in the power-law exponent, for clusters
with a factor of $\sim 200$ range in densities and a factor of $\sim 10$ range in metallicities (Massey 1998, Wyse et al. 2002).
The near-uniformity of the IMF seems to extend far beyond the Galaxy. Indeed, a measure of the low-mass to high-mass ratio is
the $[Fe/O]$ ratio, or the ratio of $\alpha$-element. This ratio has been found to be very similar in elliptical
galaxies, the intracluster medium (Wyse 1997) and QSO Ly-$\alpha$ absorption systems (Lu et al. 1996) and is consistent with a Salpeter IMF at high masses
and a more flatish IMF at low masses.
The same conclusion holds for dwarf-spheroidal galaxies, believed to be dominated by dark matter, finding LFs (and MFs) similar to
globular clusters of comparable metallicities for $m\ga 0.3\,\msol$ (Wyse et al. 2002).
This near-uniformity of the IMF
points towards a {\it dominant self-similar, scale-free process at large scales}, yielding a
power-law distribution.

\indent (iii) Star formation is a rapid process, probably more rapid than
the thermal crossing timescale $\tau_S=L/c_s \ga 10\, {\rm pc}/ 0.2\,{\rm km s^{-1}}\approx 50$ Myr for a cloud of size L and temperature $T\sim 10$ K, or
the ambipolar diffusion timescale $\tau_{ad}\ga 10$ Myr for a cloud of average density $\sim 10^2$ cm$^{-3}$ and $B\approx$ a few $ \mu$G (Ciolek \& Mouschovias 1995), and comparable to
the dynamical timescale $\tau_{dyn}=(3\pi/ 32G\rho)^{1/2} \approx 10^6\,(n/10^3\,{\rm cm}^{-3})^{-1/2}$ yr $ \approx$1-5$\times 10^5$ yr for typical star-forming molecular clouds (see e.g. Elmegreen 2000, Hartmann 2001, 2003, Onishi et al. 2002). This is illustrated in Figure \ref{pasp_CMDclusters_col.ps},
where observations of low-mass objects in young ($\la 1$ Myr) clusters are superposed to various
theoretical isochrones from Baraffe et al. (2002).
Most of the objects tend to pile up above
the $\sim$ 1-3 Myr isochrones.
Similar conclusions have been reached
for other clusters (see e.g. Luhman 2000, Najita et al. 2000, Brice\~no et al. 2002, Hartmann 2003).
The position of the few objects
located on the left hand part of the diagram, which appear to be hotter and fainter than the bulk of the data, are interpreted as an effect of strong accretion (Comer\'on et al. 2003). As shown by Hartmann et al. (1997),
the typical accretion rate in young clusters, $\dot m_{acc} \approx 10^{-8}$-$10^{-9}\,\msol$ yr$^{-1}$ (Hartmann et al. 1998),
is too small to affect appreciably the mass of the object, but it can modify its
evolution, and thus the mass-age relationship. Indeed, the object
contracts abruptly to adjust to the accreted material, reducing its radiating
surface and mimicking the luminosity of an older non-accreting object. The stronger the accretion the larger the effect.

Even though the theoretical isochrones at young ages should be taken with great caution
(see \S 2.2), it is clear from Figure \ref{pasp_CMDclusters_col.ps} that
the peaked age distribution is different from the one expected from a constant formation rate, for which the number of stars per age interval increases rapidly with time, as would result from star formation history linked to
ambipolar diffusion and crossing times (Kenyon \& Hartman 1990). Moreover, the wide dispersal of the Rosat All Sky Survey
(RASS) sources is essentially impossible to explain with a 10 Myr old population, given the low velocity dispersions ($\la 10$
km s$^{-1}$) in star forming regions (Feigelson 1996). More generally, observations of star forming regions suggest not only that
star formation is a rapid process, but that cloud dispersal is fast as well ($\la 10$ Myr) (Feigelson 1996, Hartmann 2001). 
This property of star formation
points towards a {\it process dominated by turbulent motions}, and the rapid damping of turbulence, which can not support clouds for a long time. This picture is supported by the turbulent structure of the clouds, as discussed below.

\indent (iv)
Observations of star forming clouds like
$\rho$-Ophiuchus show many pre-stellar clumps with masses around 0.1-0.3 $\msol$ (Motte et al. 1998,  Bontemps et al. 2001), similar to the
characteristic mass range determined in the present IMF calculations.
Most interestingly, the mass
spectrum of these clumps is quite similar to the IMF discussed in \S2, and thus quite
 similar to the young star mass spectrum in the
cloud, suggesting that the cloud stellar IMF derives directly from the clump mass
spectrum. Similar results have been obtained in the
Serpens cloud (Testi \& Sargent 1998) and in Taurus (Onishi et al. 2002).
This similarity of the IMF in pre-stellar condensation clouds and in isolated objects
suggests that the general shape of the IMF is determined in the original gaseous phase,
and not during the collapse process during which the gas condenses into stars.
This suggests that accretion can not be the dominant mechanism of star formation.
In other words, the initial conditions which determine the very star formation process
most likely originate from large-scale processes which dissipate toward smaller scales.
It also supports kind
of a fragmentation characteristic mass scale such as the thermal Jeans mass, as discussed below.
The large difference between the star and gas mass-distribution derived from
large-scale CO studies of molecular clouds, $N(m)\propto m^{-0.5}$ (see e.g. Kramer et al. 1998), indicates the low-efficiency in converting gas into star.
The rapid dispersal of molecular gas is likely to be one of the reasons for this low efficiency.

\indent (v)
On large scales ($\ga $ pc), the spectral line widths observed in molecular clouds indicate highly supersonic
motions (see e.g. Falgarone, Puget \& Perault 1992), largely exceeding the thermal sound speed $c_S=\sqrt{kT/\mu m_H}\approx 0.2\,\kms$ for $T=10$ K ($\mu=2.33$ is the mean molecular weight and $m_H$ the atomic mass unit). Moreover, the observations are
consistent with super-Alfv\'enic conditions, with 
Alfv\'enic Mach numbers ${\mathcal M}_A(L)=v/(B_0/\sqrt
{4\pi \rho_0})\sim 10$, where $\rho_0$ denotes the gas density (Padoan \& Nordlund 1999).
These motions are thus believed to originate from large scale supersonic and super-Alfv\'enic
turbulence (Larson 1981, 1992, Elmegreen 1997, Padoan \& Nordlund 1999).
The observed line-width component $\sigma$ obeys reasonably 
well the Larson (1981) relation $\sigma \sim 1\,\kms\,( L/1\,{\rm pc})^{0.4}$
over a large scale range $0.01<L<100$ pc (Falgarone et al. 1992). Dissipation of this non-thermal turbulent support on
small scales ($\la 0.1$ pc) is prerequisite for the formation of prestellar cores (Nakano 1998).
Note that the Larson scaling relation yields subsonic velocity dispersions at very small scales, consistent with the fact that even dense cluster-forming regions exhibit very narrow
line widths for $\la 0.1$ pc, with a non-thermal to thermal velocity dispersion ratio of H$_2$ $\sigma_{NT}/\sigma_T \sim 0.7$
(Belloche et al. 2001).

Star formation theories must now be confronted to the general results (i)-(v).
In a canonical theory for isolated star formation, low-mass stars form from the collapse of initially hydrostatic but unstable
dense cloud cores which have reached a $\rho(r) \propto r^{-2}$ density distribution of a singular isothermal spheroid
 (Shu, Adams \& Lizano 1987). In this scenario, deuterium-burning, which occurs at pre-main-sequence ages,
is a key ingredient to trigger star formation. The onset of D-burning induces
convective instability in the interior. Combined with the rapid rotation resulting from the accretion of
angular momentum with mass, this convection is believed to generate a strong magnetic field through
dynamo process. The field will drive a magnetocentrifugal wind that ultimately
sweeps away the surrounding accreting material, and determines the mass of the nescent star.
Within this picture, objects below the deuterium-burning limit can not reverse the infal and thus can not form gravitationally bound objects. Such a scenario can now be reasonably
excluded on several grounds. First of all, substellar objects are fully convective, with or without deuterium-burning,
except for the oldest ones which develop a conductive core at late ages (Chabrier \& Baraffe 2000, Chabrier et al. 2000a).
In fact, the numerous detections of free-floating objects at the limit and below the 
deuterium-burning minimum mass, and the rising mass spectrum down
to this limit in several young clusters (see Figure \ref{pasp_MFclusters.ps} and Najita et al. 2000, B\'ejar et al. 2001, Mart\'\i n et al 2001, Lucas et al. 2001) seem to exclude deuterium burning as
a peculiar process in star formation.
{\it This should close the
ongoing debate in the literature arguing that the deuterium-burning minimum mass distinguishes BDs from planets,
since such a distinction does not appear to be supported by physical considerations}.
Second, as mentioned above, star formation in young clusters appear to form over a timescale significantly shorter than the ambipolar diffusion
timescale $\ga 10$ Myr, indicating that, if magnetic field plays some role in star formation, it is unlikely to be a dominant process.
In the ambipolar diffusion scenario, the cloud must survive long enough, in near equilibrium between magnetic and gravitational pressure.
This is not consistent with observations of rapid star formation and cloud dissipation, and with the observed turbulent nature of clouds. Indeed, equipartition between kinetic, gravitational and magnetic energy fails to reproduce the observed
properties of molecular clouds, which are dominated by super-Alfv\'enic and supersonic motions, where kinetic energy dominates magnetic energy, with a decay timescale approximately equal to a dynamical timescale (see e.g. Padoan \& Nordlund 1999 and references therein).
In fact, ambipolar diffusion models require large static magnetic field strengths ($\sim$30-100 $\mu$G) exceeding
Zeeman estimates for low-mass dense cores ($\la$ 10 $\mu$G) (Crutcher \& Troland 2000,
Padoan \& Nordlund 1999, Padoan et al. 2001a,b, 2001b Bourke et al. 2001).

Another version of the wind-limited accretion model relates the gas cloud properties, sound speed and angular velocity, to the
stellar properties, mass and luminosity, through a direct relation between the accretion rate onto the star and the wind-driven
mass loss rate (Silk 1995, Adams \& Fatuzzo 1996). However, it seems difficult, in this model, to explain the nearly universal shape of the IMF, without sensitivity to the cloud parameters. More importantly, it appears rather difficult to produce free floating gravitationally bound objects of brown dwarf and Jupiter masses in large numbers, since for too small objects there will be no wind to stop accretion. Another argument against the wind-limiting of the final core mass is that
the observed outflows are relatively collimated, with opening angles $< 60^0$, making
difficult for these outflows to remove a large fraction of the protostellar core.
Therefore, although wind
regulation might play some role in determining the final object mass, in particular for large masses,
it is unlikely to play a dominant role in star formation.

An alternative process to determine the final stellar mass is the opacity-limited fragmentation (Hoyle 1953, Larson 1969, Silk 1977). In this model, the protostellar cloud keeps fragmenting under the
action of gravity until it becomes optically thick and can no longer cool. 
Coincidentally, this characteristic minimum mass for the low-temperature, chemically enriched
conditions prevailing in today molecular clouds, is similar to the deuterium-burning minimum mass, namely $\sim 0.01\,\msol$ (Silk 1977, Larson 1992).
In this scenario, one might expect an accumulation of objects near this characteristic mass, from which more massive objects
will grow. The rather smooth continuation of the stellar and substellar MF down to this mass scale, although still subject to large uncertainties, seems to exclude such an accumulation.

In fact, all these scenarios enter more or less the general models of hierarchical fragmentation, based
on a Jeans formulation, where the fragmentation process is determined essentially by comparing the effective isothermal sound crossing time $ \sim L/T^{1/2}$ and the free fall time $\sim \rho^{-1/2}$,
and where the physical process that initiates the gravitational collapse is generally believed to be gravity, yielding a formation process determined
by the local gravitational timescale $(G\rho)^{-1/2}$ (Larson 1978, Elmegreen 1997, 1999).
However, all substellar objects have a mass significantly smaller than any Jeans mass.
This suggests that gravity is not the determinant mechanism which triggers star formation
and shapes the clumps in the initial molecular clouds. Gravity more likely amplifies the
existing density fluctuations but do not create them.
The observational data thus seems to exclude gravitational fragmentation as the {\it dominant} process
which determines the characteristic mass distribution for star formation.
Moreover, in the models of Jeans-instability driven star formation, the thermal energy of the cloud
must be comparable to its gravitational content. As mentioned above, however, turbulent kinetic energy in star forming clouds superseds thermal energy by about a factor of $\sim 100$. Redifining the Jeans scenario in term of turbulent kinetic energy is flawed, because compressible turbulence dissipates in fragmenting the gas in filaments, i.e. in a highly non-homogeneous manner. Furthermore, turbulence is a highly non-linear process, opposite to the basic assumption of gravitational instability model.

Alternative models suggested that stars grow from protostar collisions and/or coalescence between gas clumps until the bound
fragment becomes optically thick (Nakano 1966, Nakano, Hasegawa \& Norman 1995). The aforementioned rapid timescale for star formation, however, is much shorter than the
typical collision time between multiple protostellar clumps, and thus seems to exclude this scenario, at least for the low-mass stars.
It is also difficult to reconcile this star formation mechanism, involving kind of feedback effects, with the universal form of the IMF,
suggesting that this latter reflects the initial conditions imposed in the cloud. Not mentioning the difficulty to reconcile coalescence process with supersonic turbulence.
A recent extension of this type of scenario, where the IMF is determined by the competitive
accretion between the various stellar cores, and a combination of mass accretion and stellar mergers, has been proposed by Bonnel, Bate and collaborators (Bonnel et al. 2001a, 2001b, Bonnell \& Bate 2002), based on hydrodynamical simulations of gas accretion onto a pre-existing cluster of 1000 stars. Based on 15 stars with $m \ga 5\, \msol$ at the
end of the calculations, the high-mass tail of the IMF, obtained with such an
accretion and merger scenario, seems to reproduce a Salpeter slope, although the reason for
such a result is not clear, whereas the lower-mass part of the
IMF yields a shallower power-law with $x\approx 0.5$. Although, as mentioned earlier, the accretion
process certainly plays some role in shaping the final stellar mass, the present scenario relies
on some assumptions for the initial conditions, e.g. an ensemble of already formed stars of
equal mass as nucleation centers and a gas reservoir apparently not supersonic,
which appear rather unrealistic.

Finally, some models suggest that star formation is not due to a dominant process, but is rather the byproduct of several independent
processes of comparable importance. 
The product of a large number of statistically-independent processes naturally points to
the central limit theorem, as initially suggested by Larson (1973), Zinnecker (1984) and Elmegreen (1983), and later on by Adams \& Fatuzzo (1996).
The final product of the central limit theorem is a gaussian distribution, i.e. a lognormal form in a logarithmic plane.
In this type of theory, however, the statistical aspect of star formation
still arises from the hierarchical structure produced by fragmentation, and thus is
linked back to the original concept of Hoyle.
Moreover, this theory is frustrating from the physics point of view, since it
relies on a
purely statistical mechanism and
prevents understanding star formation from identified physical processes.
Not mentioning delicate applications of the central limit theorem concept, which strictly
speaking implies an {\it infinity} of statistically independent variables, in real nature !
In fact, no current theory of the IMF is consistent with all the aforementioned constraints (i)-(v) and predicts in particular the formation of free-floating objects in significant numbers at very low mass.

A picture of star formation driven by compressible turbulence has been suggested recently
by various independent approaches (Padoan \& Nordlund 1999, Klessen 2001, Bate, Bonnell \& Bromm 2002, 2003, see Nordlund \& Padoan 2002 and Mac Low \& Klessen 2003 for recent reviews). Although these approaches use different methods, and differ on the details of the interplay between turbulence and gravity, they are similar in spirit and
share the same underlying dominant idea : star formation is generated initially by the (inhomogeneous) dissipation of
supersonic turbulence, forming dense cores in which eventually gravity becomes important enough for the cores to become unstable and form gravitationally bound objects. 
A comprehensive picture of such a star formation mechanism, in super-Alfv\'enic conditions, has
been derived recently by Nordlund \& Padoan and is summarized below (see Nordlund \& Padoan 2002).

The power spectrum of turbulence on a large scale $L$ in the inertial range (below the energy injection scale and above the dissipation scale) is a power law

\begin{eqnarray}
E(k) \propto k^{-\beta}, 
\end{eqnarray}

\noindent where $k=2\pi/\lambda$ is the wave number and $\lambda$ is the dynamical scale of turbulence. Recent numerical simulations of compressible
turbulence in a magnetized gas (Boldyrev, Nordlund \& Padoan 2002a, 2002b, Padoan et al. 2003a) yield a power spectrum consistent with $\beta =1.74$, between incompressible turbulence, $\beta \approx 5/3$ (Kolmogorov spectrum) and pressureless turbulence, $\beta \approx 2$ (Burgers 1974). The rms velocity $\sigma$ on the scale $L$
of the gas extension before the shock is related to this power-spectrum index by
$\sigma^2\propto L^{\beta-1}\Rightarrow \sigma \propto L^{{\beta-1 \over 2}=\alpha} \sim L^{0.37}$ which
agrees quite well with the aforementioned observed $\sigma$-$L$ Larson's relation.
In these simulations, the upstream Alfv\'enic Mach number ${\mathcal M}_A(L)$ is assumed to follow a Larson-type relation, i.e. to scale
as $L^\alpha$, where $\alpha=(\beta-1)/2\approx 0.4$ from above.
A second assumption is that self-similarity holds at different scales, so that the number of cores in the shocked gas scales with the size $L$ of the
upstream flows out of which they formed, $N\sim L^{-3}$.  

This spectrum, completed by the jump conditions for  isothermal MHD shocks,
$\rho/\rho_0 \approx L/l \approx B/B_0 \approx {\mathcal M}_A$ - where $\rho_0$,
$B_0$ and $L$ denote the gas density, magnetic field strength and gas extension before
the shock, while $l$ denotes the size of the cores in the shocked gas
and the indiceless quantities refer to the postshock situation -
and by the aforementioned scaling relations for the number of dense cores
$N(m)$ of mass $m$ formed in the shocked (filamentary) gas,
yields a mass distribution of dense cores (Padoan \& Nordlund 2002):

\begin{eqnarray}
N(m)\, d\log m \propto m^{-{3\over 4-\beta}}\,\, d\log m\propto m^{-1.33}\, d\log m,
\label{ndist}
\end{eqnarray}

\noindent similar to the Salpeter value.

In these simulations, the typical core mass formed in shocked gas reads :

\begin{eqnarray}
m(L)\sim \rho l^3 \sim \rho_0 L^3/{\mathcal M}_A(L)^2 \sim L^{3-2\alpha}\sim L^{2.2}
\end{eqnarray}

In this turbulent
picture of fragmentation, the distribution of cores arises essentially from
internal cloud turbulent dissipation. 
The collapse of these cores into protostars is then determined by the dynamical timescale
of supersonic MHD turbulence, $\tau_{dyn}=L/\sigma(L)$, rather than by
the local gravitational timescale $(G\rho)^{-1/2}$.
Sufficiently massive cores continue to collapse under selfgravity, so for large $m$ the distribution of cores is directly reflected in the distribution of stellar masses. At smaller masses, only cores with sufficient density are able to collapse further, which reduces the number of stars formed out of a given distribution of cores with mass $m$, and causes the IMF to deviate from the large $m$ behavior. Thus, in a generic sense the roll-over of the IMF happens when gravity is no longer able to cause the collapse of most cores of a given mass. At this stage, different factors such as cooling functions, equations of state, additional fragmentation during collapse etc., become important. Quantitative predictions then require detailed numerical simulations, such as those of Bate et al (2002, 2003)
or Klessen (2001).
If, for example, because of gravitational fragmentation during the collapse, each core gives rise to a distribution of stars, the IMF will be shifted to smaller masses. This will not influence the power law shape on the high mass side, but will shift the maximum mass of the IMF to a smaller value, affecting
 the expected number of low-mass stars and brown dwarfs.

Qualitatively, however, the roll-over of the IMF is displayed already for idealized assumptions with isothermal conditions.
A universal behaviour of turbulent fragmentation for an isothermal gas is that it produces a lognormal probability
distribution function (PDF) of gas density in unit of mean density $x=n/n_0$:

\begin{eqnarray}
 p(x)\,d\ln x\propto exp \Bigl\{-(\ln x-\langle \ln x \rangle)^2/2 \sigma^2 \Bigr\} d\ln x
\end{eqnarray}

\noindent where $n_0$ is the mean density, and
 $\sigma^2\approx \ln \,(1+0.25\, {\mathcal M}^2)$ (Padoan et al. 1997, Padoan \& Nordlund 1999, Ostriker et al. 1999). In the present context of star formation, this
yields no longer a unique Jeans mass
but a distribution of local Jeans masses $p(m_J)$, obtained from the PDF of gas density, assuming that the distribution of average density
of clumps of a given mass has also a lognormal distribution (Padoan et al. 1997, Padoan \& Nordlund 2002):

\begin{eqnarray}
 p(m_J)\,d\ln\, m_J= {2\over \sqrt{2\pi} \sigma }\,m_J^{-2}\, \exp\Bigl\{-{(\ln\, m_J^2-|\langle \ln\,x\rangle|)^2 \over 2\,\sigma^2}\Bigr\} d\, \ln\, m_J
\end{eqnarray}

\noindent where $m_J$ is written in units of the
thermal Jeans mass at mean density $n_0$: 

\begin{eqnarray}
M_J=1.2\,\msol\,({T\over 10\,{\rm K}})^{3/2}\,({n_0\over 10^4\,{\rm cm}^{-3}})^{-1/2}
=1.2\,\msol\,({T\over 10\,{\rm K}})^{2}\,({P_0\over 10^5\,{\rm cm}^{-3}\, {\rm K}})^{-1/2}
\end{eqnarray}

\noindent which thus ranges from $\sim 1.2$ to $\sim 0.12\,\msol$
for characteristic low-pressure to high-pressure clumps\footnote{Note the incorrect density scaling factor, $10^3\,{\rm cm}^{-3}$ in Padoan \& Nordlund (2002, Eq.(21)) (Nordlund, private communication).
Using the same expression for the Jeans mass as Bonnell et al. (2001a, Eq. 1), the scaling constant changes from 1.2 to 1.9 $\msol$}. The Jeans mass is approximately the same for spheres and for filaments (Larson 1985).

The fraction of small cores of mass $m<M_J$ to collapse to gravitationally-bound structures
 is thus given by the probability distribution $P(m)=\int_0^m p(m_J) dm_J$,
and the mass distribution of collapsing cores reads (from eq. [\ref{ndist}]):

\begin{eqnarray}
N(m)d\ln m\propto m^{-{3\over 4-\beta}} \,P(m)\,d\ln m
\end{eqnarray}

Therefore,
although the {\it average} star mass is similar to the average thermal Jeans mass of the medium, the global mass distribution extends well below this limit, with decreasing probability.
Within this picture,
star formation proceeds as follows (see Nordlund \& Padoan 2002):

\indent (i) Supersonic turbulence in the ISM, produced by large amounts of kinetic energy at large scales, dissipates in fragmenting
molecular clouds (preventing a global collapse of the cloud)
into highly anisotropic filaments, due to the random convergence of the velocity field.
These filaments form dense cores, with large density contrasts (much larger than
the maximum value $\sim 14$ for a self-gravitating, pressure-bounded Bonnor-Ebert sphere) via the action of radiative MHD shocks and thus determine the fragmentation length scale over which collapse is possible.
Cooling becomes more efficient as density increases in these dense cores, of typical dimensions $\sim 0.01$-0.1 pc, which become self-gravitating and begin to collapse.
During this stage, the star formation process itself, during which gas is converted into stars,
plays no particular role. Star formation arises essentially from dissipation of super-sonic turbulence toward small scales in molecular clouds, since there is no dissipation mechanism at large scales. This turbulent dissipation
yields a universal power spectrum,
and is thus independent of the local conditions in the star-forming clouds, a result supported by the observations. The main source of kinetic energy in the cloud is supplied by large scale motions, which produce the turbulent cascade.
The initial cloud structure is not essential, except possibly for the initial amount of available turbulent energy in the cloud, because of the universal character of turbulent structures in various environments.

\indent (ii)
 The small-scale dissipation of this large-scale turbulence follows the Larson (1981)
relation, yielding subsonic structures ($\la 0.1\,\kms$) at small scales ($\la 0.1$ pc), typical of protostellar cores. While this process is scale free for scales largely above the minimum Jeans scale, generating a power-law tail, a characteristic mass enters at
small scale, namely the minimum mass for gravitational binding energy to exceed mostly
the thermal, and to a less extend the
magnetic energy, i.e. the Jeans mass (or more exactly the Bonnor-Ebbert mass). 
Since fragmentation is driven by super-sonic turbulence, however, star-forming clumps
can no longer be regarded as equilibrium configurations, and the concept of a unique
thermal Jeans mass no longer applies.
Indeed, there is a {\it distribution}
of local Jeans masses determined by the (lognormal) probability distribution
 function of gas density set up by turbulent fragmentation. Objects below this mass scale form with a
rapidly decreasing (but not zero) probability with decreasing core mass,
since they come from the exponential tails of the density and Mach number distributions.
It should be noted that turbulence in protostellar clouds indeed appears to generate structures much smaller than the thermal Jeans
mass, down to $\sim 10^{-4}$ $\msol$ (Langer et al. 1995, Heithausen et al. 1998, Kramer et al. 1998), suggesting that a fragmentation mechanism other than a purely
Jeans gravitational instability may play an important role for the dynamics of these dense structures.

In this scenario, other processes like gravitational or opacity limited fragmentation, protostar interactions, stellar winds or accretion,
although playing some role in determining the final stellar mass distribution, e.g. by
limiting the efficiency of star formation, appear
to be of secondary importance,
the triggering process of star formation being small-scale dissipation of compressible turbulence, which forms cores. Then, {\it self-gravity drives the subsequent collapse and star formation}.
The turbulent structure of the parent cloud can be maintained either by the young stars which, because of
the short timescale of star formation, do not have time to move far away from their birth site and re-inject kinetic energy into the ISM through outflows, by supernovae or by galactic shear. In fact, given the
short timescale for star formation, turbulent energy does not have to be
resupplied constantly and cloud dissipation may occur within the star formation timescale (Elmegreen 2000).
Note that the turbulent nature of the cloud
also provides a natural explanation for the low efficiency of star formation, besides rapid dissipation of the cloud itself.
Indeed, star formation occurs only in some high density regions of
the filament intersections, and most molecular gas resides either in the
low-density interclump regions or in the dense regions too small to become
self-gravitating (Padoan 1995).
Interestingly enough, the problem of stopping accretion on the collapsing protostar
does not really arise in this picture, for the protostar mass is essentially defined
by the finite amount of mass available in the corresponding core mass.
Rotation is an other important issue. Indeed, although rotational energy in cores ($\Omega_{rot}\sim 10^{-14}$
rad s$^{-1}$) represents about 1-2 \%
of the gravitational energy, the core must get rid of its angular momentum.
The detailed simulations of Abel, Bryan \& Norman (2002),
however, indicate that a core does get rid of angular momentum sufficiently efficiently to give rise to a single star, most of the angular momentum being transported by the shock waves during the turbulent collapse.

Providing large enough density in the initial cloud (or conversely large enough Mach
number for a given density), the mass-spectrum
obtained by these calculations extends well into the BD domain (Padoan \& Nordlund 2002, Nordlund \& Padoan 2002). This provides a natural explanation for the formation
of BDs, suggesting that BDs
form from the same general IMF produced by the cloud collapse as the stars. This is consistent
with the results presented in \S2, which show that the observed population
of BDs in the Galactic field is well reproduced by the same underlying IMF as in the
stellar regime (see also Chabrier 2002). This seems to disfavor the scenario of BD formation produced by violent
dynamical ejection of small embryos from the collapsing cloud (Reipurth \& Clarke 2001, Bate et al. 2002, 2003)
as the {\it dominant} formation process for these objects. Such a scenario raises also other problems
including the BD 
radial velocity dispersion, binary frequency and circumstellar properties (see e.g. Joergens \& Guenther 2001, Bate et al. 2003, Close et al. 2003, White \& Basri 2003).
Competition between collision and interrupted accretion (Bate et al. 2002), the detailed
fragmentation distribution in the collapse from cores to objects (Klessen 2001), and the multiplicity distribution, however, can contribute to extending the IMF towards smaller scales and need to be
quantified by detailed numerical simulations.

As mentioned earlier, the initial density of the parent cloud, as well as
the initial level of turbulence is important in
determining the amount of objects well below the thermal Jeans mass, in particular in the BD regime.
Nordlund \& Padoan (2002) find that reducing the density of the cloud by a factor of 5, or reducing the alfvenic Mach number on a 10-pc size by a factor of 2, reduces the number
of BDs by about a factor of 10.
We had already noted that the fraction of substellar over stellar objects in Taurus ($n\sim 1$ pc$^{-3}$) is
about a factor of 2 smaller than in other young clusters (Brice\'no et al. 2002).
The initial amount of turbulence also leads to significant differences at
intermediate scales ($\sim 0.1$ pc) between the line widths observed in regions of
isolated star formation like Taurus, dominated by thermal motions ($\sigma \la 0.2\ \kms \sim c_S$), and the ones observed in dense
cluster-forming cores ($n\ga 10^3$ pc$^{-3}$), like e.g. Ophiuchus or Orion, dominated by turbulent motions.
In fact, protostellar clouds in Taurus, with size $\sim 0.1$ pc, exhibit properties different from the
ones observed in denser cluster forming regions, and seem to be consistent with the
standard Shu et al. (1987) quasi static isothermal collapse scenario (Motte \& Andr\'e 2001). 
If confirmed in other clouds of similar density, these results may indicate that star formation in low-density environments,
representative of isolated mode of star formation, differs from star
formation in denser clouds, representative of cluster-forming star formation, the dominant mode of star formation (see e.g. Myers 1998). These results imply that
star formation of low-mass objects depends to some extend on the
environment, (i) from the initial amount of kinetic energy imprinted by turbulence, (ii) because
of some density threshold which separates
two dominant mechanisms. In dense regions, above the threshold, star formation
is dominated by dissipation of compressible turbulence, whereas in regions below the threshold density, where the amount of turbulence is smaller, stars form in isolation and obey the
standard isothermal gravitational collapse scenario. As noted previously, however, the timescale
for star formation in these regions is much shorter than the one predicted by ambipolar diffusion
(Figure \ref{pasp_CMDclusters_col.ps}), implying that even in low-density star formation regions, the initial
collapse is triggered by large-scale turbulent dissipation, but the final cores are formed essentially by (dynamic) gravitational fragmentation in subsonic flows (see e.g. Hartmann 2002, Padoan et al. 2003b). Given the lower density in
these regions, we expect a larger mean thermal Jeans mass than in denser
regions, since $M_J\propto \rho^{-1/2}$, and thus a deficit of very-low-mass and
substellar objects.

\section{Summary and conclusion}

In this review, we have examined recent determinations of the IMF in various
components of the Galaxy, disk, spheroid, young and globular clusters.
Based on the most recent observations and state-of-the-art evolutionary models
for low-mass stars and brown dwarfs, we have determined the PDMF and IMF in these
different environments. As a general feature, we find that the IMF depends weakly
on the environment and is well described by a power-law form at $m\ga 1\,\msol$
and a lognormal form below this limit. The disk IMF, for isolated objects, has a characteristic mass around
$\sim 0.1\,\msol$ and a variance in logarithmic mass $\sigma \sim 0.7$, whereas the
disk IMF for mutiple systems has a characteristic mass $\sim 0.2\,\msol$  and a variance $\sigma \sim 0.6$.
These disk single and system MFs are consistent with a binary fraction among low-mass stars $\sim$50\%,
implying a fraction $\sim 20\%$ of BDs companions of M dwarfs,
 in agreement with present determinations. The results are consistent
with masses for the singles, primaries and companions drawn randomly from the same
underlying single IMF or similarly from a more or less uniform mass ratio distribution. The extension
of the single MF into the BD regime is in good agreement with present estimates of
L- and T-dwarfs densities, when considering all the uncertainties in these estimates.
This yields a disk BD number density comparable to the stellar one, namely $\sim 0.1$ pc$^{-3}$.
The IMF of several
young clusters is found to be consistent with this same field IMF, providing a
similar correction for unresolved binaries, confirming the fact that young
star clusters and disk field stars represent the same stellar population.
Dynamical effects, yielding depletion of the lowest-mass objects, are found to become
consequential for ages slightly older than the age of the Pleiades, i.e. $\ga$ 130 Myr.

The spheroid IMF relies on much less robust grounds. The large metallicity spread in
the photometric local sample, in particular, remains puzzling. Recent observations
suggest that there is a continuous kinematic shear between the thick-disk population,
present in the local samples, and the spheroid one, observed with the HST. This enables
us to derive only an upper limit for the spheroid mass contribution and IMF. This latter is found to be similar to the one derived for
globular clusters, and is well described also by a lognormal form, but
with a characteristic mass slightly larger than for the disk, around $\sim 0.2$-0.3 $\msol$.
Such an IMF excludes a significant population of BDs in globular clusters and in the spheroid,
i.e. in metal-depleted environments.
These results, however, remain hampered by large uncertainties such as the exact amount of dynamical evolution near the half-mass radius of a globular cluster, the exact identification of the genuine spheroid population, the exact fraction of binaries in globular cluster and spheroid populations.

The early star IMF,  representative of stellar populations formed at large redshift ($z\ga 5$), remains undetermined, but different observational constraints
suggest that it does not extend below $\sim 1\,\msol$. Whether it extends down to this mass range,
implying the existence of a primordial white dwarf population, or whether the cutoff for this primordial IMF occurs at
much larger masses remains unsettled. In any case, the baryonic content of the dark halo
represents very likely
at most a few percents of the Galactic dark matter. 

These determinations point to a characteristic mass for star formation which decreases with
time, from early star formation conditions  of temperature and metallicity to conditions characteristic of the spheroid or thick-disk environments, to present-day conditions. These results, however,
remain more suggestive than conclusive.
These IMFs
allow a reasonably robust determination of the
Galactic stellar and brown dwarf content. They have also important galactic implications beyond
the Milky Way in yielding more accurate mass-to-light ratio determinations.
The IMFs determined for the disk and the spheroid
yield mass-to-light ratios a factor of 1.8 to 1.4 smaller than for a Salpeter IMF, respectively, in agreement
with various  recent dynamical determinations.

This IMF determination is examined in the context of star formation theory. Theories
based on a pure Jeans-type mechanism, where fragmentation is due only to gravity, appear to
have difficulties explaining the determined IMF and various observational constraints on star formation.
On the other hand, recent numerical simulations of compressible turbulence, in particular in
super-Alfv\'enic conditions, reproduce qualitatively and reasonably quantitatively the
determined IMF, and thus provide an appealing solution.
In this picture, star formation is induced by the dissipation of large scale
turbulence to smaller scales, through radiative shocks, producing filamentary
structures. These shock produce local, non-equilibrium structures with large density contrasts.
Some of these dense cores then collapse eventually in gravitationally bound objects, under the
combined action of turbulence and gravity. The concept of a
single Jeans mass, however, is replaced by a distribution of local
Jeans masses, representative of the lognormal probability density function
of the turbulent gas. Cores exceeding the average Jeans mass ($\ga 1\, \msol$) naturally collapse
into stars under the action of gravity
whereas objects below this limit still have
a possibility to collapse, but with a decreasing probability, as gravity selects only the densest cores
in a certain mass range (the ones such that the mass exceeds the local Jeans mass $m_J$).
This picture, combining turbulence, as the initial mechanism for fragmentation, and gravity thus
provides a natural explanation for a scale free, power-law IMF
at large scales and a broad lognormal form below about 1 $\msol$. Additional mechanisms, such
as accretion, subfragmentation of the cores, multiplicity will not affect significantly the high-mass, power-law part
of the mass spectrum, but can modify the extension of its low-mass part. The initial level of turbulence in the cloud, and its initial density, can also affect the low-mass part of the IMF.

Future improvements, both on the theoretical and observational sides, should
confirm (or refute) this general scenario and help quantifying the details of the interaction
between turbulence and gravity, but it is encouraging to see that we are
now reaching a reasonable paradigm in our understanding of the Galactic
mass function over 5 orders of magnitudes, from very massive stars to Jupiter-like
objects, of the census of baryonic objects in the Galaxy, which can be
applied to external galaxies, and of the dominant physical mechanisms underlying the process of star formation.

{\bf Acknowledgments}:

It is a great pleasure for me to thank many colleagues who contributed to this
review. A special thank to D. Barrado y Navascues, V. B\'ejar,  J. Bouvier, A. Burgasser, F. Comer\'on, C. Dahn, P. Dobbie, N. Hambly,
H. Harris, D. Kirkpatrick, K. Luhman, E. Moraux, for sending their data and in some cases sharing unpublished results.
My profound gratitude to S. Charlot for calculating the mass-to-light ratios in \S 6. Finally, I am
deeply indebted to  I. Baraffe, J. Bouvier,  G. DeMarchi, L. Hartmann, P. Kroupa,  R. Larson, \AA. Nordlund,
for numerous discussions, highly valuable comments and for reading a preliminary version of this review.
Finally a special thank to Anne Cowley and David Hartwick, editors of PASP, for their astronomical patience
before they received the present manuscript.

\eject

\eject

\begin{table}
\caption[]{Disk initial mass function (IMF) and present-day mass function (PDMF) for single
objects. For unresolved binary systems, the coefficients are
given by eq.[\ref{IMFsys}].The normalization coefficient
$A$ is in ${\rm pc^{-3}}({\log\,\msol})^{-1}$. }
\bigskip
\begin{tabular}{lccc}
\tableline
&  $m\le 1.0\,\msol$ & $m> 1.0\,\msol$ &  \\
& $\xi(\log \,m)=A\, \,exp\{-{(\log \, m\,\,-\,\,\log \, m_c)^2\over 2\,\sigma^2}\}$ & $\xi(\log \,m)=A\,m^{-x}$ \\
\hline \\
\mbox{}\hspace{0.2cm} IMF    & A=0.158$^{+0.051}_{-0.046}$          & A=4.43$\times 10^{-2}$    \\
\mbox{}\hspace{0.2cm}        & $m_c=0.079^{-0.016}_{+0.021}$   & x=1.3 $\pm 0.3$ \\
\mbox{}\hspace{0.2cm}        & $\sigma=0.69_{+0.05}^{-0.01}$       &  \\
\mbox{}\hspace{0.2cm} PDMF  & A=0.158$^{+0.051}_{-0.046}$ & $\,\,\,\,0\le \log \,m \le 0.54\,:\,A=4.4\times 10^{-2}\,\,$, x=4.37 \\
\mbox{}\hspace{0.4cm}      & $m_c=0.079^{-0.016}_{+0.021}$  & $0.54\le \log \,m \le 1.26\,:\,A=1.5\times 10^{-2}\,\,$, x=3.53 \\
\mbox{}\hspace{0.4cm}      & $\sigma=0.69_{+0.05}^{-0.01}$ & $1.26\le \log \,m \le 1.80\,:\,A=2.5\times 10^{-4}\,\,$, x=2.11 \\
\tableline
\end{tabular}
\label{table.diskIMF}
\end{table}
\clearpage\eject

\begin{table}
\caption[]{Initial mass functions for the various components of the Galaxy.}
\bigskip
\begin{tabular}{lccc}
\tableline
&  $m\le m_{norm}$ $[\msol]$ & $m> m_{norm}$ $[\msol]$&  \\
& $\xi(\log \,m)=A\, \,exp\{-{(\log \, m\,\,-\,\,\log \, m_c)^2\over 2\,\sigma^2}\}$ & $\xi(\log \,m)=A\,m^{-x}$ \\
\hline \\
\mbox{}\hspace{0.2cm} Disk and young clusters & $m_{norm}=1.0$ & $m_{norm}=1.0$ \\
\mbox{}\hspace{0.2cm}  & A=0.158$^{+0.051}_{-0.046}$ & A=4.4$\times 10^{-2}$   \\
\mbox{}\hspace{0.2cm}  & $m_c=0.079^{-0.016}_{+0.021}$  & x=1.3 $\pm 0.3$ \\
\mbox{}\hspace{0.2cm}  & $\sigma=0.692^{-0.01}_{+0.05}$ &  \\
\mbox{}\hspace{0.2cm} Globular clusters    & $m_{norm}=0.9$ & $m_{norm}=0.9$ \\
\mbox{}\hspace{0.2cm}  &  $m_c=0.33 \pm 0.03$   &  x=1.3 $\pm 0.3$ \\
\mbox{}\hspace{0.2cm}  & $\sigma=0.34\pm 0.04$  &  \\
\mbox{}\hspace{0.2cm} Spheroid   & $m_{norm}=0.7$  &    $m_{norm}=0.7$ \\
\mbox{}\hspace{0.2cm}  & A=$(3.6\pm 2.1) \times 10^{-4}$ & A=7.1$\times 10^{-5}$   \\
\mbox{}\hspace{0.2cm}  & $m_c=0.22 \pm 0.05$   & x=1.3 $\pm 0.3$ \\
\mbox{}\hspace{0.2cm}  & $\sigma=0.33\pm 0.03$ &  \\
\tableline
\end{tabular}
\label{table.IMF}
\end{table}
\clearpage\eject

\begin{table}
\caption[]{Present day stellar and brown dwarf Galactic budget. The number densities $n$ are in [pc$^{-3}$], the mass
densities $\rho$ are in [$\mvol$].  } 
\bigskip
\begin{tabular}{lccc}
\tableline
&  Disk   & Spheroid &   Dark halo  \\
\hline \\
\mbox{}\hspace{0.2cm} $n_{BD}$  & 0.13$\pm$0.06 & $\la 3.5\times 10^{-5}$ & \\
\mbox{}\hspace{0.2cm} $\rho_{BD}$ & ($0.4\pm 0.2)\times 10^{-2}$ &  $\la 2.3\times 10^{-6}$ & \\
\mbox{}\hspace{0.2cm} $n_{\star}(\le 1\,\msol)$ & $0.13\pm 0.02$ &  $\le (2.4\pm 0.1)\times 10^{-4}$ & \\
\mbox{}\hspace{0.2cm} $\rho_{\star}(\le 1\,\msol)$ & $(3.5\pm 0.3)\times 10^{-2}$  & $\le (6.6\pm 0.7)\times 10^{-5}$ & $\ll 10^{-5}$ \\
\mbox{}\hspace{0.2cm} $n_{\star}(> 1\,\msol)$ & $0.4\times 10^{-2}$ & 0  & \\
\mbox{}\hspace{0.2cm} $\rho_{\star}(> 1\,\msol)$ & $0.6\times 10^{-2}$ & 0  & \\
\mbox{}\hspace{0.2cm} $n_{rem}$ & ($0.7\pm 0.1)\times 10^{-2}$  & $\le (2.7\pm 1.2)\times 10^{-5}$  & ? \\
\mbox{}\hspace{0.2cm} $\rho_{{rem}}$ & ($0.6\pm 0.1)\times 10^{-2}$  & $\le (1.8\pm 0.8)\times 10^{-5}$ &  \\
\mbox{}\hspace{0.2cm} $n_{{tot}}$ & $0.27\pm 0.06$ &  $\le 3.0\times 10^{-4}$ &  \\
\mbox{}\hspace{0.2cm} $\rho_{{tot}}$ &($5.1\pm 0.3)\times 10^{-2}$ &   $\le (9.4\pm 1.0)\times 10^{-5}$ & $\ll 10^{-5}$\\
\hline \
\mbox{}\hspace{0.2cm} $\mathcal{N}$(BD);  $\mathcal{M}$(BD)     & 0.48; 0.08  &  0.10; 0.03 &  \\
\mbox{}\hspace{0.2cm} $\mathcal{N}$(LMS);  $\mathcal{M}$(LMS)   &0.48; 0.68   & 0.80; 0.77 &  \\
\mbox{}\hspace{0.2cm} $\mathcal{N}$(IMS);  $\mathcal{M}$(IMS)  &0.015; 0.11 &  0.; 0. &  \\
\mbox{}\hspace{0.2cm} $\mathcal{N}$(HMS); $\mathcal{M}$(HMS) & 0; 0             & 0.; 0.&   \\
\mbox{}\hspace{0.2cm} $\mathcal{N}$(rem.); $\mathcal{M}$(rem.) & 0.025; 0.13 &  0.10; 0.20   & ? \\
\tableline
\end{tabular}
\label{table.budget}
\end{table}
\clearpage\eject

\begin{table}
\caption[]{Initial stellar and brown dwarf Galactic budget (from Table 2). } 
\bigskip
\begin{tabular}{lcccc}
\tableline
& Disk &  Spheroid  \\
\hline \\
\mbox{}\hspace{0.2cm} $n_{BD}$  & 0.13$\pm$0.06 & $\sim 3.5\times 10^{-5}$  \\
\mbox{}\hspace{0.2cm} $\rho_{BD}$ & ($0.4\pm 0.2)\times 10^{-2}$ & $\sim 2.3\times 10^{-6}$  \\
\mbox{}\hspace{0.2cm} $n_{\star}(\le 1\,\msol)$ & $0.13\pm 0.02$ & $\le (2.4\pm 0.1)\times 10^{-4}$  \\
\mbox{}\hspace{0.2cm} $\rho_{\star}(\le 1\,\msol)$ & $(3.5\pm 0.3)\times 10^{-2}$ & $\le (6.6\pm 0.7)\times 10^{-5}$  \\
\mbox{}\hspace{0.2cm} $n_{\star}(> 1\,\msol)$ & $1.5\times 10^{-2}$ & $\le 2.3 \times 10^{-5}$ \\
\mbox{}\hspace{0.2cm} $\rho_{\star}(> 1\,\msol)$ & $4.7\times 10^{-2}$ & $\le 1.0 \times 10^{-4}$ \\
\mbox{}\hspace{0.2cm} $n_{{tot}}$ &  $0.27$ & $\le 3.0 \times 10^{-4}$  \\
\mbox{}\hspace{0.2cm} $\rho_{{tot}}$ &($8.6\pm 0.3)\times 10^{-2}$ & $\le 2.0 \times 10^{-4}$  \\
\hline \
\mbox{}\hspace{0.2cm} $\mathcal{N}$(BD);  $\mathcal{M}$(BD)     & 0.48; 0.04 & 0.10; 0 \\
\mbox{}\hspace{0.2cm} $\mathcal{N}$(LMS);  $\mathcal{M}$(LMS)   & 0.48; 0.41 & 0.80; 0.48 \\
\mbox{}\hspace{0.2cm} $\mathcal{N}$(IMS);  $\mathcal{M}$(IMS)  & 0.04; 0.35 & 0.09; 0.34  \\
\mbox{}\hspace{0.2cm} $\mathcal{N}$(HMS); $\mathcal{M}$(HMS)  & 0; 0.20 & 0; 0.18  \\
\tableline
\end{tabular}
\label{table.budget}
\end{table}
\clearpage\eject

\begin{figure}
\centerline{\it Figure Legends}

\caption[]{Disk mass function derived from the local V-band LF (circles and solid line)
and K-band LF (squares and dash-line).
The solid line and the two surrounding dash-lines display the lognormal form given by
eq.[\ref{IMFdisk}], whereas the dotted line illustrates the 4-segment power-law form of Kroupa (2002).
The empty circles and squares for $\log \, m\ge -0.15$
display the MF obtained for $t=10$ Gyr and 1 Gyr, respectively, illustrating the age uncertainty on the MF
for $m>0.7\,\msol$. The empty triangles and dotted error bars display the MF obtained from the bulge LF (see text).}
\label{pasp_MF1_col.ps}
\end{figure}

\begin{figure}
\caption[]{Disk mass function derived from the system K-band LF (solid squares and solid line) and the
HST corrected MF (triangles and short-dash-line) from Zheng et al. (2001).
 The solid line and surrounding dotted lines display the lognormal form given by
eq.[\ref{IMFdisk}] for single objects, as in Figure \ref{pasp_MF1_col.ps}, whereas the dash-line illustrates
the lognormal form given by
eq.[\ref{IMFsys}]. }
\label{pasp_MFbin_col.ps}
\end{figure}

\begin{figure}
\caption[]{Luminosity functions for the Galactic disk predicted with the IMF
(eq.[\ref{IMFdisk}]) and a constant SFR. Solid: stars+BDs; long-dash:
BDs only ($m\le 0.072\,\msol$); long dash-dot: T-dwarfs only  (J-H$<$0.5 and
H-K$<$0.5); short dash-dot: objects below the D-burning minimum mass
($m\le 0.012\,\msol$). The short-dash lines illustrate the range of uncertainty
in the IMF (eq.[\ref{IMFdisk}]). The dotted line in the middle panel displays the result
obtained with a power-law IMF with $x=0$ ($\xi(\log \, m)=constant$) with the
same normalisation at 0.1 $\msol$ as IMF (eq.[\ref{IMFdisk}]).
The dotted line in the bottom panel displays the results obtained with the
system IMF (eq.[\ref{IMFsys}]).
The histogram displays the nearby LF (Henry \& McCarthy 1990).
Empty and filled squares are estimated L-dwarf densities by Gizis et al. (2000)
and Kirkpatrick (1999, 2000)+Burgasser (2001), respectively. Triangles are estimated T-dwarf densities
from Burgasser (2001).
}
\label{pasp_LFcounts.ps}
\end{figure}

\begin{figure}
\caption[]{Pleiades Mass Function calculated with the Baraffe et al. (1998)
and Chabrier et al. (2000) MMRs, from various observations : squares : Hambly et al. (1999); 
triangles : Dobbie et al. (2002b); circles : Moraux et al. (2003).
The short-dash and long-dash lines display the single (eq.[\ref{IMFdisk}]) and system
(eq.[\ref{IMFsys}]) field MFs,
respectively, arbitrarily normalized to the present data.}
\label{pasp_MFPleiades_col.ps}
\end{figure}

\begin{figure}
\caption[]{Mass Function calculated for various young clusters with the Baraffe et al. (1998)
and Chabrier et al. (2000) MMRs, from various data. Solid circles: $\sigma$-Or
(B\'ejar et al. 2001); filled squares : $\alpha$-Per (Barrado y Navascues et al. 2002); empty symbols : Pleiades (Hambly et al., 1999, squares; Moraux et al.,
2003, circles;
Dobbie et al., 2002b, triangles); filled triangles : M35 (Barrado y Navascues et
al. 2002). The ages for each cluster are indicated. The dashed line illustrates the field
system MF (eq.[\ref{IMFsys}]), while the dotted lines display
various power-law segments $\xi(m)=dN/d\log m\propto m^{-x}$, as derived by the aforementioned authors, with : $x=-0.2$ ($\sigma$-Or), $x=-0.4$ ($\alpha$-Per), $x=-0.4$
(Pleiades), $x=-0.2$ and -1.9 (M35).
}
\label{pasp_MFclusters.ps}
\end{figure}

\begin{figure}
\caption[]{V-band luminosity function of the Galactic spheroid (stellar halo).
Solid dots (solid line)
: Dahn et al. (1995) completed by Dahn \& Harris (2002, private communication);
solid squares (dash-dot line): NLTT survey (Gould 2003);
open circles (long-dash line): Bahcall \& Casertano (1986); triangles (short-dash line):
HST (Gould et al. 1998)).
All LFs have been recalculated with a completion factor based on the Casertano et al. (1990)
kinematic model (see text).}
\label{pasp_LFsph.ps}
\end{figure}

\begin{figure}
\caption[]{Color-magnitude diagram for the Dahn et al. (1995) subdwarf sample
with $v_\perp \ge 220\,\kms$. Superimposed, are the 10 Gyr isochrones of
Baraffe et al. (1997), for $\mh=$-2.0, -1.5, -1.3, -1.0 and -0.5, from left to right.
The small dots indicate the Monet et al. (1992) solar-metallicity local sample.}
\label{pasp_CMDsph.ps}
\end{figure}

\begin{figure}
\caption[]{Mass function of the Galactic spheroid, based on the NLTT (Gould 2003) LF (solid
lines)
and the Baraffe et al. (1997)
mass-$\mv$ relationship, for three metallicities: $\mh=$-1.5 (triangles), -1.0 (circles) and -0.5 (squares), respectively.
Dotted line: same calculation based on Dahn et al. (1995, 2002) LF and $\mh=-1.0$ models.
Solid line : parametrization given by eq.[\ref{IMFsph}])
}
\label{pasp_MFsph_col.ps}
\end{figure}

\begin{figure}
\caption[]{Mass Function calculated for various globular clusters with the Baraffe et al. (1997)
MMRs in several bandpasses, from the LFs of Paresce \& DeMarchi (2000), spanning a metallicity range $-2.0\la \mh \la -1.0$. The
short-dash and dotted lines display the IMF (eq.[\ref{IMFGC}]) and 
(eq.[\ref{IMFsph}]), respectively, whereas the long-dash and dot-dash lines at the top illustrate the disk
single IMF (eq.[\ref{IMFdisk}]) and system IMF (eq.[\ref{IMFsys}]), respectively.
}
\label{pasp_MFGC_col.ps}
\end{figure}

\begin{figure}
\caption[]{Mass-to-light ratios in various passbands, in units of stellar mass per solar luminosity
in the considered band, calculated wiht the Salpeter IMF (dotted line), the disk IMF (eq.[\ref{IMFdisk}])
(solid line),
the spheroid IMF (eq.[\ref{IMFsph}]) (long-dash line) and the top-heavy IMF  (eq.[\ref{IMFhalo}]) (short-dash line). The calculations correspond to
Simple Stellar Populations (SSP), i.e. a stellar birthrate parameter $b=0$ (see text).
All IMFs are normalized to $\int_{0.01}^{100} m\,(dN/dm)\,dm=1$. Courtesy of S. Charlot.}
\label{pasp_ML_col.ps}
\end{figure}

\begin{figure}
\caption[]{Hertzsprung-Russell diagram for young objects in Chameleon (circles), Lupus (triangles),
IC348 (squares) from Comer\'on et al. (2003) and in Taurus (dots) from
Brice\~no et al. (2002).
Superposed are various isochrones of Baraffe et al. (2002), for $\tau=10^6,\,2\times 10^6,\,3\times 10^6,\,5\times 10^6,\,10^7$ yr from top to bottom,
for different masses, as indicated.
}
\label{pasp_CMDclusters_col.ps}
\end{figure}


\vfill\eject
\clearpage

\begin{figure}
\begin{center}
\epsfxsize=190mm
\epsfysize=200mm
\epsfbox{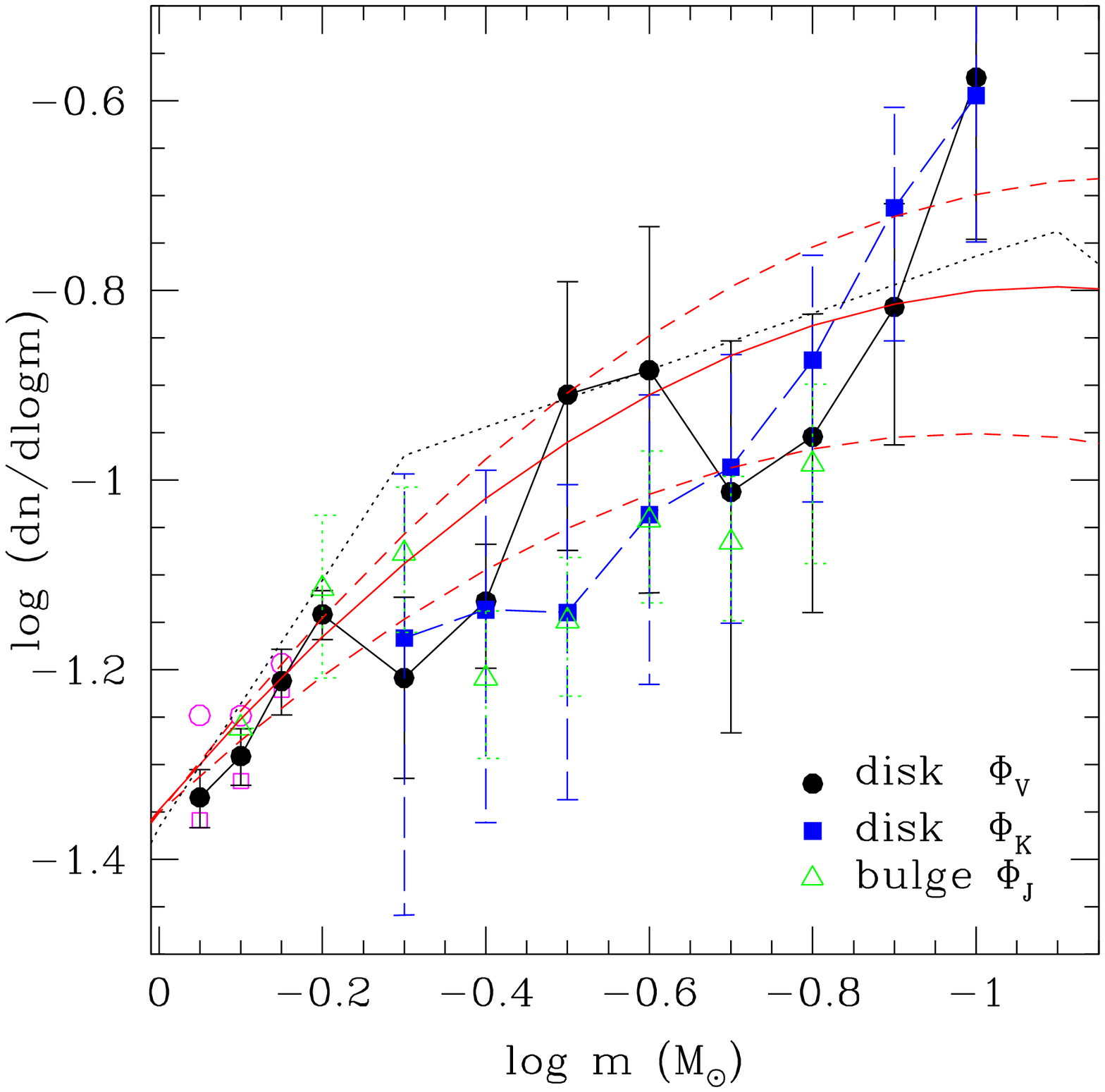}
\end{center}
\end{figure}

\vfill\eject

\begin{figure}
\begin{center}
\epsfxsize=190mm
\epsfysize=200mm
\epsfbox{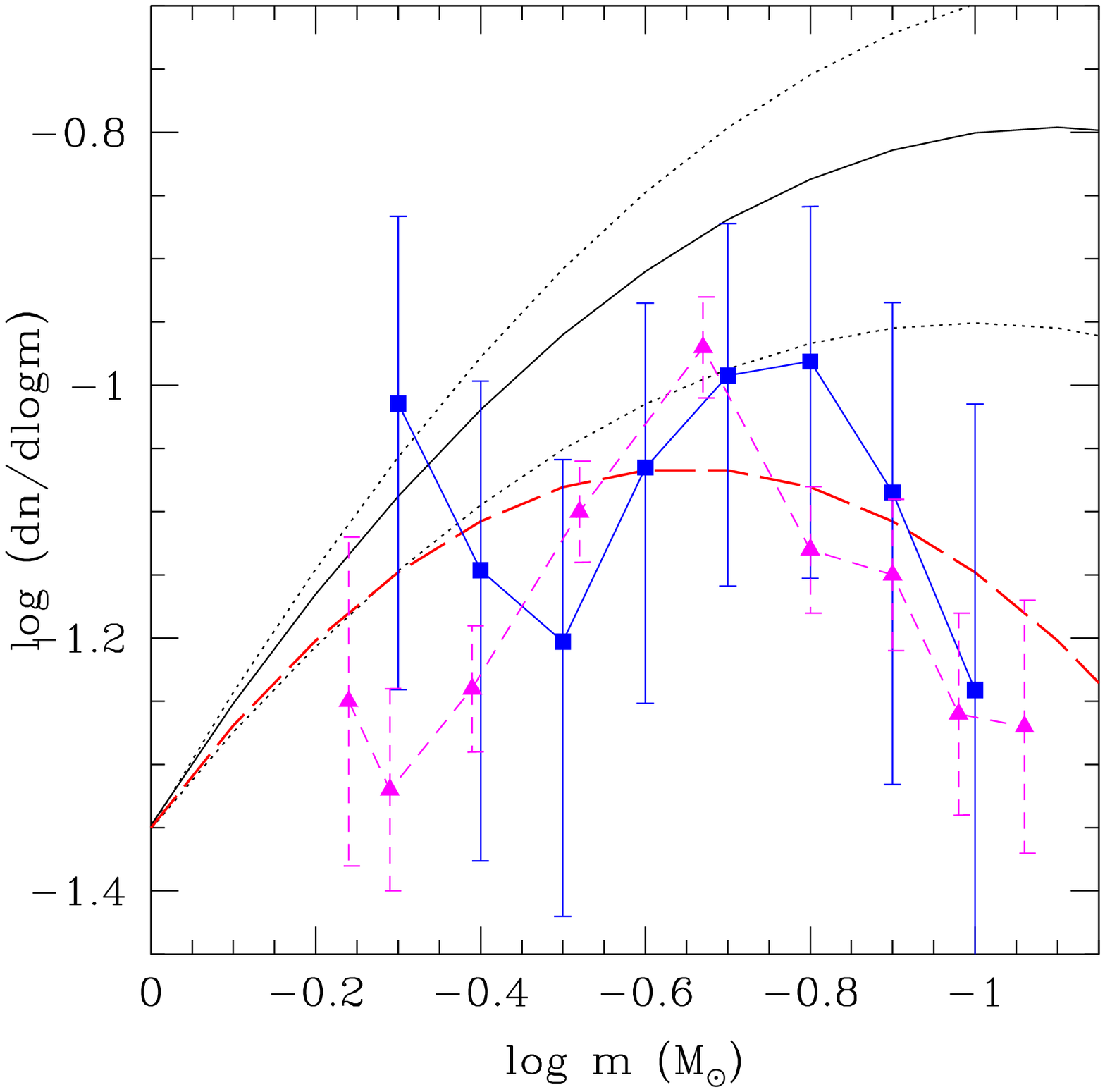}
\end{center}
\end{figure}

\vfill\eject

\begin{figure}
\begin{center}
\epsfxsize=190mm
\epsfysize=200mm
\epsfbox{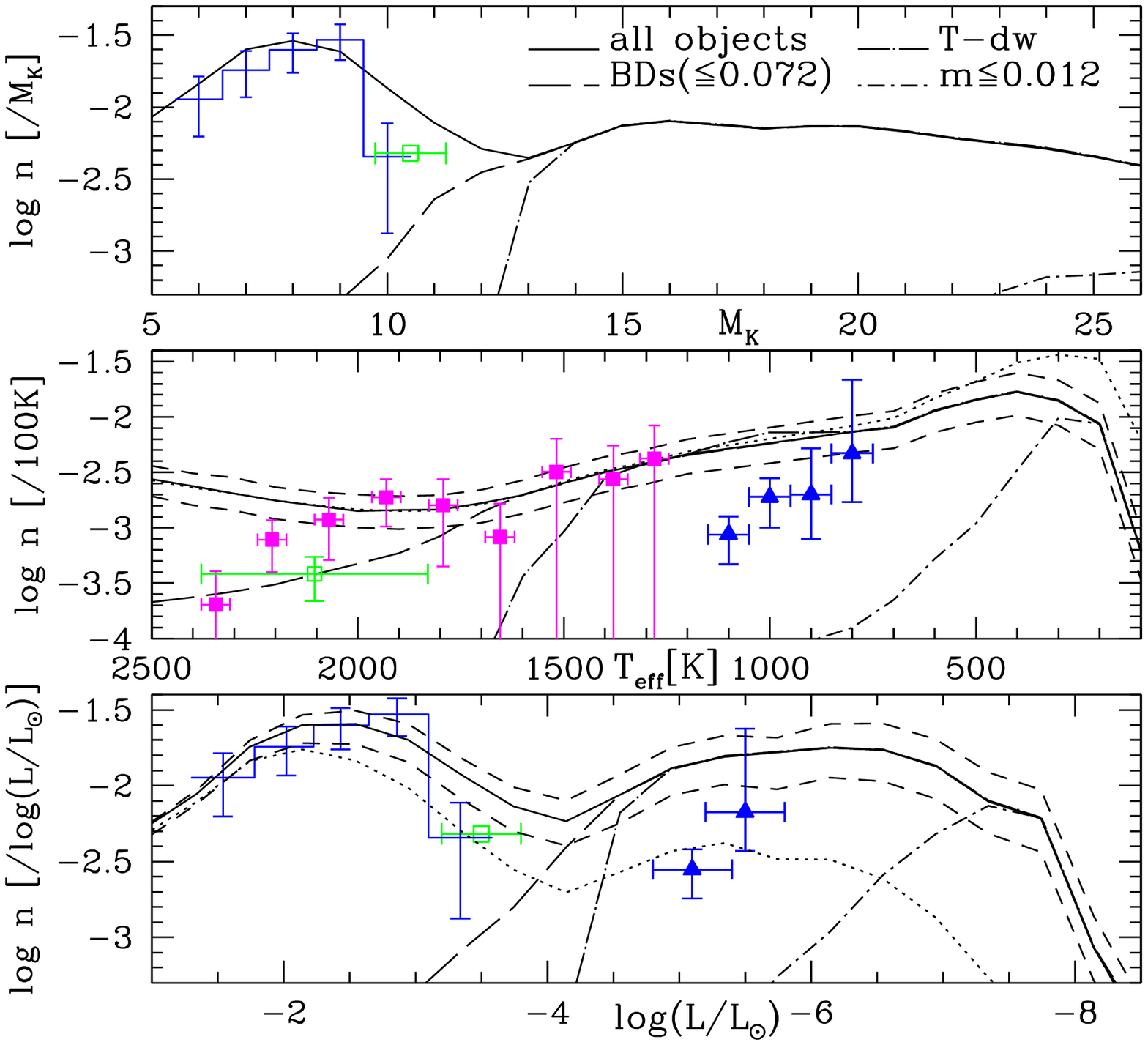}
\end{center}
\end{figure}

\vfill\eject

\begin{figure}
\begin{center}
\epsfxsize=190mm
\epsfysize=200mm
\epsfbox{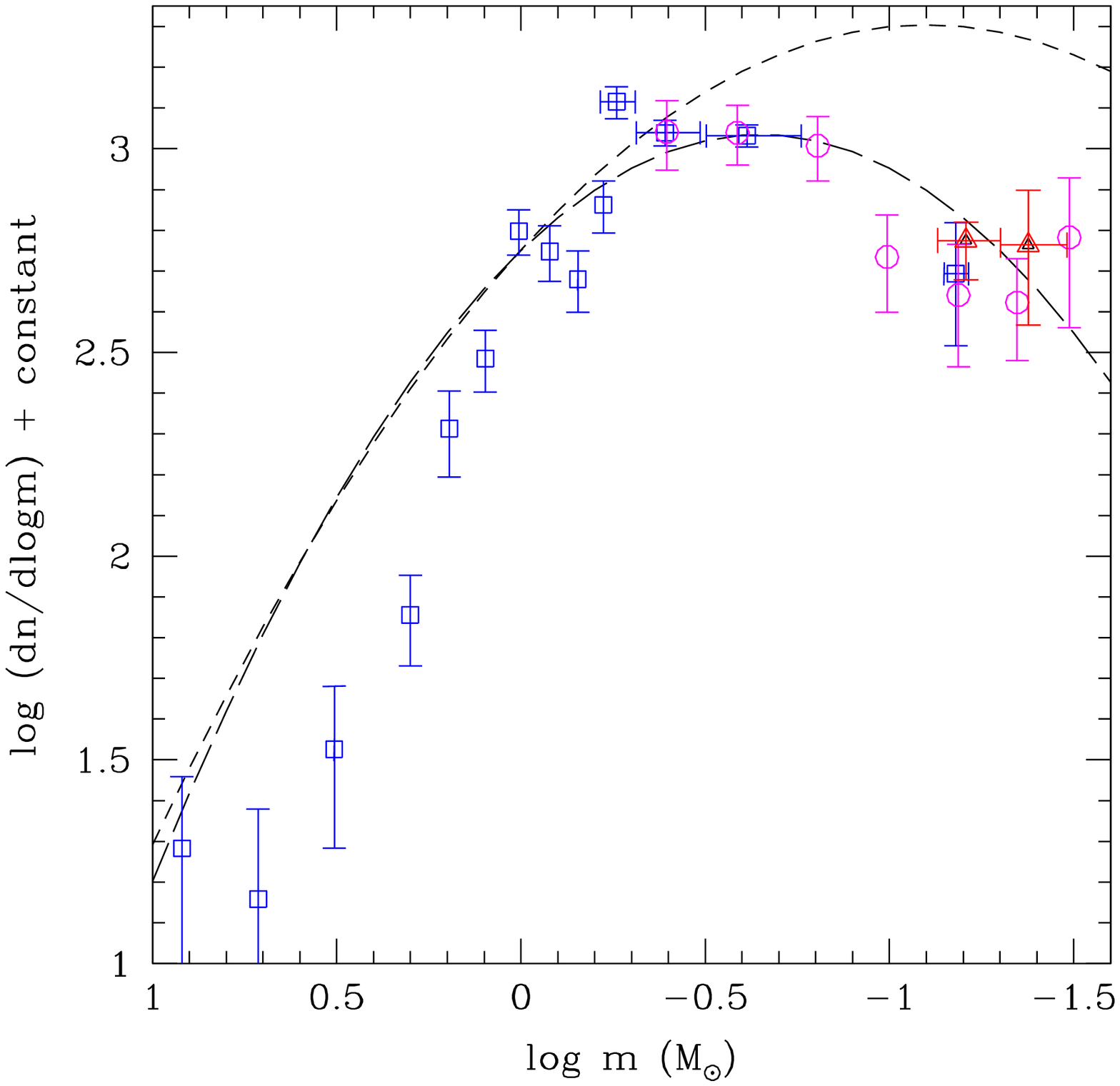}
\end{center}
\end{figure}

\vfill\eject

\begin{figure}
\begin{center}
\epsfxsize=190mm
\epsfysize=200mm
\epsfbox{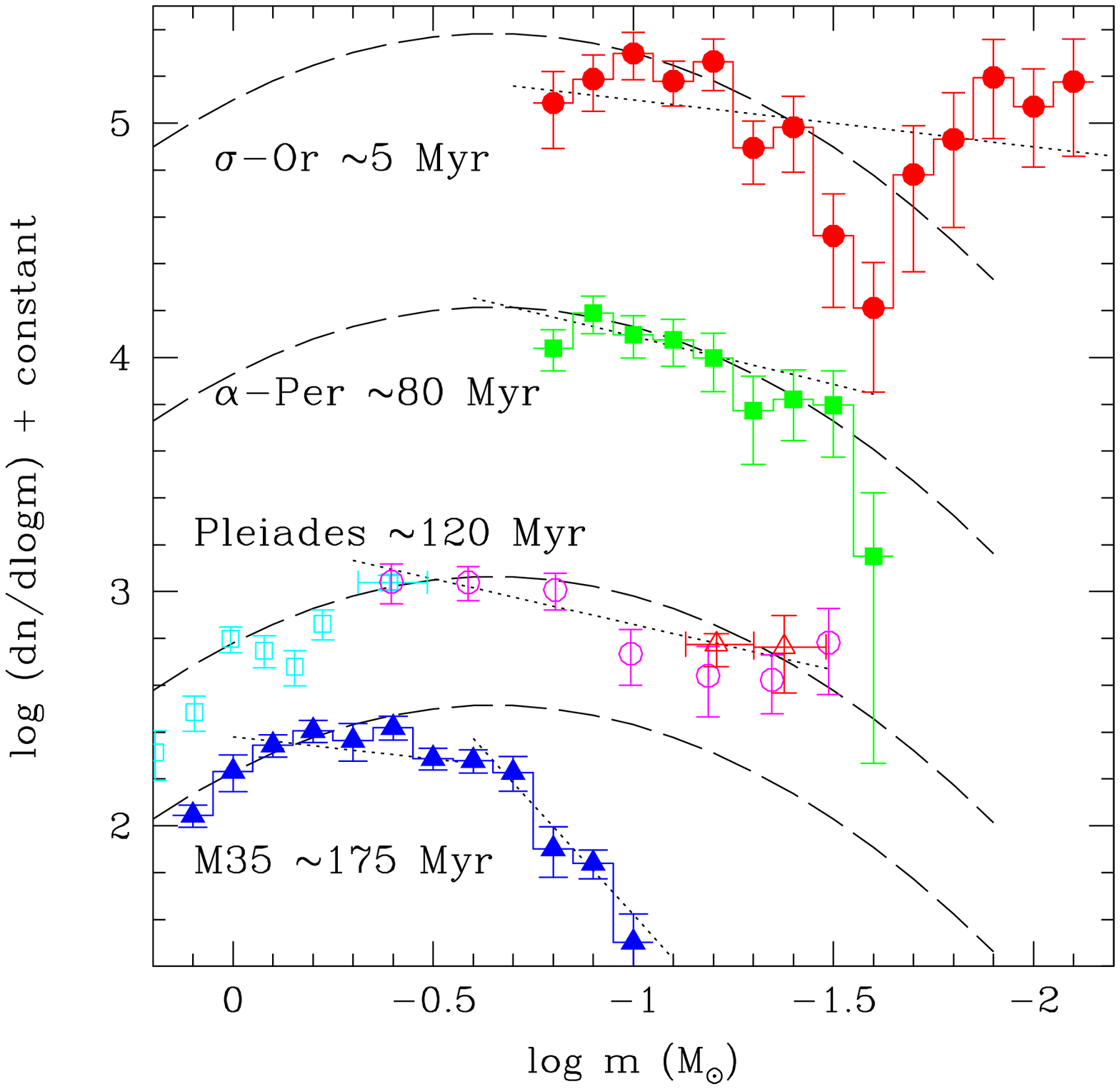}
\end{center}
\end{figure}

\vfill\eject

\begin{figure}
\begin{center}
\epsfxsize=190mm
\epsfysize=200mm
\epsfbox{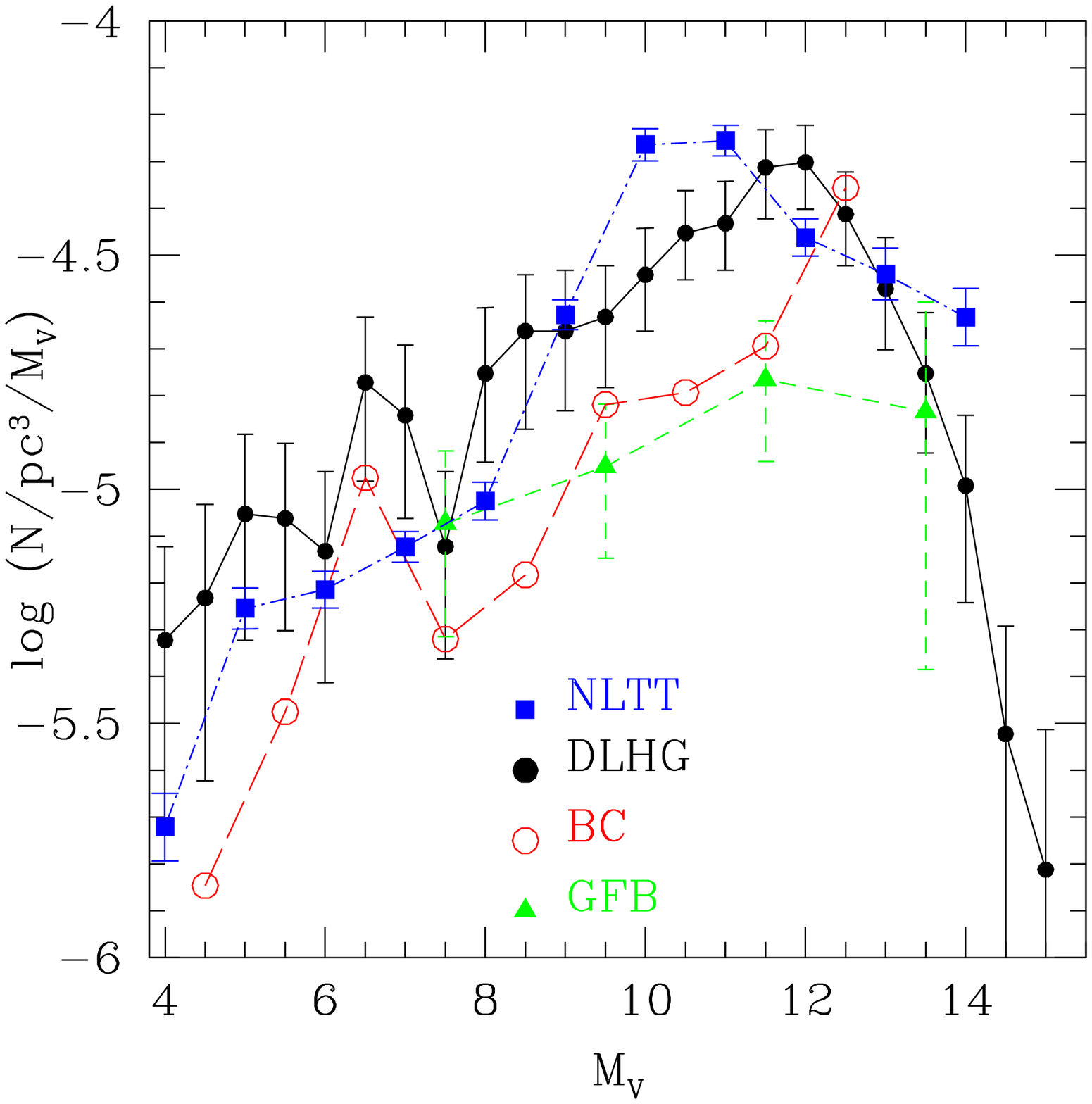}
\end{center}
\end{figure}

\vfill\eject

\begin{figure}
\begin{center}
\epsfxsize=190mm
\epsfysize=200mm
\epsfbox{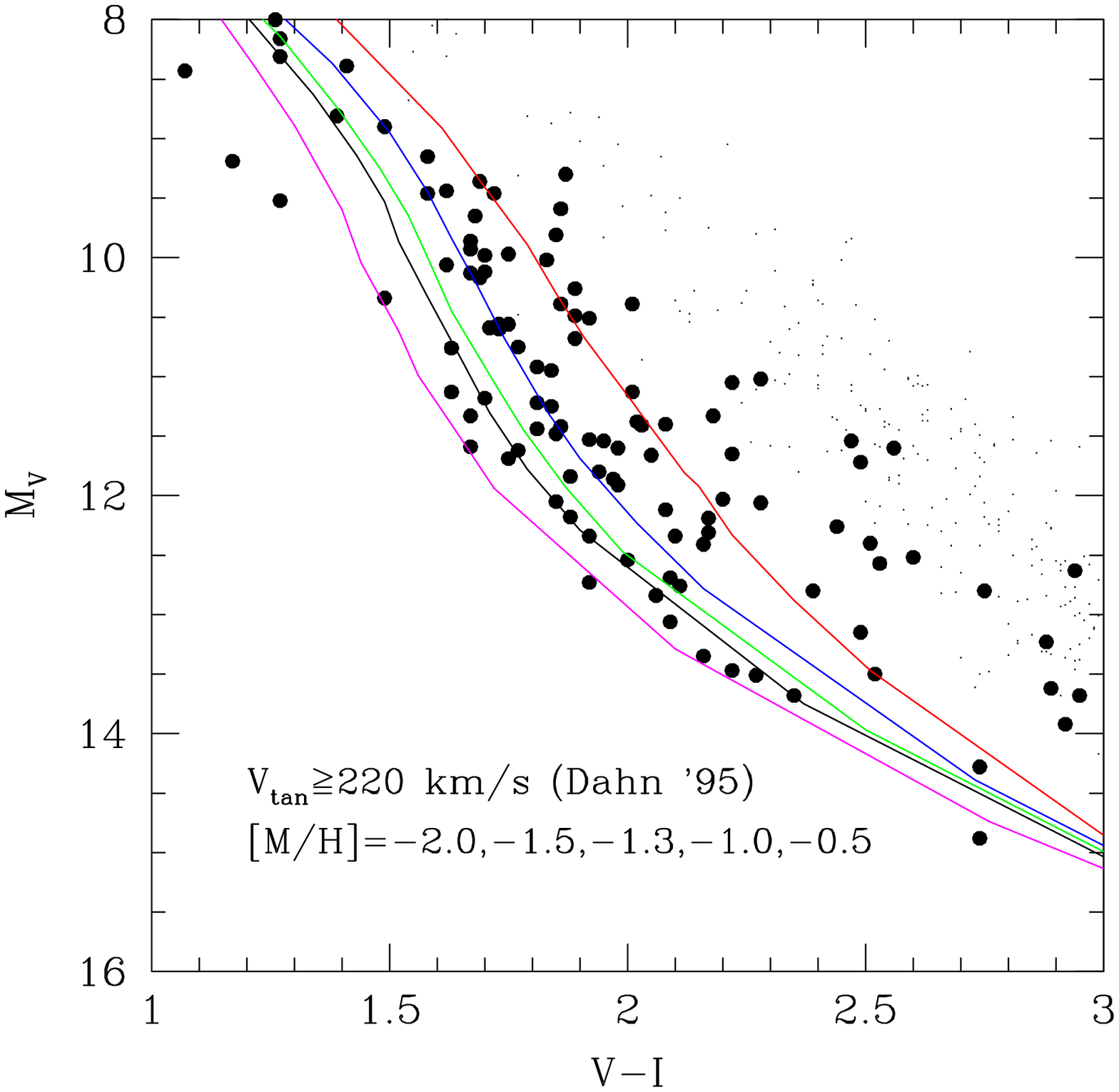}
\end{center}
\end{figure}

\vfill\eject

\begin{figure}
\begin{center}
\epsfxsize=190mm
\epsfysize=200mm
\epsfbox{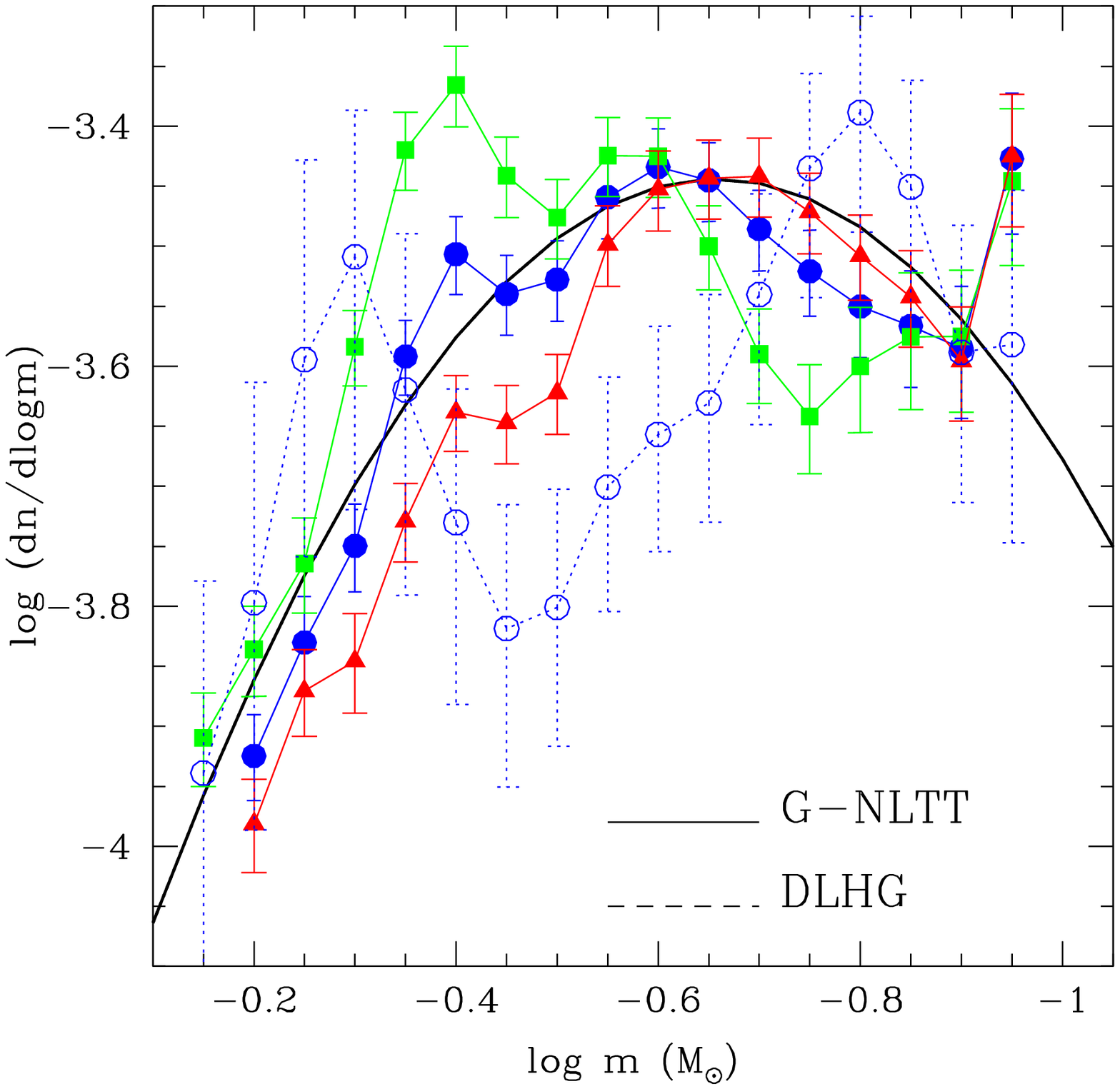}
\end{center}
\end{figure}

\vfill\eject

\begin{figure}
\begin{center}
\epsfxsize=190mm
\epsfysize=200mm
\epsfbox{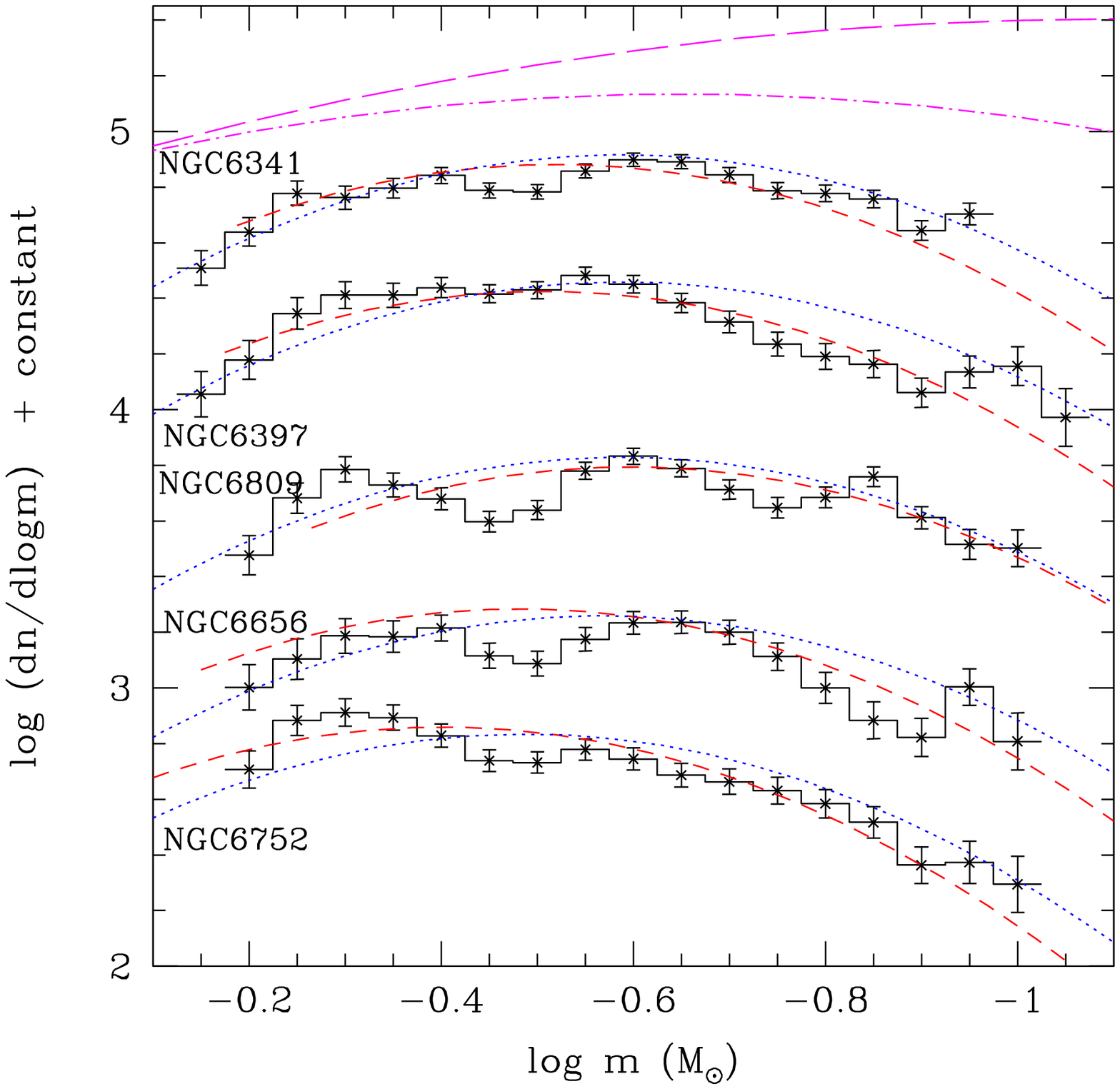}
\end{center}
\end{figure}

\vfill\eject

\begin{figure}
\begin{center}
\epsfxsize=190mm
\epsfysize=200mm
\epsfbox{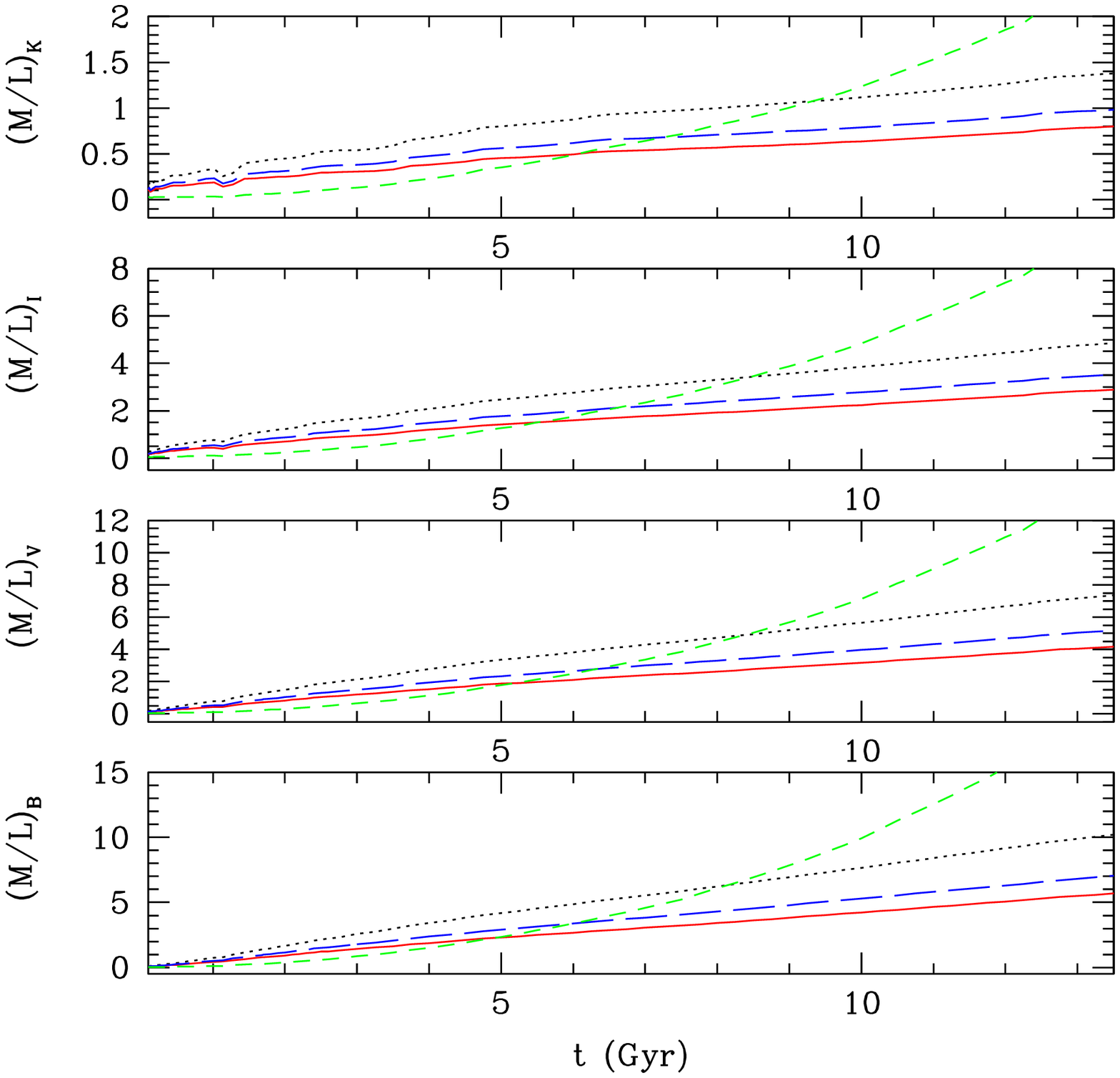}
\end{center}
\end{figure}

\vfill\eject

\begin{figure}
\begin{center}
\epsfxsize=190mm
\epsfysize=200mm
\epsfbox{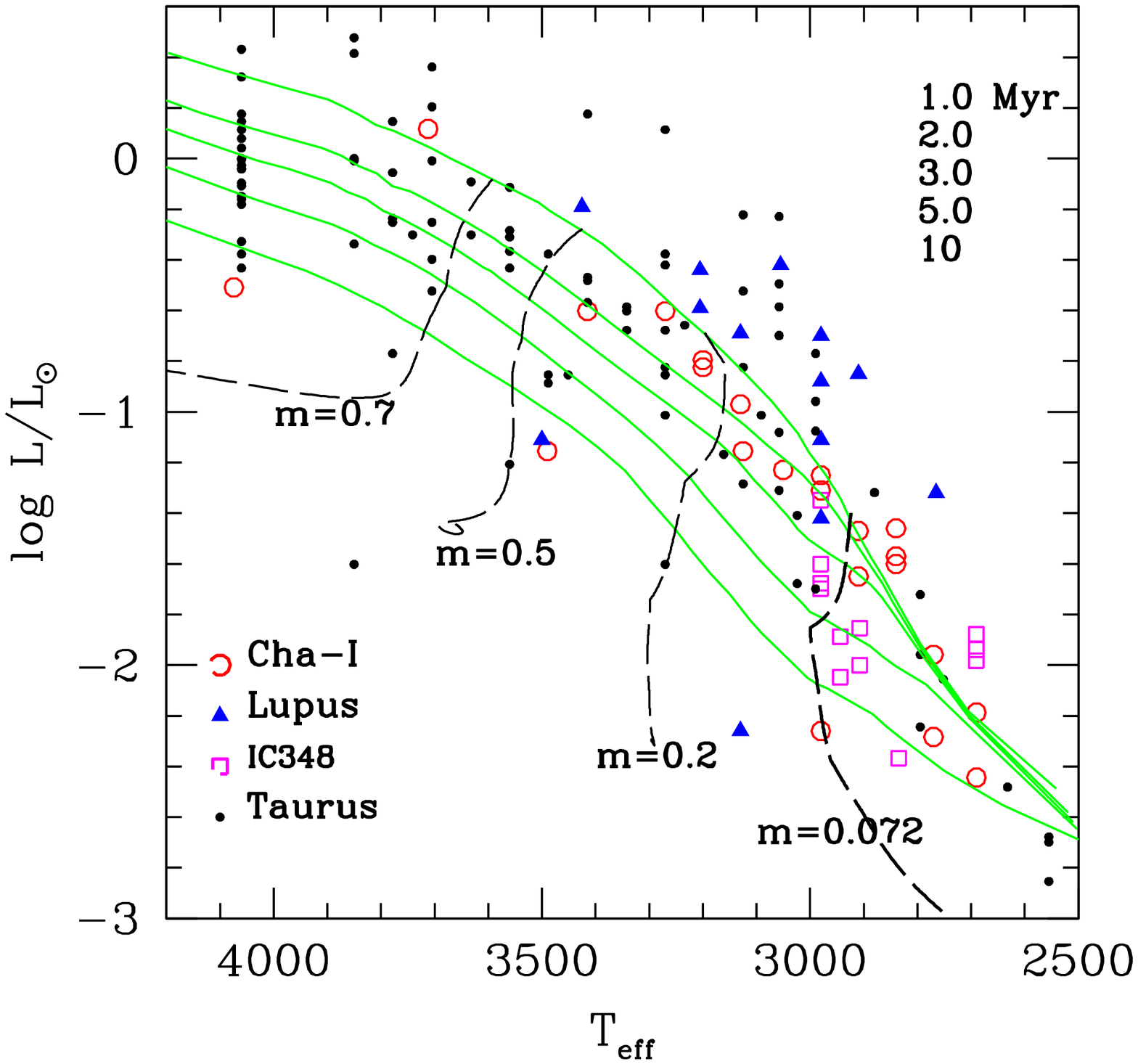}
\end{center}
\end{figure}

\end{document}